\documentclass[10pt,journal,compsoc]{IEEEtran}

  \usepackage{cite}
\usepackage[dvipsnames,monochrome]{xcolor}

\usepackage{todonotes}
\usepackage{xfrac}
\usepackage{xurl}
\usepackage{hyperref}
\usepackage{booktabs}

\usepackage{tikz}
\usepackage{tabularx}
\usepackage{listings}
\usepackage{nicefrac}
\usepackage{multirow}
\usepackage{hhline}
\usepackage{graphicx}
\usepackage{calc}
\usepackage{array}
\usepackage{soul}
\usepackage{proof}
\usepackage{algorithm}
\usepackage{algorithmic}
\usepackage[export]{adjustbox}
\usepackage{verbatim}
\usepackage{stmaryrd}
\usepackage{amsfonts}
\usepackage{mathtools}
\usepackage{caption}
\usepackage{subcaption}
\usepackage{mathabx}
\usepackage[framemethod=TikZ]{mdframed}
\usepackage[inline,shortlabels]{enumitem}
\setlist[enumerate,1]{label={(\arabic*)}}
\usepackage[normalem]{ulem}
\usepackage{transparent}
\usepackage{arydshln}
\usepackage{makecell}
\usepackage{amsmath}
\usepackage{amsthm}
\usepackage{textcomp}
\usepackage{verbatimbox}
\usepackage{wasysym}

\usepackage{fancyvrb}
\usepackage{comment}

\definecolor{keywordcolor}{rgb}{0.5,0,0.1}
\definecolor{stringcolor}{rgb}{0,0,1}
\definecolor{emphColor}{rgb}{0,0,0}
\definecolor{codecommentcolor}{rgb}{0.03,0.6,0}
\definecolor{codegreen}{rgb}{0,0.6,0}
\definecolor{codegray}{rgb}{0.5,0.5,0.5}
\definecolor{codepurple}{rgb}{0.58,0,0.82}
\definecolor{mediumgray}{rgb}{0.80, 0.80, 0.80}
\definecolor{listinggray}{rgb}{0.9,0.9,0.9}

\lstdefinelanguage{vql}{
    morekeywords={@QueryBasedFeature,@Constraint,count,pattern,private,neg,find,import,true,false,or,check,job,action,state,severity,
    message,oclIsKindOf,self,exists,includes,invariant,class},
    sensitive=true, 
    morecomment=[l]{//},
    morecomment=[s]{/*}{*/},
	  morestring=[b]{"},
}

\lstdefinestyle{mystyle}{
  breaklines=true,
  showstringspaces=false,
  basicstyle=\ttfamily\scriptsize,
  identifierstyle=\color{black},
  stringstyle=\color{orange},
  columns=flexible,
  keywordstyle=[0]*\bfseries\color{blue!70!black},
  keywordstyle=[1]\bfseries\color{blue!70!black}, 
  commentstyle=\color{codecommentcolor},
  backgroundcolor=\color{black!2},
  frame=single, 
  numbers=left,
  numbersep=5pt,
  numberstyle=\ssmall\color{gray},
  captionpos=b,
  xrightmargin=3pt,
  xleftmargin=3pt,
  keepspaces=true,
  showspaces=false,
  breakatwhitespace=false,
  tabsize=2,
  showtabs=false,
}
\surroundwithmdframed[
  topline=false,
  rightline=false,
  bottomline=false,
  leftmargin=\parindent,
  skipabove=\medskipamount,
  skipbelow=\medskipamount,
  linewidth=4pt,
  linecolor={black},
  backgroundcolor={black!5},
  leftmargin=0cm
]{proposition}

\surroundwithmdframed[
  topline=false,
  rightline=false,
  bottomline=false,
  leftmargin=\parindent,
  skipabove=\medskipamount,
  skipbelow=\medskipamount,
  linewidth=4pt,
  linecolor={black!50},
  backgroundcolor={black!5},
  leftmargin=0cm
]{definition}

\surroundwithmdframed[
  topline=false,
  rightline=false,
  bottomline=false,
  leftmargin=\parindent,
  skipabove=\medskipamount,
  skipbelow=\medskipamount,
  linewidth=4pt,
  linecolor={gray!50},
  leftmargin=0cm
]{example}

\newmdenv[
  linewidth=1pt,
  roundcorner=5pt 
]{findingRuled}

\newmdenv[
  topline=false,
  bottomline=false,
  rightline=false,
  skipabove=\topsep,
  skipbelow=\topsep,
  leftmargin=-10pt,
  rightmargin=-10pt,
  innertopmargin=6pt,
  innerbottommargin=6pt,
  linewidth=4pt,
  linecolor={blue!50},
  backgroundcolor={blue!5},
]{thesisQuestionRuled}

\newmdenv[
  topline=false,
  bottomline=false,
  rightline=false,
  skipabove=\topsep,
  skipbelow=\topsep,
  leftmargin=-10pt,
  rightmargin=-10pt,
  innertopmargin=6pt,
  innerbottommargin=6pt,
  linewidth=4pt,
  linecolor={black},
  backgroundcolor={black!5},
]{thesisResultRuled}

\newmdenv[
  topline=false,
  bottomline=false,
  rightline=false,
  skipabove=\topsep,
  skipbelow=\topsep,
  leftmargin=-10pt,
  rightmargin=-10pt,
  innertopmargin=6pt,
  innerbottommargin=6pt,
  linewidth=4pt,
  linecolor={orange!80!red},
  backgroundcolor={orange!5},
]{hypothesisRuled}

\definecolor{viatraEmphColor}{RGB}{0,80,125}
\definecolor{keywordcolor}{rgb}{0.5,0,0.1}
\definecolor{commentcolor}{rgb}{0,0.3,0.1}
\definecolor{stringcolor}{rgb}{0,0,1}

\lstdefinelanguage{viatra}
{
morekeywords={@QueryBasedFeature,@Constraint,count,pattern,package,neg,find,import,true,false,or,check,job,action,state,severity,location,message,oclIsKindOf,self,exists,includes,invariant,class,private,epackage,java},
emph={Pseudostate,Vertex,Region,Transition,Entry,Synchronization,State,RegularState,CompositeElement,Trigger,Guard,Action,Statechart,vertices,regions,source,target,incomingTransitions,trigger,guard,action,elements,newElements,PartialModel,open,must,may,var,element,transitions,FamilyTree,members,Member,parents,age,VisionBlocked,blockedBy,Actor,xPos,yPos,ySpeed,placedOn,Lane_Horizontal,length,width},
emphstyle={\color{viatraEmphColor}},
sensitive=true, morecomment=[l]{//}, morecomment=[s]{/*}{*/},
morestring=[b]{"}
}
\lstdefinestyle{viatrasmall}{
	basicstyle=\scriptsize\ttfamily,
	commentstyle=\color{commentcolor}\ttfamily,
	stringstyle=\color{stringcolor}\ttfamily,
	captionpos=b,
	keywordstyle=\color{keywordcolor}\bfseries\ttfamily,
	showstringspaces=false,
	tabsize=2,
	language=viatra,
	escapeinside={(*@}{@*)}
}
\lstdefinestyle{viatrabig}{
	basicstyle=\ttfamily,
	commentstyle=\color{commentcolor}\ttfamily,
	stringstyle=\color{stringcolor}\ttfamily,
	captionpos=b,
	keywordstyle=\color{keywordcolor}\bfseries\ttfamily,
	showstringspaces=false,
	tabsize=2,
	language=viatra,
	escapeinside={(*@}{@*)}
}

\definecolor{emphColor}{rgb}{0.1,0.1,0.1}  
\newcommand{\dslStyle}[1]{\texttt{\color{emphColor}{#1}}}

\newcommand{\symbolClass}[2]{\dslStyle{#1}_{#2}}

\newcommand{\logicModelTwo}[2]{\langle {#1}, {#2} \rangle}

\newcommand{\dslObjectSet}[0]{\mathcal{O}}
\newcommand{\dslObjectSetI}[1]{\dslObjectSet_{#1}}

\newcommand{\kw}[1]{\textup{\texttt{\color{magenta}#1}}}
\newcommand{\true}[0]{\kw{true}}
\newcommand{\false}[0]{\kw{false}}
\newcommand{\unk}[0]{\kw{unknown}}
\newcommand{\err}[0]{\kw{error}}

\definecolor{logsemColor}{RGB}{0,80,125}

\newcommand{\interpretationFun}[1]{\mathcal{I}_{#1}}

\definecolor{numsemColor}{RGB}{204,100,0}

\newcommand{\fss}[0]{FSS}
\newcommand{\moopt}[0]{MHS}
\newcommand{\mom}[0]{MIN}

\newcommand{\qa}[1]{{\footnotesize\textit{{\textsf{#1}}}}}

\newcommand{\qaLeft}[0]{\qa{left}}
\newcommand{\qaLeftPred}[2]{\qaLeft(#1, #2)}
\newcommand{\qaRight}[0]{\qa{right}}
\newcommand{\qaRightPred}[2]{\qaRight(#1, #2)}
\newcommand{\qaFront}[0]{\qa{ahead}}
\newcommand{\qaFrontPred}[2]{\qaFront(#1, #2)}
\newcommand{\qaBehind}[0]{\qa{behind}}
\newcommand{\qaBehindPred}[2]{\qaBehind(#1, #2)}

\newcommand{\qaClose}[0]{\qa{close}}
\newcommand{\qaClosePred}[2]{\qaClose(#1, #2)}
\newcommand{\qaMed}[0]{\qa{medDist}}
\newcommand{\qaMedPred}[2]{\qaMed(#1, #2)}
\newcommand{\qaFar}[0]{\qa{far}}
\newcommand{\qaFarPred}[2]{\qaFar(#1, #2)}
\newcommand{\qaVeryFar}[0]{\qa{veryFar}}
\newcommand{\qaVeryFarPred}[2]{\qaVeryFar(#1, #2)}

\newcommand{\qaCanSee}[0]{\qa{canSee}}
\newcommand{\qaCanSeePred}[2]{\qaCanSee(#1, #2)}

\newcommand{\qaNoColl}[0]{\qa{noColl}}
\newcommand{\qaNoCollPred}[2]{\qaNoColl(#1, #2)}

\newcommand{\qaOnRoad}[0]{\qa{onAnyRd}}
\newcommand{\qaOnRoadPred}[1]{\qaOnRoad(#1, #1)}

\newcommand{\qaCar}[0]{\qa{Car}}

\definecolor{blue0}{HTML}{47BCFF}
\definecolor{blue1}{HTML}{009BF5}
\definecolor{blue2}{HTML}{0067A3}
\definecolor{blue3}{HTML}{003452}
\definecolor{car1}{RGB}{128,125,123}  
\definecolor{car2}{RGB}{194,92,85}  
\definecolor{car3}{RGB}{75,119,157}  

\newcommand{\scenic}{\textsc{Scenic}}
\newcommand{\conType}[1]{\textbf{#1}}

\newcommand{\actorNM}[1]{\vec{\texttt{a}}_#1}
\newcommand{\actor}[1]{$\actorNM{#1}$}

\newcommand{\actorR}[0]{\textcolor{car2}{\actor{R}}}
\newcommand{\actorBl}[0]{\textcolor{car3}{\actor{B}}}

\newcommand{\actorA}[0]{\textcolor{car2}{\actor{A}}}
\newcommand{\actorB}[0]{\textcolor{gray}{\actor{B}}}

\newcommand{\actorSampleNM}[0]{\actorNM{i}}
\newcommand{\actorSample}[0]{\actor{i}}

\newcommand{\justParam}[1]{\texttt{#1}}
\newcommand{\justParamB}[1]{\textcolor{gray}{\justParam{#1}}}
\newcommand{\param}[2]{\actorNM{#2}.\justParam{#1}}

\newcommand{\paramR}[1]{\textcolor{car2}{\param{#1}{R}}}
\newcommand{\paramBl}[1]{\textcolor{car3}{\param{#1}{B}}}

\newcommand{\paramA}[1]{\textcolor{car2}{\param{#1}{A}}}
\newcommand{\paramB}[1]{\textcolor{gray}{\param{#1}{B}}}

\newcommand{\actorFunNoMath}[1]{\texttt{o}_#1}
\newcommand{\actorFunNMG}[0]{\textcolor{car1}{\actorFunNoMath{G}}}
\newcommand{\actorFunNMR}[0]{\textcolor{car2}{\actorFunNoMath{R}}}
\newcommand{\actorFunNMBl}[0]{\textcolor{car3}{\actorFunNoMath{B}}}
\newcommand{\actorFunNMA}[0]{\textcolor{car2}{\actorFunNoMath{A}}}
\newcommand{\actorFunNMB}[0]{\textcolor{gray}{\actorFunNoMath{B}}}

\newcommand{\actorFun}[1]{$\actorFunNoMath{#1}$}
\newcommand{\actorFunG}[0]{\textcolor{car1}{\actorFun{G}}}
\newcommand{\actorFunR}[0]{\textcolor{car2}{\actorFun{R}}}
\newcommand{\actorFunBl}[0]{\textcolor{car3}{\actorFun{B}}}

\newcommand{\actorFunA}[0]{\textcolor{car2}{\actorFun{A}}}
\newcommand{\actorFunB}[0]{\textcolor{gray}{\actorFun{B}}}

\newcommand{\funtolog}[0]{\oldstylenums{f2l}}
\newcommand{\logtofun}[0]{\oldstylenums{l2f}}

\newmdenv[topline=false,bottomline=false,rightline=false,innertopmargin=1pt,innerbottommargin=1pt,innerrightmargin=0pt,innerleftmargin=0.65ex,skipabove=0.65ex,skipbelow=0.65ex,linewidth=0.75pt]{exlineEnv}
\newcounter{exline}
\newenvironment{exline}[1][]{\refstepcounter{exline}
\begin{exlineEnv}\noindent\textit{Example~\theexline: #1}\rmfamily}
{\end{exlineEnv}}

\newcolumntype{Z}{>{\centering\let\newline\\\arraybackslash\hspace{0pt}}X}

\newcommand{\rquestion}[1]{\textbf{\textsc{RQ}#1}}
\newcommand{\ranswer}[2]{\boxedVal{\rquestion{#1:}}{\emph{#2}}}
\newcommand{\contribution}[1]{\textbf{\textsc{C}#1}}
\newcommand{\contribox}[2]{\boxedVal{\contribution{#1 }}{#2}}
\newcommand{\theorem}[1]{\textbf{\textsc{Theorem} #1}}
\newcommand{\theorembox}[2]{%
\vskip 0.5\baselineskip
\noindent
\begin{tabularx}{\linewidth}{X}
\theorem{#1:} 
#2\\
\end{tabularx}}

\newcommand{\boxedVal}[2]{%
\vskip 0.5\baselineskip
\noindent
\begin{tabularx}{\linewidth}{|X|}
\hline
#1 
#2\\\hline
\end{tabularx}}

\newcounter{enumi-saved}

\definecolor{completed}{RGB}{51,153,102}

\definecolor{inProgress}{RGB}{196,196,35}

\definecolor{notYetStarted}{RGB}{255,0,0}

\NewDocumentCommand{\rot}{O{45} O{1em} m}{\makebox[#2][l]{\rotatebox{#1}{#3}}}%
\newcommand{\halfcheck}{X\kern-1.1ex\raisebox{.7ex}{\rotatebox[origin=c]{125}{--}}}

\newcommand{\yes}{\newmoon}
\newcommand{\maybe}{\LEFTcircle}
\newcommand{\no}{\fullmoon}

\begin{document}






\title{Concretization of Abstract Traffic Scene Specifications Using Metaheuristic Search}

\author{Aren~A.~Babikian,
        Oszk\'{a}r~Semer\'{a}th
        and~D\'{a}niel~Varr\'{o}
\IEEEcompsocitemizethanks{
\IEEEcompsocthanksitem Aren A. Babikian and D\'{a}niel Varr\'{o} are with the Department of Electrical and Computer Engineering, McGill University, Canada.\protect\\
E-mail: aren.babikian@mail.mcgill.ca
\IEEEcompsocthanksitem Oszk\'{a}r Semer\'{a}th and D\'{a}niel Varr\'{o} are with the Department of Measurement and Information Systems, Budapest University of Technology and Economics, Hungary.\protect\\
E-mail: semerath@mit.bme.hu
\IEEEcompsocthanksitem D\'{a}niel Varr\'{o} is with the Department of Computer and Information Science, Link\"{o}ping University, Sweden.\protect\\
E-mail: daniel.varro@liu.se}
\thanks{This work has been submitted to the IEEE for possible publication. Copyright
may be transferred without notice, after which this version may no longer be
accessible.}}

\markboth{Submitted to IEEE Transactions on Software Engineering}%
{}

\IEEEtitleabstractindextext{%
\begin{abstract}

Existing safety assurance approaches for autonomous vehicles (AVs) perform system-level safety evaluation by placing the AV-under-test in challenging traffic scenarios captured by abstract scenario specifications and investigated in realistic traffic simulators. 
As a first step towards scenario-based testing of AVs, the initial scene of a traffic scenario must be concretized.
In this context, the scene concretization challenge takes as input a high-level specification of abstract traffic scenes and aims to map them to concrete scenes where exact numeric initial values are defined for each attribute of a vehicle (e.g. position or velocity). 
In this paper, we propose a traffic scene concretization approach that places vehicles on realistic road maps such that they satisfy an extensible set of abstract constraints defined by an expressive scene specification language which also supports static detection of inconsistencies.
Then, abstract constraints are mapped to corresponding numeric constraints, which are solved by metaheuristic search with customizable objective functions and constraint aggregation strategies.
We conduct a series of experiments over three realistic road maps to compare eight configurations of our approach with three variations of the state-of-the-art \scenic~tool, and to evaluate its scalability.
\end{abstract}

}

\maketitle
\IEEEdisplaynontitleabstractindextext
\IEEEpeerreviewmaketitle

\section{Introduction}
\label{sec:intro}



The increasing popularity of autonomous vehicles (AVs) has resulted in a rising interest in their safety assurance.
As such, rigorous certification criteria 
must be met to ensure the safe, widespread use of AVs from a societal perspective.
A high-level certification objective may be formulated for AVs as follows: if an AV-under-test is intending to execute \textit{any valid target maneuver}, at \textit{any valid location} on Earth and alongside \textit{any valid placement} of external actors performing \textit{valid maneuvers}, the AV-under-test executes the maneuver \textit{safely} 
(e.g. without getting into an accident).  

In this context, existing safety assurance approaches \cite{Abdessalem2018TestingFeatureInteractionsBriandNsgaForConcreteScenes,Babikian2021dReal} test AVs by placing them in challenging traffic scenarios to evaluate their system-level safety. 
Graph models (e.g. scene graphs) are frequently used to define such test scenarios
along qualitative abstractions (relations) of concrete values and positions of scenario actors. 
Such a formal representations particularly allows for the analysis of various properties at the level of test scenario suites through high-level metrics such as situation coverage \cite{Alexander2015SituationCoverage,Babikian2020DocSymp}.

On the one hand, safety experts and standards typically express scenarios at a high-level of abstraction using abstract relations between various actors to evaluate situation coverage of a test suite. 
On the other hand, modern traffic simulators (like CARLA \cite{Dosovitskiy2017CarlaSimulator} or DriveSim \cite{nvidiadrivesim}) necessitate concrete scenarios with exact numeric values provided for the various actors in order to evaluate the safety compliance of each test scenario.
To derive such concrete test scenarios, first, an initial concrete scene needs to be derived from the abstract scenario representation.
The initial scene is then augmented with concrete behavior before being run in simulation.
As a key challenge, \emph{automated concretization of an initial scene} takes an abstract scene specification with numerous high-level constraints as input, and automatically derives concrete scenes by providing concrete  parameter values for each actor. 
Since the relevance of certain test scenarios may depend on the physical location (e.g. in case of  geofencing for AVs), \emph{scene concretization parameterized within a designated geographical location} is particularly challenging.



Graph model generation has been used extensively in research to derive models that satisfy high-level (abstract) constraints.
Existing approaches may rely on logic solvers \cite{Sen2009AutomaticModelGeneration}, metaheuristic search \cite{Soltaba2017SDG} or a dedicated graph solver \cite{Semerath2018AGraphSolver}.
However, such approaches derive abstract graph models as output without any numeric information, which is insufficient for scene concretization. 
To derive numeric solutions, modern model generators \cite{Soltana2020PLEDGE,Babikian2021dReal} propose a hybrid search technique that integrates a back-end numeric reasoning tool.
However, their use in traffic scene concretization is limited to generating instances of a specific type of traffic scene (selected a priori) over a simple pre-defined map.

Specialized traffic scene concretization approaches such as the state-of-the-art \scenic~tool \cite{Fremont2019ScenicLanguage}
have been developed  with a custom scene specification language to \emph{capture arbitrary abstract constraints} over a \emph{custom road map} (both given as input).
However, the limited expressiveness of these languages (e.g. related to the allowed constraint structures) prevents adequate measurement of situation coverage as necessitated in safety standards for road vehicles \cite{ISO26262FunctionalSatefy,ISO21448SOTIFRoadVehicles}.
Furthermore, as shown in our paper, the underlying exploration strategies are not scalable enough to provide effective assurance for AVs aligned with these safety standards.

In this paper, we propose a scene concretization approach that automatically derives concrete scenes in accordance with an abstract scene specification as input. The specific contributions of the paper are as follows:
\begin{itemize}
    \item (\contribution{1}) \textbf{Scene specification language:}
    We propose an expressive (abstract) functional scene specification language with 4-valued partial model semantics that generalizes the \scenic~language \cite{Fremont2019ScenicLanguage} and enables static detection of inconsistencies at specification time.
    \item (\contribution{2}) \textbf{Mapping for abstract constraints:} 
    We define an extensible mapping from abstract (relational) constraints to corresponding numeric constraints to derive a numeric scene concretization problem.
    \item (\contribution{3}) \textbf{Integration of metaheuristic search:}
    We formalize the scene concretization problem as a customizable optimization problem which we solve using metaheuristic search algorithms.
    \item (\contribution{4}) \textbf{Extensive evaluation:}
    We evaluate eight configurations of our proposed approach over three realistic road maps to assess success rate, runtime and scalability compared to the \scenic~tool.
\end{itemize}





\section{Preliminaries}
\label{sec:preliminaries}








\subsection{Traffic scenes and scenarios}
\label{sec:pre-scene}

\textit{Traffic scene} is defined by Ulbrich, \textit{et al.} \cite{Ulbrich2015Defining} as a snapshot of the environment, including the \textit{scenery} and \textit{dynamic elements}, as well as the \textit{relations} between those entities.
\begin{itemize}
    \item The \textit{scenery} is comprised of the lane network, stationary components such as traffic lights and curbs, vertical elevation of roads and environmental conditions.
    \item \textit{Dynamic elements} (or \textit{actors}) include the various vehicles and pedestrians involved in a scene such as the ego vehicle. A scene may contain information about the state (e.g. position and speed) and attributes (e.g. vehicle color, whether a car door is open) of actors.
    \item \textit{Relations} are defined between scenery elements and actors. For example, 
    two vehicles may be \textit{far from each other}, or a vehicle may be placed \textit{on} a specific lane. See \autoref{sec:pre-lang} for further details. 
\end{itemize}

A sequence of consecutive traffic scenes together with related temporal developments corresponds to a \textit{scenario}.
A scenario is defined by an \textit{initial scene}, followed by a sequence of \textit{actions and events} performed by the actors according to individual \textit{goals and values}.
\textit{Actions and events} may refer to traffic maneuvers (e.g. a lane change maneuver), while \textit{goals and values} may be transient (e.g. reaching a certain area on a map) or permanent (e.g. driving in a safe manner).

Existing safety standards  (e.g. ISO 26262-1 \cite{ISO26262FunctionalSatefy} and  SOTIF \cite{ISO21448SOTIFRoadVehicles}) place system-level safety requirements and restrictions on autonomous vehicles (AVs) under test.
Such requirements are often formalized as high-level constraints between actors. 
Adherence to such safety requirements is often evaluated by using sophisticated traffic simulators like CARLA \cite{Dosovitskiy2017CarlaSimulator} or DRIVE Sim \cite{nvidiadrivesim} which can only handle a lower-level representation of the investigated scenarios.






\begin{figure}[htp]
    \begin{center}
      \includegraphics[width=.8\linewidth]{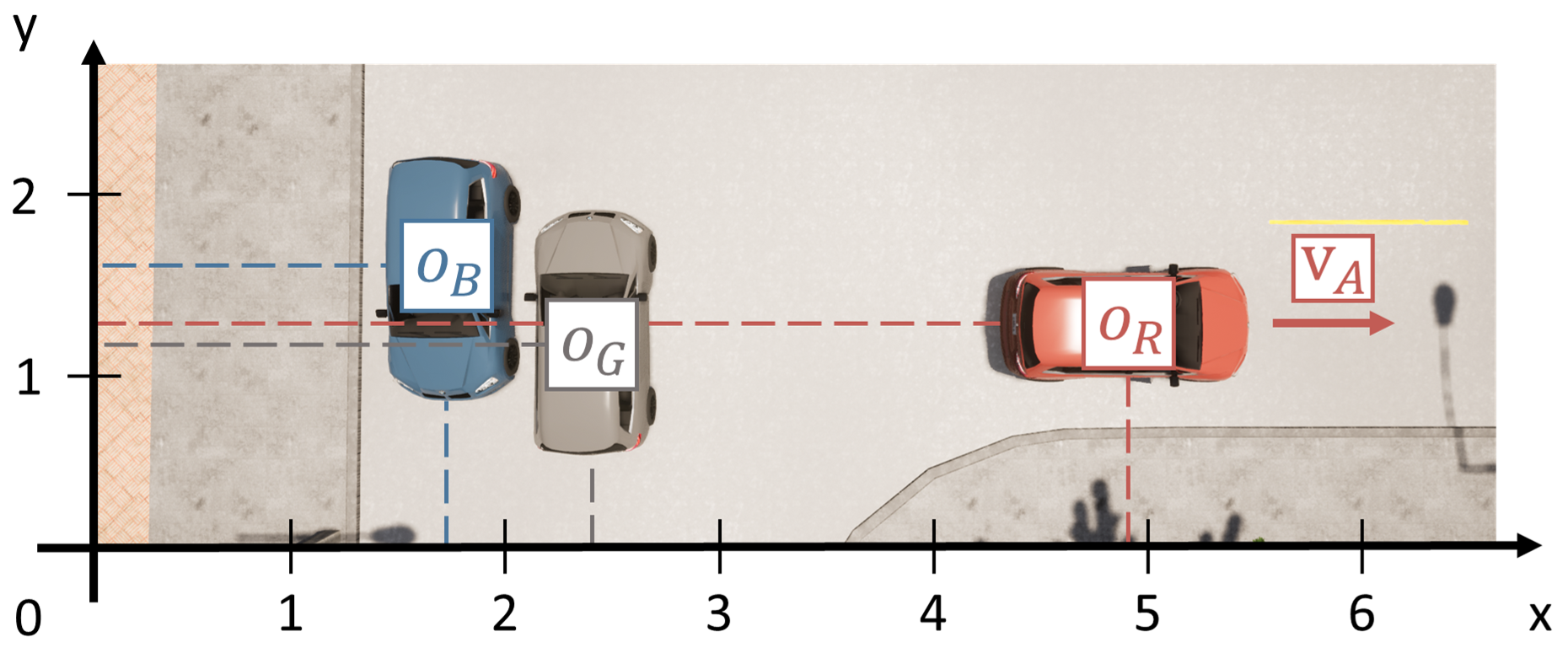}
      \caption{A traffic scene involving three vehicles}
      \label{fig:scene}
    \end{center}
\end{figure}

\begin{exline}
\autoref{fig:scene} depicts a scene with three \textit{actors}  (i.e. vehicles) \actorFunBl, \actorFunG~and \actorFunR~at an intersection, which composes the \textit{scenery}.
Their respective concrete positions are \textcolor{car3}{$\langle 1.7, 1.6 \rangle$}, \textcolor{car1}{$\langle 2.4, 1.2 \rangle$} and \textcolor{car2}{$\langle 4.9, 1.3 \rangle$} according to the indicated coordinate system.
\actorFunBl~and \actorFunG~are positioned behind \actorFunR, and to the left of each other.
They have \textit{opposite} headings and they are placed on \textit{adjacent}, but \textit{opposite} lanes.
\actorFunR~faces in the direction of the x axis and has a forward speed 
{\textcolor{car2}{$\texttt{v}_A$}}
of 0.5 units, depicted by \textcolor{car2}{a red arrow}, thus it is moving \textit{away from} \actorFunBl~and \actorFunG, both of which are static.
\end{exline}

\subsection{Levels of abstraction in traffic scenarios}
\label{sec:pre-levels}


Menzel, \textit{et al.} \cite{Menzel2018ScenariosForDevelopment} define three abstraction levels to adequately describe traffic scenarios for simulating AVs \cite{Majzik2019TowardsSystemLevel}.
\begin{enumerate}
    \item \emph{Functional Scenarios} include abstract (qualitative) constraints pertaining to traffic concepts. For example, such abstract constraints may be used to describe geospatial concepts (e.g. two vehicles are \emph{close to} or \emph{far from} each other), causal concepts (e.g. vehicle $A$ stopped moving \emph{because} it encountered a red light) and temporal concepts (e.g. event $A$ occurred \emph{before} or \emph{after} event $B$).
    \item \emph{Logical Scenarios} refine the abstract constraints of functional scenarios into constraints over parameter ranges or intervals, optionally accompanied by probability distributions. For example, geospatial functional constraints may be refined to areas on a map, and temporal functional constraints may be refined to time intervals.
    \item \emph{Concrete Scenarios} substitute concrete numeric values from the parameter ranges/intervals defined in a logical scenario. For example, concrete scenarios contain exact values for the position coordinates of actors, as well as exact times and durations for event executions.
\end{enumerate}

Given a specific concrete scenario executable in a traffic simulator, any abstract relation in functional scenarios can be derived by (i) identifying relevant logical constraints (e.g. geospatial, temporal), and (ii) assigning the truth value of abstract relations accordingly by predicate abstraction.




\begin{figure}[htp]
\noindent
\captionsetup{justification=centering}
\begin{tabularx}{\linewidth}{|c|}
\hline

Functional (initial) scene specification \\
\hdashline
\setlength{\tabcolsep}{0pt}
\begin{tabular}{l}
\actorFunR~: \qaCar. \actorFunBl~: \qaCar.\\
\textcolor{OliveGreen}{\qaFrontPred{\actorFunR}{\actorFunBl}}.
\textcolor{LimeGreen}{\qaMedPred{\actorFunBl}{\actorFunR}}.
\texttt{!}\qaRightPred{\actorFunBl}{\actorFunR}.
\end{tabular}
\\
\hline
\hline

\

Logical (initial) scene specification \\
\hdashline
\setlength{\tabcolsep}{0pt}
\begin{tabular}{l}
\actorFunR $\mapsto$ \actorR = $\langle \paramR{x}, \paramR{y}, \paramR{h}\rangle$\\

\actorFunBl $\mapsto$ \actorBl = $\langle \paramBl{x}, \paramBl{y}, \paramBl{h}\rangle$ \\

\textcolor{OliveGreen}{$\exists r > 0: \genfrac{}{}{0pt}{}
{\paramBl{x} = \paramR{x} + r \cos{\paramR{h}} \wedge}
{\paramBl{y} = \paramR{y} + r \sin{\paramR{h}}}$}\\

\textcolor{LimeGreen}{$c_l \leq \sqrt{(\paramR{x} + \paramBl{x})^2 + (\paramR{y} + \paramBl{y})^2} \leq c_u$}\\

$\neg \Big( \exists r > 0: \genfrac{}{}{0pt}{}{\paramR{x} = \paramBl{x} + r \cos{(\paramBl{h} - \sfrac{\pi}{2})} \wedge}{\paramR{y} = \paramBl{y} + r \sin{(\paramBl{h} - \sfrac{\pi}{2})}} \Big)$
\end{tabular}
\\
\hline
\hline

Concrete (initial) scene \\
\hdashline
\setlength{\tabcolsep}{0pt}
\begin{tabular}{l}
\actorR = $\langle \paramR{x}=56.20, \paramR{y}=188.48, \paramR{h}=1.57\rangle$ \\
\actorBl = $\langle \paramBl{x}=42.44, \paramBl{y}=188.48, \paramBl{h}=1.57\rangle$
\end{tabular}
\\
\hline
\hline

Visualisation \\
\hdashline
\includegraphics[valign=c, margin=0pt 1ex 0pt 1ex, width=\linewidth-2\tabcolsep-1.3333\arrayrulewidth]{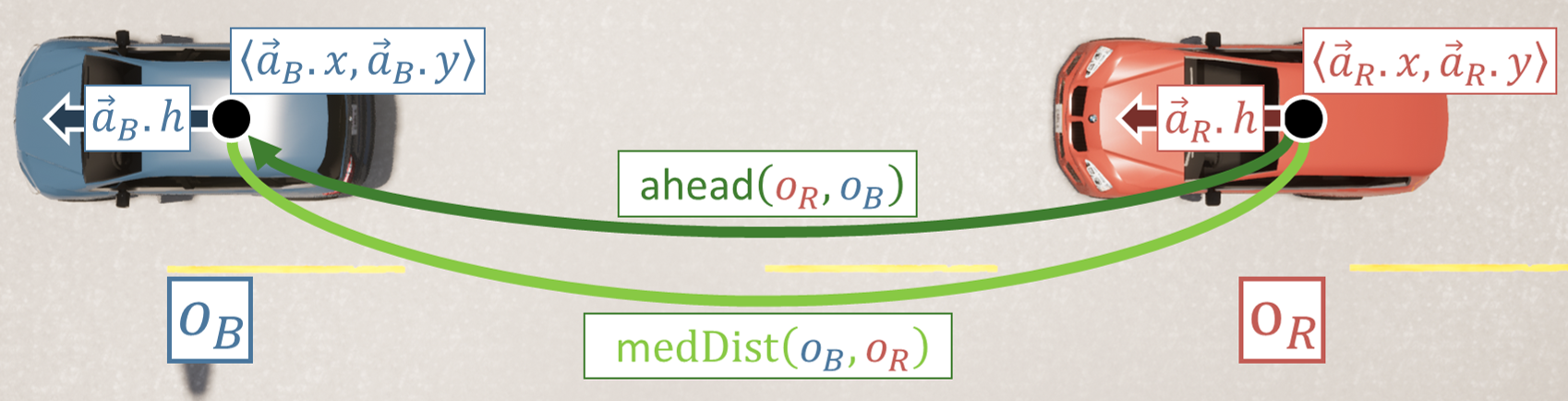}
\\
\hline
\end{tabularx}
\caption{An initial traffic scene described at various\\levels of abstraction}
\label{fig:sceneAbsLevels}
\end{figure}

\begin{exline}
An initial traffic scene containing two actors \actorFunR~and \actorFunBl~is described at various levels of abstraction in \autoref{fig:sceneAbsLevels} (\textcolor{LimeGreen}{$c_l$} and \textcolor{LimeGreen}{$c_u$} are constants).
Functional and logical scenes define constraints (i.e. the \textit{problem}), while concrete scene defines an instance (i.e. a \textit{solution}) that satisfies these constraints.
The problem is exclusively comprised of geospatial constraints, since no dynamic behavior is involved in the initial scene.
The functional scene is defined using the abstract language proposed in \autoref{sec:pre-lang}.
The logical and concrete scenes are handled in \autoref{sec:elab-func-to-log}, while \autoref{sec:elab-concretization} describes our solution to such problems.
\end{exline}
\subsection{Scenario-based testing by simulation}
\label{sec:pre-intial}


Scenario-based testing by simulation \cite{Fremont2019ScenicLanguage,Abdessalem2018TestingVisionBased,Haq2022,Riccio2020} is commonly used to evaluate the adherence of AVs under test to traffic safety requirements.
In line with the definition of a traffic scenario proposed in \autoref{sec:pre-scene} \cite{Ulbrich2015Defining}, scenario-based test cases are composed of  an abstract \textit{Initial Scene Specification} (ISS), abstract \textit{Behavioral/Temporal Constraints} (BTCons) over actors along with \textit{Evaluation Criteria} (ECrit), often defined as oracles \cite{Jahangirova2021Oracle} based on safety requirements.

For a test case to be executable in simulation, its abstract constraints must be concretized. As a first step, (1) the abstract ISS is refined into a concrete scene.
Then, (2) concrete behaviors, such as exact trajectories to follow, are assigned to each actor in accordance to BTCons.
Finally, (3) the concrete scenario is simulated, and its success (pass/fail) is evaluated according to ECrit.
For instance, a test may be considered successful if the ego vehicle can navigate its assigned trajectory without colliding with any other actor.

    

In this paper, we exclusively focus on \emph{Step (1)}, i.e., the automated concretization of abstract ISSs into realistic \textit{initial scenes}.
This is an important aspect of AV testing as the initial scene may \emph{have a direct impact on the outcome of  test execution while assuming identical behavior for all actors}. 
For instance, consider two test scenarios with different initial scenes (ISSs) but identical actor behaviors. 
In both cases, the ego actor \actorFun{{ego}} and a non-ego actor \actorFunBl{} are driving at a high speed inside their lane (\actorFun{{ego}} is following \actorFunBl{}) while another actor \actorFunR{} abruptly cuts in front \actorFunBl{}, forcing \actorFunBl{} to do an abrupt brake.
If, according to the ISS, \actorFun{{ego}} is close behind \actorFunBl{}, these actor will collide.
However, if the ISS specifies that \actorFunBl{} and \actorFun{{ego}} are initially far from each other, collision will be avoided since \actorFunBl{} will have time to slow down.
Videos depicting the two test scenarios are included in an online publication page\footnote{\url{https://doi.org/10.5281/zenodo.6345282}} dedicated to this paper.

\subsection{Scene concretization in scenario-based testing }
\label{sec: complement}

Our paper proposes a \textit{ scene concretization} approach where a functional-level (initial) scene specification is given as input and the concrete positioning of vehicles is constructed as output.
For that purpose, abstract constraints specifying the functional scene are first mapped into an equivalent numeric problem (i.e. the logical scene) along the mapping of \autoref{sec:elab-mapping}.
Then, metaheuristic search (\moopt) is used to derive a numeric solution (i.e. a concrete scene) that satisfies all related constraints.

Search-based test generation techniques \cite{Abdessalem2016TestingADAS,Abdessalem2018TestingVisionBased,calo2020GeneratingAvoidableCollision,Wu2021,Riccio2020} have been actively used to provide potentially dangerous concrete test scenarios as input for traffic simulators. As general assumption of these approaches, a single search process is conducted to find concrete scene parameters and actor behavior that leads to potential danger. 
Our paper investigates \emph{scene concretization as a standalone subproblem} of the complex challenge of scenario-based testing, which complements existing work in three key aspects. 
\begin{itemize}
    \item Our abstract scene representation enables to \emph{evaluate the coverage of arbitrary automatically generated test scenarios} with formal precision by qualitative abstraction for similar behavior of actors.
    \item When a potentially dangerous scenario is found by existing test generators, our approach can provide \emph{what-if} analysis by deriving a diverse set of initial scenes to investigate similar behavior of actors. Such analysis can help better understand what contextual parameters of the scene itself can contribute to potential danger.  
    \item Our approach enables to investigate \emph{traffic scenarios in a realistic context by concretizing scenes in concrete map locations}. Demonstrating safe behavior of AV at a specific location can help \emph{geofencing}, e.g. an AV is allowed to take a particular route on the map but not other routes.
\end{itemize}



\section{Functional scene specification}
\label{sec:pre-lang}

Functional scene specifications (\fss{}s) are often captured by an abstract constraint language \cite{Fremont2019ScenicLanguage} that leverages qualitative abstractions \cite{Bagschik2018,Menzel2019} of concrete scene attributes. 
In this paper, we adapt (4-valued) partial graph models as a \fss{} language using the syntax and formal semantics defined in \cite{Marussy2020JOT}.
As a key benefit over state-of-the-art traffic scene concretization approaches, e.g.  \scenic~ \cite{Fremont2019ScenicLanguage}, partial models enable the detection of inconsistencies at the \fss{} level. 


\subsection{Scene specification language}
\label{sec:func-lang}
\textit{Vocabulary:}
Objects in a partial model correspond to actors of a scene.
The relations between actors
are captured by a finite set of \textit{relation symbols}
$\Sigma = \{ \symbolClass{R}{pos} \cup \symbolClass{R}{dist} \cup \symbolClass{R}{vis} \cup\symbolClass{R}{coll} \cup \symbolClass{R}{road} \}$ grouped into
5 \textit{geospatial relation categories}:


\begin{itemize}

    \item
    $\symbolClass{R}{pos} = \{ \qaLeft, \qaRight, \qaFront, \qaBehind\}$ 
    are \conType{positional relations} denoting the relative position of the target actor with respect to the heading of the source actor.
    
    \item
    $\symbolClass{R}{dist} = \{ \qaClose, \qaMed, \qaFar\}$ 
    is a set of \conType{distance relations} which qualitatively characterize the Euclidean distance between two actors (using $x$ and $y$ coordinates).
    
    
    \item
    $\symbolClass{R}{vis} = \{ \qaCanSee \}$ 
    is the \conType{visibility relation} to capture if the target actor is in the field of view of the source actor. 

    \item
    $\symbolClass{R}{coll} = \{ \qaNoColl \}$ 
    is the \conType{collision avoidance relation} which denotes that two actors are positioned such that they are not overlapping (colliding).
    
    \item
    $\symbolClass{R}{road} = \{ \qaOnRoad \}$ 
    represents the unary \conType{road placement relation} which denotes that an actor is placed on any driveable road segment of the map (i.e. any segment which can be used by vehicles).

\end{itemize}

The abstract relations listed above are adapted from the \scenic~specification languages \cite{Fremont2019ScenicLanguage}, and similar abstract relations have been proposed in  \cite{Bagschik2018,Menzel2019,Urbieta2021}.
For simplictiy, abstract relations are restricted to binary relations as they are the most common in traffic scene specifications.
Thus, the unary \qaOnRoad~relation is represented as a (binary) self-loop relation. However, the proposed formalism can be generalized to \textit{n-ary} relations and constraints, such as a ternary constraint specifying that the line of sight between actors \actorFun{A} and \actorFun{B} is obstructed by an actor \actorFun{C}.

Our approach assumes that these relations can be derived from concrete scenes by qualitative abstractions. 
Note that our approach is independent from the included concrete relations, therefore we can extend the set of relations by various sorts of abstractions from the concrete scenes.
Moreover, we can also adjust relation categories accordingly.


\textit{Syntax and semantics:}
Given a vocabulary of geometric relations \(\Sigma\), a \fss{} is a partial model $P = \logicModelTwo{\dslObjectSetI{P}}{\interpretationFun{P}}$,
where $\dslObjectSetI{P}$ is the finite set of objects (each object corresponds to an actor),
and $\interpretationFun{P}$ gives a 4-valued logic interpretation for each symbol $\dslStyle{r}\in\Sigma$ as 
$\interpretationFun{P}(\dslStyle{r}) \colon
\dslObjectSetI{P} \times \dslObjectSetI{P} \rightarrow
\{\false,\true,\unk~\text{(unspecified)},\err~\text{(inconsistent)}\}$.


A \fss{} consists of \textit{relation assertions}, which explicitly assign a 4-valued truth-value to a binary relation over a pair of objects \cite{sagiv2002parametric,Gopan04summarized}.
Syntactically, a relation assertion can be prefixed by the $\texttt{?}$ (\unk) and $\texttt{!}$ (\false) symbols, while no prefix represents \true.

When multiple assertions to the same relation instance exist, the interpretation value is obtained by the 4-valued \emph{information merge} operator \(\oplus\) \cite{Marussy2020JOT}, where contradictory information results in \err~while unspecified relations result in \unk.
For instance, if a \fss~ contains both 
\qaRightPred{\actorFunA}{\actorFunB} and 
\texttt{!}\qaRightPred{\actorFunA}{\actorFunB}, then the relation will be interpreted as \(\interpretationFun{P}(\qaRight)(\actorFunNMA, \actorFunNMB) \)
= \(\true \oplus \false \)
= \(\err\).

Such \err~values detect inconsistencies in the \fss, i.e. they detect sets of constraints that cannot be satisfied by any concrete scene.
For instance, 
\qaRightPred{\actorFunA}{\actorFunB} and \texttt{!}\qaRightPred{\actorFunA}{\actorFunB} cannot hold at the same time for any pair of actors $\langle$\actorFunA, \actorFunB $\rangle$.
\err~values also arise from more complex inconsistencies when enforcing domain-specific validity rules.




\newcommand{\axiom}[1]{\textit{#1}}
\textit{Validity rules}:
For a partial model $P = \logicModelTwo{\dslObjectSetI{P}}{\interpretationFun{P}}$ to represent a valid \fss, $P$ must be refined according to five validity rules ($V_{road}, V_{loop}, V_{sym}, V_{pos}, V_{dist}$).
\axiom{$V_{road}$}
states that all \qaOnRoad~relations between two distinct actors are known to be \false, since only self-looping \qaOnRoad~relations are valid.
\axiom{$V_{loop}$} states that
any self-loop relation (other than \qaOnRoad) is known to be \false.
\axiom{$V_{sym}$} states that
if a \textbf{distance} or \textbf{collision avoidance} relation \qa{r} holds, then the same relation in the opposite direction is known to be \true.
\axiom{$V_{pos}$} states that
if a \textbf{positional relation} \qa{r1} holds between a given directed pair of actors, then all other \textbf{positional relations} \qa{r2} between those actors are known to be \false.
\axiom{$V_{dist}$} is analogous to \axiom{$V_{pos}$}, but is applied to \textbf{distance relations}. Formally:

\begin{itemize}
\footnotesize

    \item[\axiom{$V_{road}$}]
    \( \vcenter{\infer
    {\interpretationFun{P}(\qaOnRoad)(a, b) \coloneqq \interpretationFun{P}(\qaOnRoad)(a, b) \oplus \false}
    {a, b \in \dslObjectSetI{P} \wedge a \neq b}}\)
    
    \item[\axiom{$V_{loop}$}]
    \( \vcenter{\infer
    {\interpretationFun{P}(\qa{r})(o, o) \coloneqq \interpretationFun{P}(\qa{r})(o, o) \oplus \false}
    {o \in \dslObjectSetI{P}}}\); for $\qa{r} \in \Sigma\setminus\{\qaOnRoad\}$
    
    \item[\axiom{$V_{sym}$}]
    \( \vcenter{\infer
    {\interpretationFun{P}(\qa{r})(b, a) \coloneqq \interpretationFun{P}(\qa{r})(b, a) \oplus \true}
    {a, b \in \dslObjectSetI{P} \wedge \interpretationFun{P}(\qa{r})(a, b) = \true }}\); for $\qa{r} \in \symbolClass{R}{dist} \cup \{\qaNoColl\}$

    \item[\axiom{$V_{pos}$}]
    \( \vcenter{\infer
    {\interpretationFun{P}(\qa{r2})(a, b) \coloneqq \interpretationFun{P}(\qa{r2})(a, b) \oplus \false}
    {a, b \in \dslObjectSetI{P} \wedge \interpretationFun{P}(\qa{r1})(a, b) = \true }}\); for $\qa{r1} \in \symbolClass{R}{pos}, \qa{r2} \in \symbolClass{R}{pos}\setminus \qa{r1}$
    
    \item[\axiom{$V_{dist}$}]
    \( \vcenter{\infer
    {\interpretationFun{P}(\qa{r2})(a, b) \coloneqq \interpretationFun{P}(\qa{r2})(a, b) \oplus \false}
    {a, b \in \dslObjectSetI{P} \wedge \interpretationFun{P}(\qa{r1})(a, b) = \true }}\); for $\qa{r1} \in \symbolClass{R}{dist
    }, \qa{r2} \in \symbolClass{R}{dist}\setminus \qa{r1}$

\end{itemize}

In this paper, we provide a sound but incomplete set of validity rules.
If the enforcement of these rules produces an error, then the scene specification is surely inconsistent.
However, if no error is produced, the scene specification is not ensured to be consistent.

The above validity rules are provided in accordance with the included abstract relations and they can easily be extended with new relations in the future.
Additionally, custom validity rules may be defined to prevent semantic inconsistencies (i.e. physically infeasible specifications) or to enforce additional requirements (e.g. traffic laws).
For instance, a custom validity rule $V_{cust}$ can capture that if a vehicle $b$ is behind a vehicle $a$, then $a$ cannot see $b$.
\begin{itemize}
\footnotesize

    \item[\axiom{$V_{cust}$}]
    \( \vcenter{\infer
    {\interpretationFun{P}(\qaCanSee)(a, b) \coloneqq \interpretationFun{P}(\qaCanSee)(a, b) \oplus \false}
    {a, b \in \dslObjectSetI{P} \wedge \interpretationFun{P}(\qaBehind)(a, b)}}\)

\end{itemize}

\newcommand{\semitransp}[2][35]{\color{!#1}#2}
\begin{figure}[htp]
     \centering
        \begin{subfigure}[b]{0.49\columnwidth}
            \centering
            \begin{tabularx}{\linewidth}{|X|}\hline
                \textcolor{commentcolor}{\# road placement}\\
                \qaOnRoadPred{\actorFunG}\\
                \qaOnRoadPred{\actorFunBl}\\
                \qaOnRoadPred{\actorFunR}\\
                \textcolor{commentcolor}{\# collision avoidance}\\
                \qaNoCollPred{\actorFunG}{\actorFunBl}\\
                \qaNoCollPred{\actorFunG}{\actorFunR}\\
                \qaNoCollPred{\actorFunBl}{\actorFunR}\\
                \textcolor{commentcolor}{\# visibility}\\
                \textcolor{JungleGreen}{\qaCanSeePred{\actorFunG}{\actorFunBl}}\\
                \textcolor{JungleGreen}{\qaCanSeePred{\actorFunBl}{\actorFunG}}\\
                \textcolor{commentcolor}{\# position}\\
                \textcolor{OliveGreen}{\qaLeftPred{\actorFunG}{\actorFunBl}}\\
                \textcolor{OliveGreen}{\qaRightPred{\actorFunG}{\actorFunR}}\\
                \textcolor{OliveGreen}{\qaLeftPred{\actorFunBl}{\actorFunG}}\\
                \textcolor{OliveGreen}{\qaLeftPred{\actorFunBl}{\actorFunR}}\\
                \textcolor{OliveGreen}{\qaBehindPred{\actorFunR}{\actorFunG}}\\
                \textcolor{OliveGreen}{\qaBehindPred{\actorFunR}{\actorFunBl}}\\
                \textcolor{commentcolor}{\# distance}\\
                \textcolor{LimeGreen}{\qaClosePred{\actorFunG}{\actorFunBl}}\\
                \textcolor{LimeGreen}{\qaMedPred{\actorFunG}{\actorFunR}}\\
                \textcolor{LimeGreen}{\qaMedPred{\actorFunBl}{\actorFunR}}\\
                \hline
            \end{tabularx}
            \caption[]{Full scene specification}
            \label{fig:func-full}
        \end{subfigure}
        \begin{subfigure}[b]{0.49\columnwidth}
            \centering
            \begin{tabularx}{\linewidth}{|X|}\hline
                \textcolor{commentcolor}{\# road placement}\\
                \qaOnRoadPred{\actorFunG}\\
                \qaOnRoadPred{\actorFunBl}\\
                \qaOnRoadPred{\actorFunR}\\
                \textcolor{commentcolor}{\# collision avoidance}\\
                \qaNoCollPred{\actorFunG}{\actorFunBl}\\
                \qaNoCollPred{\actorFunG}{\actorFunR}\\
                \qaNoCollPred{\actorFunBl}{\actorFunR}\\
                \textcolor{commentcolor}{\# visibility}\\
                \textcolor{JungleGreen}{\qaCanSeePred{\actorFunG}{\actorFunBl}}\\
                \textcolor{JungleGreen}{\qaCanSeePred{\actorFunBl}{\actorFunG}}\\
                \textcolor{commentcolor}{\# position}\\
                \textcolor{OliveGreen}{\qaLeftPred{\actorFunG}{\actorFunBl}}\\
                \# \qaRightPred{\actorFunG}{\actorFunR}\\
                \# \qaLeftPred{\actorFunBl}{\actorFunG}\\
                \textcolor{OliveGreen}{\qaLeftPred{\actorFunBl}{\actorFunR}}\\
                \# \qaBehindPred{\actorFunR}{\actorFunG}\\
                \# \qaBehindPred{\actorFunR}{\actorFunBl}\\
                \textcolor{commentcolor}{\# distance}\\
                \textcolor{LimeGreen}{\qaClosePred{\actorFunG}{\actorFunBl}}\\
                \textcolor{LimeGreen}{\qaMedPred{\actorFunG}{\actorFunR}}\\
                \textcolor{LimeGreen}{\qaMedPred{\actorFunBl}{\actorFunR}}\\
                \hline
            \end{tabularx}
            \caption[]{\scenic -expressible subset}
            \label{fig:func-scenic}
        \end{subfigure}

        \begin{subfigure}[b]{0.90\columnwidth}
            \centering
            \includegraphics[width=\columnwidth]{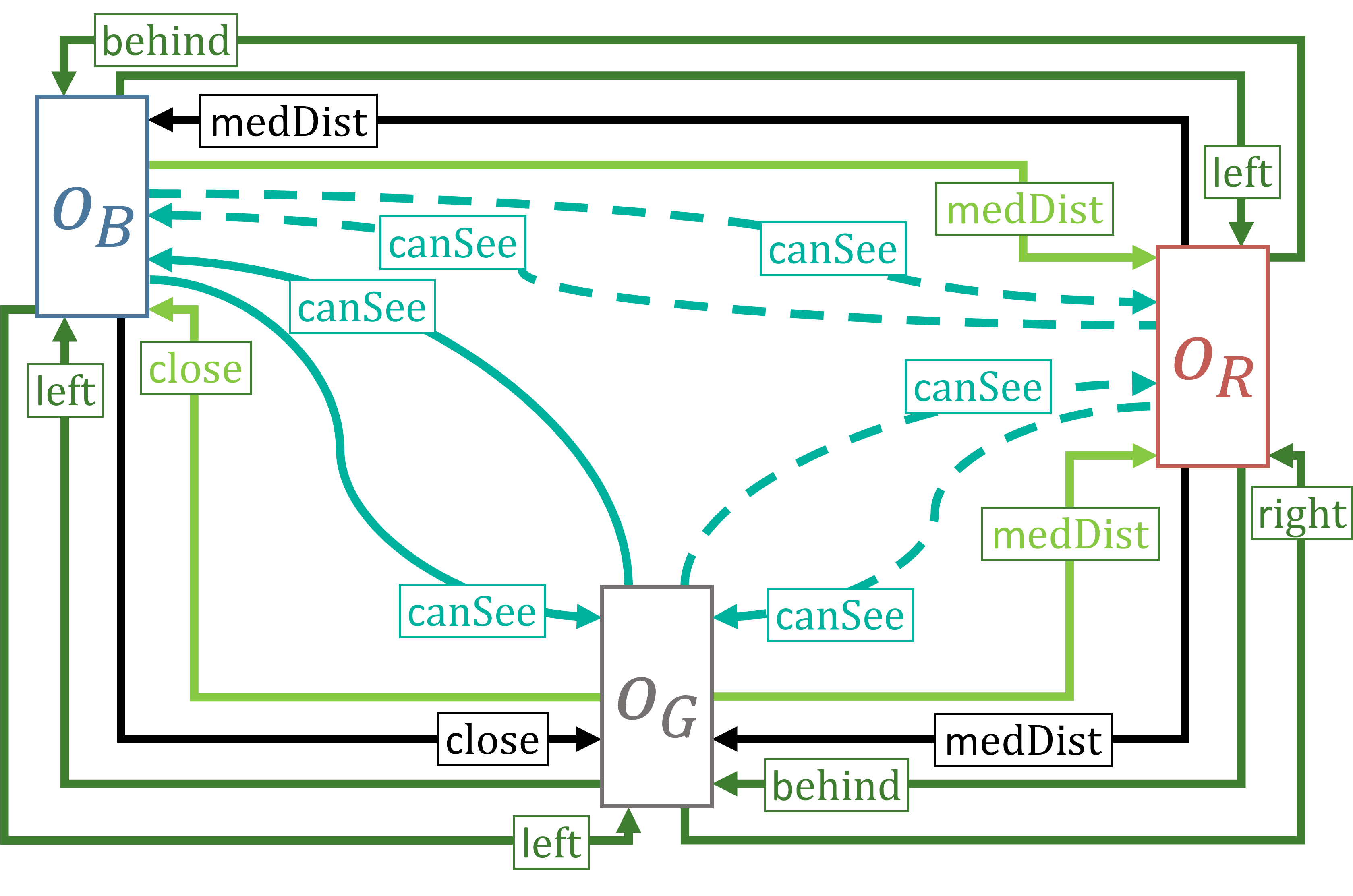}
            \caption[]{Graph representation of the functional scene}%
            \label{fig:elab-graph}
        \end{subfigure}

        \caption{A functional-level scene specification} 
        \label{fig:func-both}
\end{figure}



\begin{exline}
A functional-level scene specification is defined as a partial model $P = \logicModelTwo{\dslObjectSetI{P}}{\interpretationFun{P}}$ by the relation assertions in \autoref{fig:func-full}.
The scene ($P$) contains three car objects ($\actorFunNMG, \actorFunNMBl, \actorFunNMR \in \dslObjectSetI{P}$), which are placed on a road and do not collide with each other.

All unspecified relations are originally interpreted as $\unk$.
Additionally, positional and distance relations are included for every pair of actors (every oriented pair, in the case of positional relations).
As such, according to the validity rules, all unspecified positional and distance relations are refined to $\true$ or to $\false$.

For instance, the inclusion of \textcolor{LimeGreen}{\qaClosePred{\actorFunG}{\actorFunBl}} implies:
\begin{itemize}
\footnotesize
    \item
    $\interpretationFun{P}(\qaClose)(\actorFunNMBl, \actorFunNMG) = \true$; \textit{(from \axiom{$V_{sym}$})},
    
    \item 
    $\interpretationFun{P}(\qaMed)(\actorFunNMG, \actorFunNMBl) = \false$, $\interpretationFun{P}(\qaFar)(\actorFunNMG, \actorFunNMBl) = \false$; \textit{(from \axiom{$V_{dist}$})}, and
    
    \item 
    $\interpretationFun{P}(\qaMed)(\actorFunNMBl, \actorFunNMG) = \false$,  $\interpretationFun{P}(\qaFar)(\actorFunNMBl, \actorFunNMG) = \false$; \textit{(from both \axiom{$V_{sym}$} and \axiom{$V_{dist}$})}.
\end{itemize}

Validity rules enable the detection of inconsistencies in the \fss.
For instance, had the \fss~also included \qaFarPred{\actorFunG}{\actorFunBl}, the application of \axiom{$V_{dist}$} would result in an inconsistency:
\[\footnotesize{\infer
{\begin{tabular}{@{}r@{ }c@{ }l@{}}
$\interpretationFun{P}(\qaFar)(\actorFunNMG, \actorFunNMBl)$ & $\coloneqq$ & $\interpretationFun{P}(\qaFar)(\actorFunNMG, \actorFunNMBl) \oplus \false$\\
&$\coloneqq$& $\true \oplus \false = \err$\\
\end{tabular}}
{\actorFunNMG, \actorFunNMBl  
\in \dslObjectSetI{P} 
\wedge \interpretationFun{P}(\qaClose)(\actorFunNMG, \actorFunNMBl) = \true}}\]



\end{exline}

\begin{exline}
The visibility, positional and distance relations of the partial model are represented as a graph in \autoref{fig:elab-graph}, where nodes represent actors.
Road placement and collision avoidance relations are not represented for simplicity.
Solid and dashed lines represent relations that are interpreted as $\true$ and as $\unk$, respectively.
$\false$ relations are not depicted, and there are no $\err$ relations.
Relations and nodes are colored according to \autoref{fig:func-full}.
Relations derived from validity rules are shown in black.
\end{exline}


\subsection{Static analysis of scene specifications}
\label{sec:prel-staticAnal}

\textit{Inconsistency detection}: 
\scenic~compiles a functional scene specification into logical constraints, most of which define probability distributions over areas of the road map, and attempts to solve it by using rejection sampling \cite{Fremont2019ScenicLanguage}.
This involves randomly sampling the desired distributions until all constraints are satisfied.
A known drawback of this approach is that when a \fss~is inconsistent (e.g. if both \qaRightPred{\actorFunA}{\actorFunB} and \texttt{!}\qaRightPred{\actorFunA}{\actorFunB} are included), such an inconsistency can only be suspected when the solver repeatedly fails to provide a solution. 

A main benefit of the 4-valued semantics of partial models introduced in this paper for \fss{}s is to \emph{detect such semantic inconsistencies statically} (at specification time). When contradictory assignments are given to a particular relation, they are merged automatically into the \err~value, which can be detected easily. Note that such a static detection of inconsistent specifications is a unique feature of our technique compared to related \fss~approaches (e.g. \scenic).

A sound qualitative abstraction from logical constraints to abstract relations (to be discussed in \autoref{sec:elab-func-to-log}) ensures that whenever there is a (concrete) solution to the logical constraints, then the respective abstract relations will also evaluate to \true. Consequently, our 4-valued static analysis technique also guarantees that if an inconsistency is detected in the \fss~(by the \err~value), then no concrete solutions may exist for the logical constraints. In such a case, the underlying solver does not need to be called at all, which can result in significant time savings.
Unsurprisingly, our static analysis technique does not have completeness guarantees, i.e. there may be a set of inconsistent logical constraints, which cannot be detected on the abstract level. 

\textit{Restrictions of the \scenic~\fss~language}:
While our \fss~language builds on the scene specification language of \scenic, it is important to note that the original \scenic~language has limitations wrt. (i) error detection capabilities, as discussed above, as well as (ii) soundness.

Due to limitations of the underlying scene concretization approach, the \scenic~language rejects certain valid (consistent) constraint structures, which poses limitations to the soundness of the approach. For example, actors may be the target of at most one positional relation, pointing from an already instantiated actor. As such, positional relations may only form a tree structure (and not an arbitrary graph), which, in certain cases, is not enough to formally distinguish two semantically different scenes.

\begin{figure}[htp]
    \begin{center}
    \captionsetup{justification=centering}
      \includegraphics[width=0.9\linewidth]{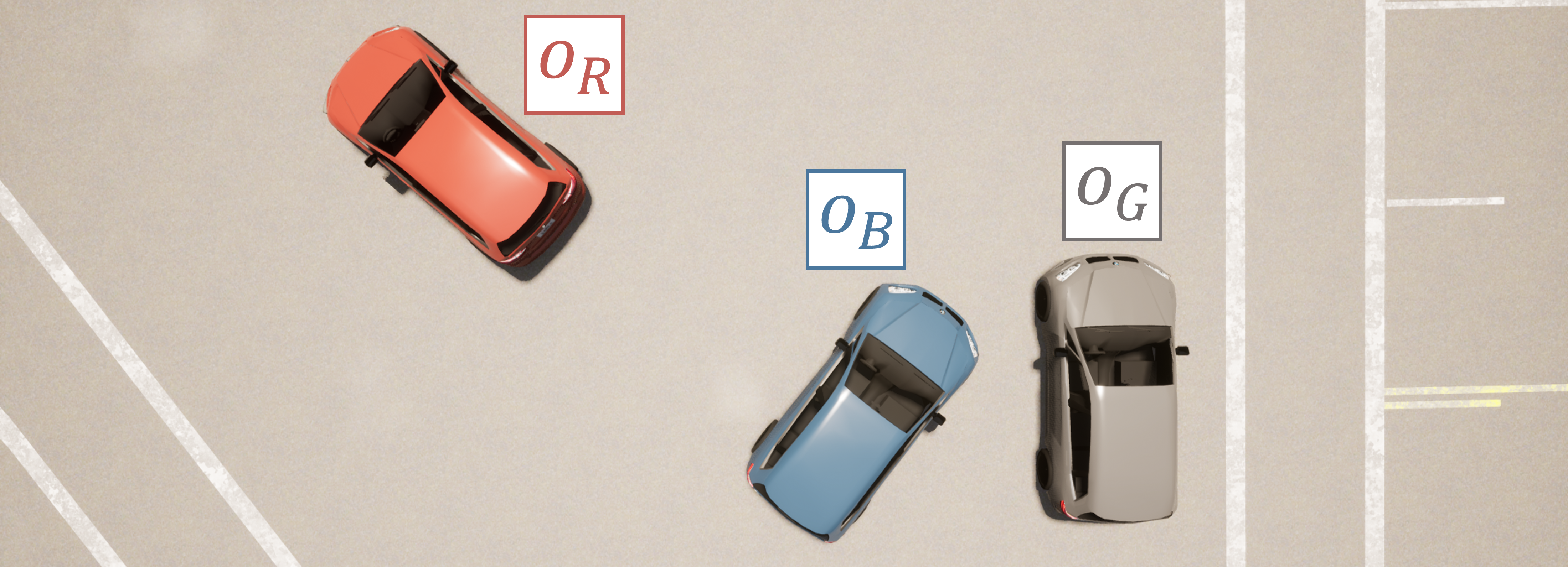}
      \caption{Concrete scene that does not satisfy \autoref{fig:func-full}, but satisfies the corresponding \scenic-expressible subset}
      \label{fig:scene-scenic}
    \end{center}
\end{figure}

\begin{exline}
\autoref{fig:func-scenic} presents a \scenic-expressible subset of the relation assertions of  \autoref{fig:func-full} (inexpressible constraints are crossed out).
Of the 6 positional relations, at most two can be expressed by \scenic.
Note that the order of actor definitions (\actorFunG, \actorFunBl, then \actorFunR) influences which relations can be included. Along this ordering, \textcolor{OliveGreen}{\qaRightPred{\actorFunG}{\actorFunR}} could be included instead of \textcolor{OliveGreen}{\qaLeftPred{\actorFunBl}{\actorFunR}}.

\autoref{fig:scene-scenic} depicts a concrete scene that satisfies the \scenic-expressible subset.
When compared to \autoref{fig:scene}, which satisfies the full \fss~proposed in \autoref{fig:func-full}, the benefits of a more expressive language become apparent.
\textit{None} of the relations excluded from the \scenic-expressible subset are satisfied by the the scene depicted in \autoref{fig:scene-scenic}.
As such, both of these scenes, which are semantically different, cannot be formally distinguished using the default \scenic~\fss~language as they would both be represented by the \fss~shown in \autoref{fig:func-scenic}.
\end{exline}





\contribox{1}{We propose a high-level traffic scene representation language with 4-valued partial model semantics. This enables static detection of inconsistencies at specification time, which is not offered by related \fss~approaches.}


\section{Logical scenes as numeric problems}
\label{sec:elab-func-to-log}

The  scene concretization problem can be represented as a numeric constraint satisfaction problem over actors on a logical-level scene.
We introduce a formalization for logical-level  scenes, and propose a novel mapping from a \fss~ to a corresponding numeric constraint satisfaction problem.

As key benefit, our mapping is \textit{extensible} to take any abstract functional relation as input and yields \textit{customizable} numeric constraints as output.
Additionally, our approach can be contextualized in   \textit{any underlying road map}.
Existing such mappings often either provide restricted numeric constraints as output, or they are  limited to approach-specific input constraints defined over a simplistic road map.



\newcommand{\mapping}[0]{map}
\newcommand{\actorSet}[0]{\mathcal{A}}
\newcommand{\actorSetI}[1]{\actorSet_{#1}}
\newcommand{\constrSet}[0]{\mathcal{C}}
\newcommand{\constrSetI}[1]{\constrSet_{#1}}
\newcommand{\roadmapI}[1]{m_{#1}}
\newcommand{\dimensionsI}[1]{D_{#1}}

\begin{figure*}[htp]
\noindent
\begin{tabular}{
  |m{\dimexpr.16\linewidth-2\tabcolsep-1.3333\arrayrulewidth}
  |m{\dimexpr.45\linewidth-2\tabcolsep-1.3333\arrayrulewidth}
  |>{\centering\arraybackslash}m{\dimexpr.39\linewidth-2\tabcolsep-1.3333\arrayrulewidth}|
  }
\hline

\centering Functional rel. & \centering Logical constraint (with \actorFunA $\mapsto$ \actorA, \actorFunB $\mapsto$ \actorB) & Visualisation \\
\hline

\textcolor{blue0}{\qaLeftPred{\actorFunA}{\actorFunB}} &
\textcolor{blue0}{$\exists r_l \geq 0, \paramA{h} + \sfrac{\pi}{2}-\textcolor{black}{\theta_l} \leq \alpha_l \leq \paramA{h} + \sfrac{\pi}{2}+\textcolor{black}{\theta_l}:
\paramB{x} = \paramA{x} + r_l \cos{\alpha_l} \wedge \paramB{y} = \paramA{y} + r_l \sin{\alpha_l}$} &
\multirow{4}{*}{
\includegraphics[valign=c,width=\linewidth-\tabcolsep]{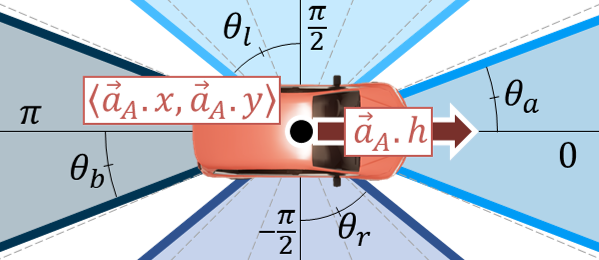}
}\\
\cdashline{1-2}

\textcolor{blue1}{\qaRightPred{\actorFunA}{\actorFunB}} &
\textcolor{blue1}{$\exists r_r \geq 0, \paramA{h} - \sfrac{\pi}{2}-\textcolor{black}{\theta_r} \leq \alpha_r \leq \paramA{h} - \sfrac{\pi}{2}+\textcolor{black}{\theta_r}:
\paramB{x} = \paramA{x} + r_r \cos{\alpha_r} \wedge \paramB{y} = \paramA{y} + r_r \sin{\alpha_r}$} &
\\
\cdashline{1-2}

\textcolor{blue2}{\qaFrontPred{\actorFunA}{\actorFunB}} &
\makecell[l]{\textcolor{blue2}{$\exists r_a \geq 0, \paramA{h}-\textcolor{black}{\theta_a} \leq \alpha_a \leq \paramA{h}+\textcolor{black}{\theta_a}:$} \\
\textcolor{blue2}{$\paramB{x} = \paramA{x} + r_a \cos{\alpha_a} \wedge \paramB{y} = \paramA{y} + r_a \sin{\alpha_a}$}}&
\\
\cdashline{1-2}

\textcolor{blue3}{\qaBehindPred{\actorFunA}{\actorFunB}} &
\makecell[l]{\textcolor{blue3}{$\exists r_b \geq 0, \paramA{h} + \pi -\textcolor{black}{\theta_b} \leq \alpha_b \leq \paramA{h} + \pi+\textcolor{black}{\theta_b}:$} \\
\textcolor{blue3}{$\paramB{x} = \paramA{x} + r_b \cos{\alpha_b} \wedge \paramB{y} = \paramA{y} + r_b \sin{\alpha_b}$}} &
\\
\hline

\rule{0pt}{18pt}
\textcolor{blue0}{\qaClosePred{\actorFunA}{\actorFunB}} &
0 \textcolor{blue0}{$ \leq \sqrt{(\paramA{x} + \paramB{x})^2 + (\paramA{y} + \paramB{y})^2} < $} $d_c$ &
\multirow{3}{*}{
\includegraphics[valign=c,width=\linewidth-\tabcolsep]{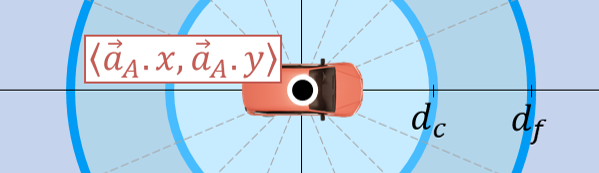}
}\\
\cdashline{1-2}
\rule{0pt}{18pt}
\textcolor{blue1}{\qaMedPred{\actorFunA}{\actorFunB}} &
$d_c$ \textcolor{blue1}{$ \leq \sqrt{(\paramA{x} + \paramB{x})^2 + (\paramA{y} + \paramB{y})^2} <$} $d_f$ &
\\
\cdashline{1-2}
\rule{0pt}{18pt}
\textcolor{blue2}{\qaFarPred{\actorFunA}{\actorFunB}} &
$d_f$ \textcolor{blue2}{$ \leq \sqrt{(\paramA{x} + \paramB{x})^2 + (\paramA{y} + \paramB{y})^2}$} &
\\
\hline

\textcolor{blue1}{\qaCanSeePred{\actorFunA}{\actorFunB}} &
\makecell[l]{\textcolor{blue1}{$\exists 0 \leq r_v \leq \textcolor{black}{d_v}, \paramA{h}-\textcolor{black}{\theta_v} \leq \alpha_v \leq \paramA{h}+\textcolor{black}{\theta_v},$} \\
\textcolor{blue1}{$\langle \justParamB{cx}, \justParamB{cy} \rangle \in \paramB{corners}:$} \\
\textcolor{blue1}{$\justParamB{cx} = \paramA{x} + r_v \cos{\alpha_v} \wedge \justParamB{cy} = \paramA{y} + r_v \sin{\alpha_v}$}}&
\includegraphics[valign=m,margin=0pt 1ex 0pt 1ex,width=0.9\linewidth]{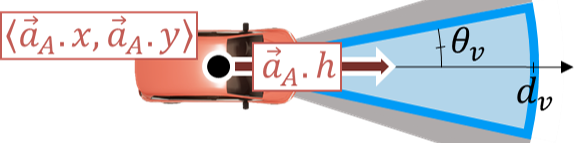}
\\
\hline

\textcolor{blue1}{\qaNoCollPred{\actorFunA}{\actorFunB}} &
\textcolor{blue1}{$\neg$ \textit{intersects}(\actorFunA, \actorFunB)}&
\includegraphics[valign=c,margin=0pt 1ex 0pt 1ex,width=0.8\linewidth]{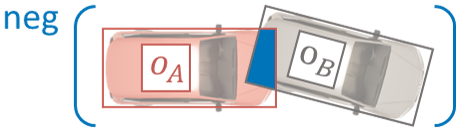}
\\
\hline

\textcolor{blue1}{\qaOnRoadPred{\actorFunA}} &
\makecell[l]{\textcolor{blue1}{$\forall \paramA{c} \in \paramA{corners}: \exists r_{map} \in \roadmapI{i} :$} \\
\textcolor{blue1}{\textit{contains}($r_{map}$, $\paramA{c}$)}
} &
\includegraphics[valign=c,margin=0pt 1ex 0pt 1ex,width=\linewidth]{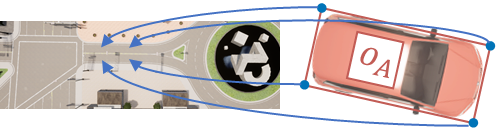}
\\
\hline

\end{tabular}
\caption{Mapping from functional relations to logical constraints}
\label{fig:func-to-log}
\end{figure*}

\subsection{Numeric concretization problem}
\label{sec:elab-num-formal}




\textit{Formalization}:
A logical  scene defines a numeric rectangle layouting problem that yields a concrete scene as solution.
Formally, such a numeric problem $N$ corresponds to a tuple $N = \langle \actorSetI{N}, \constrSetI{N}, \roadmapI{N}, \dimensionsI{N} \rangle$, where:

\begin{itemize}
    \item $\actorSetI{N}$ is a finite set of actors where each $\actorSampleNM{} \in \actorSetI{N}$ is an oriented rectangle defined as \textit{5-tuples} (see below),
    \item $\constrSetI{N}$ is a set of binary numeric (geometric) constraints $c_i (\actorNM{0}, \actorNM{1})$ over actors (oriented rectangles) $\actorSampleNM{} \in \actorSetI{N}$,
    \item $\roadmapI{N}$ is a road map that restricts the range of position variables for actors in $\actorSetI{N}$, and 
    \item $\dimensionsI{N}$ contains valid bounding box sizes (width, length) of actors as a finite set of floating-point pairs $\langle \texttt{w}_i, \texttt{l}_i \rangle$.
\end{itemize}
Note that we only handle binary numeric constraints in this paper to stay consistent with the functional-level binary relations proposed in \autoref{sec:func-lang}.
Nevertheless, our formalization can be generalized to \textit{n-ary} constraints.


We approximate actors $\actorSampleNM{} \in \actorSetI{N}$ as oriented rectangles over a map $\roadmapI{N}$ given as input.
Formally, an actor $\actorSampleNM{}$ is represented by a tuple $\actorSampleNM{} = \langle \justParam{x}, \justParam{y}, \justParam{h}, \justParam{w}, \justParam{l}\rangle$, where:
\begin{itemize}
    \item $\justParam{x}$ and $\justParam{y}$ are (floating point) variables that represent the center point of \actorSample~(where $\justParam{x} \in [0, \roadmapI{N}.x]$ and $\justParam{y} \in [0, \roadmapI{N}.y]$).
    Here, $\roadmapI{N}.x$ and $\roadmapI{N}.y$ respectively represent the width and length of the map $\roadmapI{N}$,
    \item $\justParam{h}$ is a (floating point) variable for the heading angle of \actorSample~(in \textit{radians}, i.e.  $\justParam{h} \in [-\pi, \pi]$),
    \item $\justParam{w}, \justParam{l}$ are (floating point) variables that represents the width and length of \actorSample~(where $\langle\justParam{w}, \justParam{l} \rangle \in \dimensionsI{N}$).
\end{itemize}

\textit{Deriving a numeric problem}:
A numeric problem
$N = \langle \actorSetI{N}, \constrSetI{N}, \roadmapI{N}, \dimensionsI{N} \rangle$
is derived from a partial model
$P = \logicModelTwo{\dslObjectSetI{P}}{\interpretationFun{P}}$
of a \fss~through a mapping $\funtolog: P \mapsto N$ between functional relations and numeric constraints such as the one proposed in \autoref{sec:elab-mapping}.
We derive $\funtolog$ by 
\begin{enumerate}
    \item providing the map $\roadmapI{N}$ and possible actor bounding box sizes $\dimensionsI{N}$ as external inputs (parameters),
    \item mapping each abstract object $\actorFunNoMath{i} \in \dslObjectSetI{P}$ to a corresponding numeric actor $\actorSampleNM{} \in \actorSetI{N}$,
    \item populating $\constrSetI{N}$ such that
    \begin{itemize}
        \item for every positive relation $\dslStyle{r} \in \Sigma$ over objects ${\actorFunNoMath{i}} \in \dslObjectSetI{P}$ (i.e. where $\interpretationFun{P}(\dslStyle{r})(\actorFunNoMath{0}, \actorFunNoMath{1}) = \true$), a corresponding numeric constraint
        $c_i (\actorNM{0}, \actorNM{1})$ (as defined in \autoref{sec:elab-mapping})
        is included in $\constrSetI{N}$.
        \item for every negative relation $\dslStyle{r} \in \Sigma$ over objects ${\actorFunNoMath{i}} \in \dslObjectSetI{P}$ (i.e. where $\interpretationFun{P}(\dslStyle{r})(\actorFunNoMath{0}, \actorFunNoMath{1}) = \false$), the \textit{negation} of $c_i (\actorNM{0}, \actorNM{1})$
        is included in $\constrSetI{N}$.
    \end{itemize}
\end{enumerate}

\textit{Defining a numeric solution}:
Given a numeric problem $N = \langle \actorSetI{N}, \constrSetI{N}, \roadmapI{N}, \dimensionsI{N} \rangle$ derived from a partial model $P = \logicModelTwo{\dslObjectSetI{P}}{\interpretationFun{P}}$ of a \fss, a solution $s_N: \actorSetI{N} \rightarrow \mathbb{R}^5$ is a value assignment of the variables associated to all actors $\actorSampleNM{} \in \actorSetI{N}$ (within the respective ranges) such that all constraints $c_i \in \constrSetI{N}$ are satisfied.
The numeric values assigned to variables represent a concrete  scene which is a concretization of the \fss~corresponding to $P$.

The concrete solution $s_N$ of a numeric problem $N$ can be abstracted into the partial model $P_s$ (of a functional scene) using the same set of logical constraints. 
The numeric constraint corresponding to each instance of relation $\dslStyle{r} \in \Sigma$ is evaluated on $s_N$.
Each constrain evaluation yields a Boolean truth value (true/false).
These Boolean values derived from such a mapping $\logtofun: s_N \mapsto P_s$ define a concrete partial model that does not contain \unk~values.

\newcommand{\theoremA}[0]{\theorembox{1}{
For a numeric problem $N = \funtolog(P)$ derived from a \fss, the concrete partial model $P_s$ abstracted from a solution $s_N$ of $N$ as $P_s = \logtofun(s_N)$ satisfies all relations in $P$ (formally, $P_s$ refines $P$, i.e. $P \sqsubseteq P_s$ \cite{Marussy2020JOT}). 
}}

\textit{Soundness of a numeric solution}:
In the appendix, we prove the soundness of our approach, as captured in \theorem{1}, given that every individual relation implementation soundly captures the geometrical requirements of the corresponding constraint.

\theoremA{}


\begin{exline}
The scene concretization problem in \autoref{fig:sceneAbsLevels} is defined for 2 actors \actorFunR~and \actorFunBl, and 3 functional relations
\textcolor{OliveGreen}{\qaFrontPred{\actorFunR}{\actorFunBl}} and \textcolor{LimeGreen}{\qaMedPred{\actorFunBl}{\actorFunR}} (visualised as arrows), and a negative relation 
\texttt{!}\qaRightPred{\actorFunBl}{\actorFunR}.
For better presentation, identical colors relate actors and constraints on different levels, and  numeric actor representations on logical and concrete levels exclude width and length variables.
A concrete scene (solution) is defined as an assignment of actor variables to satisfy the logical constraints, e.g. {\footnotesize \textcolor{LimeGreen}{$c_l \leq \sqrt{(\paramR{x} + \paramBl{x})^2 + (\paramR{y} + \paramBl{y})^2} \leq c_u$}}  in case of \textcolor{LimeGreen}{\qaMedPred{\actorFunBl}{\actorFunR}}.
\end{exline}

\subsection{Mapping functional relations to logical constraints}
\label{sec:elab-mapping}

In this section, we describe the formal mapping from functional relations in a partial model (i.e. \textit{relation symbols} in the abstract vocabulary $\Sigma$) to corresponding logical (numeric) constraints, alongside a visualisation depicted over actors is proposed in \autoref{fig:func-to-log}.
Our mapping builds on but also generalizes previous work by Menzel, \textit{et al.} \cite{Menzel2019}.  

In our notation, \actorFunA~and \actorFunB~define actors at the functional level (i.e. partial model objects), which are respectively mapped to actors \actorA~and \actorB~at the logical level to find a concrete solution (see \autoref{sec:elab-num-formal}).
Corresponding relations and numeric attributes are color-coded.
Customizable constants, such as $\theta_l$ and $d_c$, are depicted in black.

\textbf{Positional relations:}
When actor \actorFunA~is connected to actor \actorFunB~via a positional relation $c \in \symbolClass{C}{pos}$, this signifies that the center point $\langle \paramB{x}, \paramB{y} \rangle$ of \actorB~is located in a circular sector centered at $\langle \paramA{x}, \paramA{y} \rangle$, with infinite radius.
The orientation and central angle of the sector is defined according to the specific positional relation.
For example, the \textcolor{blue0}{\qaLeftPred{\actorFunA}{\actorFunB}} relation denotes a circular sector that covers the region located at the left of \actorA~(with respect to its heading $\paramA{h}$).
With that respect, \textcolor{blue0}{\qaLeftPred{\actorFunA}{\actorFunB}} means that \actorB~is positioned \textit{to the left of} \actorA.


\textbf{Distance relations:}
A distance relation between actors \actorFunA~and \actorFunB~signifies that the Euclidean distance between the center point of each actor falls within a certain numeric range.
For example, \textcolor{blue1}{\qaMedPred{\actorFunA}{\actorFunB}} requires that the Euclidean distance between $\langle \paramA{x}, \paramA{y} \rangle$ and $\langle \paramB{x}, \paramB{y} \rangle$ is a value in $[d_c, d_f[$.
The thresholds of distance relations can be customized, e.g., 
the definition of \textit{closeness} can be different for pedestrians and vehicles on a highway. 


\begin{figure}[htp]
    \footnotesize
    \setlength{\tabcolsep}{4pt}
    \centering
    \begin{tabular}{
      m{\dimexpr.6\linewidth-2\tabcolsep-1.3333\arrayrulewidth}
      |m{\dimexpr.4\linewidth-2\tabcolsep-1.3333\arrayrulewidth}
      }
      
        \makecell[l]{\textcolor{blue1}{$\param{corners}{A}$} $\equiv$ \\
        $\Big\{\big\langle \paramA{x} +  L \cdot \cos{\paramA{h}} - W \cdot \sin{\paramA{h}},$\\
        $ \paramA{y} + L \cdot \sin{\paramA{h}} + W \cdot\cos{\paramA{h}} \big\rangle$,\\
        where $L = \pm\frac{\paramA{l}}{2}, W = \pm\frac{\paramA{w}}{2}\Big\}$
        }
        &
        \includegraphics[valign=c,width=\linewidth]{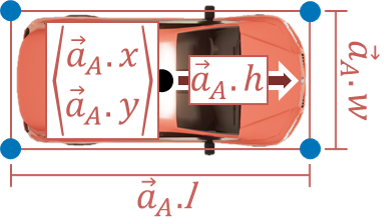}
            
    \end{tabular}
    \caption{Formalization for the \textcolor{blue1}{$\param{corners}{A}$} parameter}
    \label{fig:elab-corners}
\end{figure}


\textbf{Visibility relations:}
Visibility relations consider that the shape of an actor \actorB~is approximated as a rectangle (bounding box) with width $\paramB{w}$ and length $\paramB{l}$.
As such, \actorB~has four corners (i.e. the four corners of the rectangular approximation).
Their coordinates are collected in $\paramB{corners}$ according to the definition proposed in \autoref{fig:elab-corners}
For example, relation \textcolor{blue1}{\qaCanSeePred{\actorFunA}{\actorFunB}}  requires for at least one of the corners of \actorB~to be in a circular sector (i.e. the field of view of \actorA) centered at $\langle \paramA{x}, \paramA{y} \rangle$ in the direction of $\paramA{h}$.
Unlike the \textcolor{blue3}{\qaFrontPred{\actorFunA}{\actorFunB}} relation, the circular sector defined by visibility relations has a finite radius $r_v$.



\textbf{Collision avoidance relations:}
We reuse the definition of collision avoidance proposed in the \scenic~framework \cite{Fremont2019ScenicLanguage}.
Informally, the \textcolor{blue1}{\qaNoCollPred{\actorFunA}{\actorFunB}} relation states that the area within the bounding boxes of \actorA~and of \actorB~do not intersect.
Bounding box analysis is outside the scope of this paper, therefore we define collision avoidance through a call to the \scenic~library function \textcolor{blue1}{\textit{intersects}(\actorA, \actorB)}.

\textbf{Road placement relations:}
We also rely on the \scenic~library functions to define road placement relations.
Informally, \textcolor{blue1}{\qaOnRoadPred{\actorFunA}} states that the four corners of \actorA~(i.e. $\paramA{corners}$) must be located on (contained by) a road that is part of the scene map $\roadmapI{i}$ provided as input.
The scene map is segmented into multiple connected roads which are represented as complex polygons (e.g curved roads, or roads with varying width).
Given a road segment \textcolor{blue1}{$r_{map}$} that is part of the scene map, the \textcolor{blue1}{\textit{contains}($r_{map}$, $\paramA{c}$)} function call checks whether the point $\paramA{c}$, which is a corner point of \actorA, is placed within the bounds of the polygon represented by \textcolor{blue1}{$r_{map}$}.
Once again, polygon detection is outside the scope of this paper, therefore we refer to \scenic~library functions.

\subsection{Benefits of the mapping}
\label{sec:elab-benefits}

While our mapping is based on existing research \cite{Menzel2019}, it conceptually extends this baseline in three different aspects. 

First, thanks to the use of partial models as \fss{}s, our mapping defined in   \autoref{fig:func-to-log} is \textit{extensible} to arbitrary qualitative abstractions from a concrete scene to an abstract scene on the functional level. As such, we can seamlessly incorporate additional relations proposed by safety experts that can be observed and measured over a concrete model.
Furthermore, the implementation of additional relations may be iteratively validated by (1) creating a simple \fss{} containing only the newly implemented relation, (2) concretizing the \fss{}, (3) visually detecting implementation issues, if any, and (4) adjusting the implementation accordingly.


Moreover, our approach maps abstract functional relations to numerical constraints over 2-dimensional space with no geometric assumptions. Thus, our mapping is 
\textit{independent from the underlying road map}, and it can be contextualized to any real physical location on a map. Existing $\funtolog$ mappings \cite{Babikian2021dReal,Abdessalem2018TestingFeatureInteractionsBriandNsgaForConcreteScenes} are typically hard-coded to the specific geometry of a particular (often simplistic) road map.

Finally, our mapping is \textit{customizable} to the needs of a given scene.
For instance, parameters such as the thresholds for closeness, or the angle for the field of view of an actor can be adjusted according to its types (e.g. pick-up truck or pedestrians).
Such flexibility allows the generation of more realistic scenes compared to existing approaches that map functional relations to numeric constraints.




\contribox{2}{We provide an \textit{extensible} and  \textit{customizable} mapping from abstract functional relation to numeric constraints, which can incorporate arbitrary qualitative abstractions, and can be \emph{contextualized in any physical locations of a map}. 
}
\section{Using \moopt~to derive concrete scenes}
\label{sec:elab-concretization}




 To derive a concrete scene from a \fss, the numeric scene concretization problem defined  in  \autoref{sec:elab-func-to-log} needs to be solved.
Since exact algorithms, such as quadratic solvers, have failed to provide scalable results for similar problems \cite{Babikian2021dReal}, in this paper, we adopt metaheuristic search (\moopt) algorithms, which are commonly used in the domain of AV testing 
\cite{Abdessalem2018TestingFeatureInteractionsBriandNsgaForConcreteScenes,Abdessalem2016TestingADAS,Abdessalem2018TestingVisionBased,calo2020GeneratingAvoidableCollision,Wu2021,Gambi2019AsFaultFullPaper}.

\newcommand{\moovars}[0]{\mathcal{V}}
\newcommand{\moovarsI}[1]{\moovars_{#1}}
\newcommand{\mooobjs}[0]{\mathcal{OF}}
\newcommand{\mooobjsI}[1]{\mooobjs_{#1}}



\textit{Formalization}:
A numeric scene concretization problem $N$ is mapped to a metaheuristic minimization (\mom) problem that can be solved by a \moopt~algorithm.
A \mom{} problem is formalized as \(M_{min} = \langle \moovarsI{M}, \mooobjsI{M} \rangle\), where:
\begin{itemize}
    \item $\moovarsI{M}$ represents a set of variables \( \{ v_{1, 1}, \dots, v_{m, 5}\} \), where $m$ is the number of actors (represented as \textit{5-tuples}) in $N$. The domain of each variable $v_i \in \moovarsI{M}$ is defined by a  numeric range $[l_i, b_i]$.
    \item $\mooobjsI{M}$ represents a set of objective functions \( \{ {OF}_1, \dots, {OF}_n\} \) defined over variables in $\moovarsI{M}$ that return \textit{non-negative} numeric values.
\end{itemize}


At each iteration of the \moopt~algorithm, a set of \textit{candidate solutions} is derived, which are assignments for each $v_i \in \moovarsI{M}$ to a value within the corresponding range.
A candidate solution is a valid \textit{solution} $s_{min}$ to $M_{min}$ iff all objective functions are minimal (i.e. zero). To resolve potential issues with precision of floating point variables, we require that the value of an objective function should be below a given threshold $\epsilon > 0$.

As is the case for existing research \cite{Cai2006,Wang2008,Ray2009}, we propose an \textit{unconstrained} \mom{} problem formalization: all logical-level constraints are directly incorporated into objective functions.
As such, no additional handling of constraints is required by the underlying \moopt{} algorithm.

\textit{Deriving a \mom~problem}:
Given a numeric problem $N = \langle \actorSetI{N}, \constrSetI{N}, \roadmapI{N}, \dimensionsI{N} \rangle$ representing a logical scene as input, we define a corresponding metaheuristic minimization problem \(M_{min} = \langle \moovarsI{M}, \mooobjsI{M} \rangle\) as follows:
\begin{enumerate}
    \item $\moovarsI{M}$ collects the variables $v \mid \param{v}{i}$ that define all actors $\actorSampleNM \in \actorSetI{N}$ 
    \item $\mooobjsI{M}$ is a set of objective functions derived from distance functions associated to numeric constraints $\constrSetI{N}$.
\end{enumerate}

\begin{figure}[htp]
    \footnotesize
    \captionsetup{justification=centering}
    \centering
    \setlength{\tabcolsep}{4pt}
    \begin{tabular}{
      |m{\dimexpr.20\linewidth-2\tabcolsep-1.3333\arrayrulewidth}
      |m{\dimexpr.80\linewidth-2\tabcolsep-1.3333\arrayrulewidth}|
      }
        
        \hline
        \centering \makecell[c]{ Relation \\category} & \centering\arraybackslash Distance functions $({DF}_{pos},\dots,{DF}_{road})$ \\
        \hline
        \makecell[c]{$r(\actorFunNMA, \actorFunNMB)|$ \\ $r \in \symbolClass{R}{pos}$ } & (${DF}_{pos}$): If \actorFunB~is within the circular sector defined by $r$, ${DF}_{pos}$ returns 0. Otherwise, ${DF}_{pos}$ returns the angle relative to \actorFunA~between the $\overline{\actorFunNMA\actorFunNMB}$ segment and the closest edge of the circular sector defined by $r$.\\
        \hline
        \makecell[c]{$r(\actorFunNMA, \actorFunNMB)|$ \\ $r \in \symbolClass{R}{dist}$ } & (${DF}_{dist}$): If both actors are at an appropriate distance, ${DF}_{dist}$ returns 0. Otherwise, ${DF}_{dist}$ denotes the shortest distance that \actorFunB~must traverse to be positioned within the distance bounds required by $r$.\\
        \hline
        \makecell[c]{$r(\actorFunNMA, \actorFunNMB)|$ \\ $r \in \symbolClass{R}{vis}$ } & (${DF}_{vis}$): If \actorFunA~can see \actorFunB, ${DF}_{vis}$ returns 0. Otherwise, ${DF}_{vis}$ denotes the shortest distance that \actorFunB~must traverse for at least one of its corners to be in the field of view of \actorFunA.\\
        \hline
        \makecell[c]{$r(\actorFunNMA, \actorFunNMB)|$ \\ $r \in \symbolClass{R}{coll}$ }& (${DF}_{coll}$): ${DF}_{coll}$ returns 0 if \actorFunA~and \actorFunB~are positioned such that they do not collide, 1 otherwise.\\
        \hline
        \makecell[c]{$r(\actorFunNMA, \actorFunNMA)|$ \\ $r \in \symbolClass{R}{road}$ } & (${DF}_{road}$): If all four corners of \actorFunA~are placed on a road contained in \textit{RoadMap}, ${DF}_{road}$ returns 0. Otherwise, ${DF}_{road}$ denotes the shortest distance that \actorFunA~must traverse for \actorFunA~to be placed on a road.\\
        \hline
    \end{tabular}
    \caption{Informal overview of distance functions\\derived from positive functional relations }
    \label{fig:distance-functions}
\end{figure}

\textit{Distance functions}:
Each numeric constraint $c \in \constrSetI{N}$ has a corresponding distance function ${DF}_i(c)$ which is derived according to the relation category of $c$ (e.g. different functions for visibility and positional constraints).
A distance function ${DF}_i(c)$ returns a non-negative number that represents how far a candidate solution $CS$ is from satisfying $c$.
${DF}_i = 0$ iff the corresponding numeric constraint $c$ holds for the variable assignment defined by $CS$.

In case of positive constraints (i.e. derived from positive functional relations), ${DF}_i$ is computed according to \autoref{fig:distance-functions}.
Negative constraints are defined similarly, where $0$ is returned if the negation of the corresponding (positive) constraint is satisfied. The set of distance functions is easily extensible by other relations defined by experts without further change in the underlying \mom~problem.


\textit{Objective functions}:
Many \moopt{} algorithms are designed to handle a reduced number of objective functions.
As such, fitness values of multiple distance functions may be combined into an aggregate objective function ${OF}_{P_i} \in \mooobjsI{M}$ according to various aggregation strategies, such as:

\begin{itemize}
    \item \textbf{Global aggregation:} All constraints are aggregated into a single objective function.
    \item \textbf{Category aggregation:} All constraints corresponding to the same functional relation category are aggregated.
    \item \textbf{Actor aggregation:} All constraints applied to the same source actor are aggregated.
\end{itemize}

Each aggregation strategy partitions the set of numeric constraints $\constrSetI{N}$ into distinct subsets $P_i \subseteq \constrSetI{N}$ that contain all constraints associated to a specific objective function. 
For instance, \textbf{category aggregation} would create a subset for each functional relation category, and \textbf{actor aggregation} would create a subset for each actor.

The choice of objective functions influences the appropriate search algorithms.  
For instance, single-objective optimization algorithms (i.e. genetic algorithm) are ideal for \textbf{global aggregation}, while multi-objective (e.g. NSGA-II \cite{Deb2002NSGAII}), many-objective (e.g. NSGA3 \cite{Deb2014NSGA3}) or custom optimization algorithms are ideal for other aggregation strategies.

Formally, given a subset $P_i$ of numeric constraints, an objective function is defined as:

{\footnotesize
\[
{OF}_{P_i} = f_{P_i}\left(\sum_{c \in P_i } {DF}_{i}(c)
\right)
\]}


\begin{itemize}
    \item $f_{P_i}()$ is an arbitrary (non-negative) weight function that does not modify the minima.
    \item ${DF}_{i}(c)$ is the output of a distance function measured over a single constraint $c$, as detailed in \autoref{fig:distance-functions}.
\end{itemize}

 

Our objective function is a weighted aggregation of distance functions.
Different weights can help fine-tune different characteristics of solutions (e.g. realisticness, diversity). 


\newcommand{\theoremB}[0]{\theorembox{2}{
For a \mom~problem $M_{min}$ derived from a numeric problem $N$, a solution $s_{min}$ to $M_{min}$ is also a solution to $N$ (i.e. all numeric constraints are satisfied). 
}}

\textit{Soundness}:
The soundness of our approach is captured by Theorem 2 (with a formal proof in the appendix):

\theoremB{}


\contribox{3}{We solve scene concretization as a metaheuristic minimization problem where objective functions are derived from various aggregation strategies.
%
}
\section{Evaluation}
\label{sec:evalution}

\newcommand{\dmn}[1]{\textbf{\textsc{#1}}}

\newcommand{\domcarla}[0]{\dmn{Carla}}
\newcommand{\domtram}[0]{\dmn{Tramway}}
\newcommand{\domzala}[0]{\dmn{Zalazone}}

\newcommand{\apr}[1]{\textcolor{darkgray}{\textbf{#1}}}
\definecolor{ev_r}{HTML}{A50205}
\definecolor{ev_b}{HTML}{0876B5}
\definecolor{ev_o}{HTML}{CC6400}
\definecolor{ev_p}{HTML}{5813B7}
\newcommand{\ansga}{\apr{MHS}}
\newcommand{\ascenic}{\apr{Scenic}}
\newcommand{\asceA}{\apr{SceDef}} 
\newcommand{\asceB}{\apr{SceHyb}} 
\newcommand{\asceC}{\apr{SceReg}} 

\newcommand{\aggStrat}[1]{\texttt{#1}}
\newcommand{\aggGlob}{\aggStrat{G}}
\newcommand{\aggCat}{\aggStrat{C}}
\newcommand{\aggAct}{\aggStrat{A}}
\newcommand{\aggWeiImp}{\aggStrat{WD}}
\newcommand{\aggWeiCat}{\aggStrat{WC}}
\newcommand{\aggNOne}{\aggStrat{\o}}

\newcommand{\evolApp}[1]{\texttt{#1}}
\newcommand{\evolGA}{\evolApp{GA}}
\newcommand{\evolNTwo}{\evolApp{N2}}
\newcommand{\evolNThree}{\evolApp{N3}}

We conducted various measurements to address the following research questions:


\begin{itemize}
  \item[\rquestion{1}:] Which \moopt{} configuration provides the best scene concretization results in terms of success rate and runtime?
  \item[\rquestion{2}:] How does our approach compare to state-of-the-art scene concretization approaches with respect to success rate (\rquestion{2.1}), runtime (\rquestion{2.2}) wrt. different maps, and success rate wrt. increasing number of actors (\rquestion{2.3})?
  \item[\rquestion{3}:] How does our approach scale/fail wrt. an increasing number of constraints?
  \item[\rquestion{4}:] How does our approach scale/fail when concretizing scenes with large number of actors? 
\end{itemize}

\subsection{Case studies}
To answer these questions, we execute scene concretization campaigns over three road maps.

\domcarla: The \textsc{Carla} simulator framework \cite{Dosovitskiy2017CarlaSimulator} includes multiple road networks alongside realistic depictions of the surrounding environment.
We perform experiments over the \textit{Town02} road map (215$\times$217 units$^2$) included in \textsc{Carla}, which represents a simple town consisting of "T junctions".

\domzala: The \textit{ZalaZONE Automotive Proving Ground} \cite{Szalay2018ZalaZoneMap} is a physical test track located in Zalaegerszeg, Hungary designed to conduct experiments related to the safety assurance of AVs.
The test track consists of various subsections adapted to the testing of different facets of AVs.
We perform experiments over a digital twin of the \textit{Smart city} portion of the test track (270$\times$474 units$^2$), which features a focused urban layout with multiple complex intersections, roundabouts, parking lots and curved roads.

\domtram: We also perform experiments over a real-life road network (258$\times$181 units$^2$) in Budapest, Hungary.
A digital twin for this road network is derived together with an industrial partner using the public \emph{OpenStreetMap} database.
This road network features many multi-lane, complex intersections located over a dedicated tramway lane.

\subsection{Compared approaches}

\newcommand{\stat}[1]{\textbf{\textsc{#1}}}
\newcommand{\rven}{\stat{veneer}}
\newcommand{\rreg}{\stat{region}}
\newcommand{\rreq}{\stat{require}}
\newcommand{\rnative}{\stat{native}}

\newcommand{\y}{\stat{Yes}}
\newcommand{\n}{\stat{No}}

We compare our proposed approach with three variations of the baseline \ascenic{} approach.
Our evaluation does not consider manual or semi-automatic scene concretization for comparative evaluation considering that, despite being conceptually simple for humans, manually concretizing scenes into simulation-ready representations (1) is very time-consuming, as it relies on trial and error, and (2) does not ensure formal correctness.
Automated scene concretization approaches, such as the ones presented below, address both difficulties by (1) automatically synthesizing simulator-friendly scenes (2) that are guaranteed (by the underlying algorithm) to satisfy the mathematical definitions of the abstract constraints.

\ascenic: As a baseline reference, we concretize \fss{}s represented through the \scenic \cite{Fremont2019ScenicLanguage} specification language.
For this purpose, we use the integrated approach based on rejection sampling proposed by the framework.
However, \ascenic~has expressiveness limitations (i.e. certain scene specifications cannot be expressed, see \autoref{sec:prel-staticAnal}).
Therefore, we identify three variations of the \ascenic~approach which use different constructs to represent abstract relations. 
In our experiments, we remove the minimum number of relations to make the resulting scene expressible.




\begin{itemize}
    \item \asceA~(Default):
    \textbf{Positional} (and accompanying \textbf{distance}) relations are represented by the built-in constructs (referred to in the framework as \rven{}s) at actor definition time which restricts the sample space to a line segment on the map relative to the position of other actors.
    An example of the numeric constraint corresponding to a \rven{} is given in \autoref{fig:sceneAbsLevels} for the \textcolor{OliveGreen}{\qaFrontPred{\actorFunR}{\actorFunBl}} relation.
    
    When distance relations cannot be expressed, they are represented by \rreq~clauses, which are handled as acceptance conditions.
    \textbf{Visibility} relations are also handled through \rreq~clauses.
    Neither cyclic positional relations nor actors with multiple dependencies can be handled in this variation.
    
    \item \asceC~(Regions):
    \textbf{Positional}, \textbf{distance} and \textbf{visibility} relations are represented as \rreg{} instances, which restrict the sample space to certain regions of the map relative to the position of other actors.
    In this variation, functional relations are mapped to numeric constraints as proposed in this paper.
    This variation cannot handle cycles of positional and/or visibility relations.
    
    \item \asceB~(Hybrid):
    This variation combines the previous variations by handling \textbf{positional} relations as \rreg~instances, and representing \textbf{distance} and \textbf{visibility} relations as  acceptance conditions (\rreq~ clauses).
    As such, this variation stays consistent with the $\funtolog$ mapping proposed in this paper.
    Additionally, the only expressiveness restriction is that cycles of positional relations cannot be represented.
    
\end{itemize}
Once a \fss~is defined, the functional scene is concretized by rejection sampling.
\textbf{Collision avoidance} and \textbf{road placement} relations are included as additional acceptance conditions.
However, to ensure that the approach can solve the entire scene concretization problem (i.e. not only the expressible subset of the problem), we check whether a derived concretization satisfies the removed, inexpressible relations.
This is implemented as an additional acceptance condition for the concretized scene that is evaluated after the termination of the default sampling approach.

\newcommand{\undForm}[2]{$\textit{#1}_\textit{#2}$}
\begin{table*}[htp]
    \footnotesize
    \captionsetup{justification=centering}
    \caption{Overview of the hyperparameters and genetic operators used for each evaluated \moopt{} algorithm. \\
        The table refers to \textit{Simulated Binary Crossover} (SBX) \cite{Deb2007SBXPM}, \textit{Polynomial Mutation} (PM) \cite{Deb2007SBXPM}, the \textit{Das-Dennis} (DD) \cite{Das2006DD} approach for determining the number of reference directions \undForm{n}{refDirs}, \textit{Binary Tournament} (BT), and \textit{Non-Dominated Sorting} (NDS) . \\
        Furthermore, \textit{pr} refers to probability and $\eta$ refers to the distribution index, while in our implementation, \undForm{n}{var} = $2\times$\undForm{n}{actors}.}
    \noindent
    \centering
        \begin{tabularx}{\linewidth}{c||c|c ||Z|Z|c|c|}
              & \makecell[c]{\textbf{Pop.} \\ \textbf{Size}} & \textbf{\undForm{n}{offsprings}}  & \textbf{Crossover op.}  & \textbf{Mutation op.} & \textbf{Selection op.} & \textbf{Survival op.} \\
             \hline
             
             \evolGA{} & 5 & 5  & 
             SBX(\textit{pr}=0.9, $\eta$=3) & 
             PM(\textit{pr}=$\frac{1}{\textit{n}_\textit{var}}$, $\eta$=5) & 
             Fitness BT & 
             Fitness \\
             \hdashline
             
             \evolNTwo{} & 5 & 5 & 
             SBX(\textit{pr}=0.9, $\eta$=15)& 
             PM(\textit{pr}=$\frac{1}{\textit{n}_\textit{var}}$, $\eta$=20) &
             \makecell[c]{Fitness Domination and \\ Crowding Distance BT} & 
             \makecell[c]{NDS Rank and \\Crowding Distance} \\
             \hdashline
             
             \evolNThree{} & \undForm{n}{refDirs} & \undForm{n}{refDirs} & 
             SBX(\textit{pr}=1.0, $\eta$=30)& 
             PM(\textit{pr}=$\frac{1}{\textit{n}_\textit{var}}$, $\eta$=20) &
             Random BT &
             \makecell[c]{NDS Rank and \\ Reference Direction \\ DD(\undForm{n}{dimensions}=\undForm{n}{objectives}, \undForm{n}{partitions}=1)} \\
        \end{tabularx}
    \label{fig:evol-params}
\end{table*}

\ansga: We implement the \moopt-based approach proposed in this paper using various objective function aggregation strategies.
Aside from the \textit{global} \aggGlob{}, \textit{category} \aggCat{} and \textit{actor} \aggAct{} strategies proposed in \autoref{sec:elab-concretization}, we also implement:
\begin{itemize}
    \item \textit{weighted category aggregation} \aggWeiCat{} (a variant of \aggCat{})
    \item \textit{weighted dependency aggregation} \aggWeiImp{} (two objective functions are defined according to the dependency structure of constraints type (see \autoref{sec: new-RQ3}): \textbf{collision avoidance} and \textbf{road placement} combine for one objective function, while the remaining constraint types form the other)
    \item \textit{no aggregation} \aggNOne{} (each constraint in the \fss{} defines a separate objective function ).
\end{itemize}
In case of \aggWeiCat{} and \aggWeiImp{}, higher weight 
is given to constraint categories that have less dependencies (i.e. \textbf{collision avoidance} and \textbf{road placement} constraints).
Specifically, the objective functions for \aggWeiCat{} are:
\begin{itemize}
    \item ${OF}_{road}, {OF}_{coll} = {(\sum_{ c \in \  \symbolClass{R}{i} } {DF}_i (c))}^3$, \\ where $i$ is $road$ and $coll$ respectively, and
    \item ${OF}_{pos}, {OF}_{dist}, {OF}_{vis} = {(\sum_{ c \in  \  \symbolClass{R}{i} } {DF}_i (c))}^2$, \\ where $i$ is $pos$, $dist$ and $vis$ respectively.
\end{itemize}
The objective functions for \aggWeiImp{} are:
\begin{itemize}
    \item ${OF}_{1} = {(\sum_{ c \in \  \symbolClass{R}{road} \cup \symbolClass{R}{coll}  } {DF}_j (c))}^3$, and
    \item ${OF}_{2} = {(\sum_{ c \in \  \symbolClass{R}{pos} \cup \symbolClass{R}{dist} \cup \symbolClass{R}{vis} } {DF}_j (c))}^2$,
\end{itemize}
where $j$ represents the relation category of $c$, and 
$c \in \symbolClass{R}{i}$ is shorthand for $c \in  \constrSetI{N} | c  = \funtolog(r(\actorFunNMA, \actorFunNMB)) \wedge r \in \symbolClass{R}{i} $.

We evaluate our approach using three underlying \moopt{} algorithms through the \textsc{PyMoo} Python library \cite{Blank2020PyMoo}:
\begin{itemize}
    \item a single-objective \textit{genetic algorithm} \evolGA{} 
    \item a multi-objective \textit{NSGA-II} algorithm \evolNTwo{} \cite{Deb2002NSGAII}
    \item a many-objective \textit{NSGA-3} algorithm \evolNThree{} \cite{Deb2014NSGA3}
\end{itemize}
\autoref{fig:evol-params} provides an overview of the relevant hyperparameters and genetic operators used as part of our experimentation.
Genetic operators are selected according to the default \textsc{PyMoo} settings.
Population size and number of offsprings are selected according to preliminary measurements, which are included on the publication page. 
While small population sizes are rather unusual for \moopt{} algorithms, we believe their good performance is attributed to the particularities of our experimental setup (e.g. \moopt{} approaches are often designed to provide multiple partial solution, whereas in our experimentation a single, complete solution is retrieved).

The implemented \moopt{} configurations handle all functional constraints of our scene specification language, and handle logical constraints of \autoref{sec:elab-func-to-log}.
As such, input scenes do not need to be adjusted, as is the case for \ascenic~approaches.
A comparison of relation representations, and the handling of relation structures is provided in \autoref{fig:scenic-relation-handling}.
Furthermore, to stay consistent with the baseline \ascenic{} approaches, (1) our experimentation runs are terminated once \textit{a single solution} to the problem defined by the \fss{} is found, and (2) particular handling for same-scene concretization tasks is not implemented.

\begin{table}[htp]
    \footnotesize
    \captionsetup{justification=centering}
    \caption{Comparison of relation representation and handling of relations by concretization approaches}
    \centering
        \begin{tabularx}{\linewidth}{|r||Z|c|Z|}
          \hline
          \multirow{2}{*}{Approach} & \multicolumn{3}{c|}{Representing relations}\\
          \cline{2-4}
          & Positional & Distance & Visibility\\
          \hline
          \asceA & \rven & \rven/\rreq & \rreq\\
          \asceC & \rreg & \rreg & \rreg\\
          \asceB & \rreg & \rreq & \rreq\\
          \ansga & \rnative & \rnative & \rnative\\
          \hline
          \hline
          \multirow{3}{*}{Approach} & \multicolumn{3}{c|}{Handled relation structures}\\
          \cline{2-4}
          & Positional & Visibility & Multiple\\
          & cycles & cycles & dependencies\\
          \hline
          \asceA & \n & \y & \n\\
          \asceC & \n & \n & \y\\
          \asceB & \n & \y & \y\\
          \ansga & \y & \y & \y\\
          \hline
        \end{tabularx}
    \label{fig:scenic-relation-handling}
\end{table}

\subsection{General measurement setup}
\label{sec:res-genmeasset}

To evaluate various concretization approaches, we randomly generated \fss{}s to be used as input:

\begin{enumerate}
    \item Given a number of required actors, we create a preliminary \fss{} $P_{pre}$ used to derive the input \fss{}s for our experiments.
    $P_{pre}$ contains road placement constraints for each actor and collision avoidance constraints for each pair of actors, to yield realistic concrete scenes.
    Additionally, a maximum distance $r$ between actors is established to avoid randomly generating scenes where no relations exist between actors.
    \item We use the default sampling-based scene concretization approached proposed in \scenic \cite{Fremont2019ScenicLanguage} to derive a numeric solution $s_{pre}$ for a numeric problem $N_{pre} = \funtolog(P_{pre})$.
    Note that different runs yield different numeric solutions for a same $P_{pre}$ given as input.
    \item We derive a \fss{} $P_{in}$ by applying qualitative abstractions over $s_{pre}$ (i.e. $P_{in} = \logtofun(s_{pre})$).
    Qualitative abstractions are applied for all relation categories.
    As such, any \textbf{positional}, \textbf{distance} and \textbf{visibility relations} that hold in $s_{pre}$ are included in $P_{in}$.
    Additionally, we know that $P_{in}$ is not contradictory, as it has at least one feasible solution ($s_{pre}$).
    \item We use $P_{in}$ as input for our measurements.
\end{enumerate}


Throughout our measurements, for simplicity, we consider actors with pre-defined length and width.
Additionally, we refer to the road network to determine the expected heading at a given position in the map.
We assume that each position has a single expected heading (we avoid non-determinism at intersections, where vehicles may take multiple paths that cross each other).
As such, our scene concretization runs derive the position coordinates of each actor, and the heading is determined accordingly.



We performed the measurements on an enterprise server\footnote{12\,\(\times\)\,2.2\,GHz CPU, 64\,GiB RAM, CentOS 7, Java 1.8, 12\,GiB Heap}.
Measurements are run in a Python environment, and the garbage collector is called explicitly between runs.
All generated FSSs, concrete scenes along with related measurements and figures are included in the publication page.

\begin{figure}[htp]
\noindent
    \captionsetup{justification=centering}
    \centering
    \begin{tabular}{c}
        \includegraphics[valign=c, margin=0pt 1ex 0pt 1ex, width=\linewidth-2\tabcolsep-1.3333\arrayrulewidth]{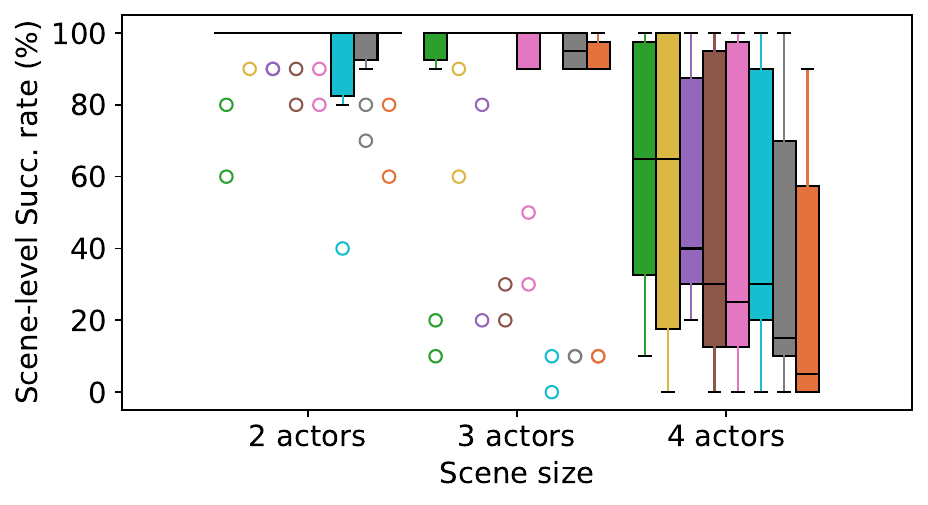}    \\
        \includegraphics[valign=c, margin=0pt 1ex 0pt 1ex, width=\linewidth-2\tabcolsep-1.3333\arrayrulewidth]{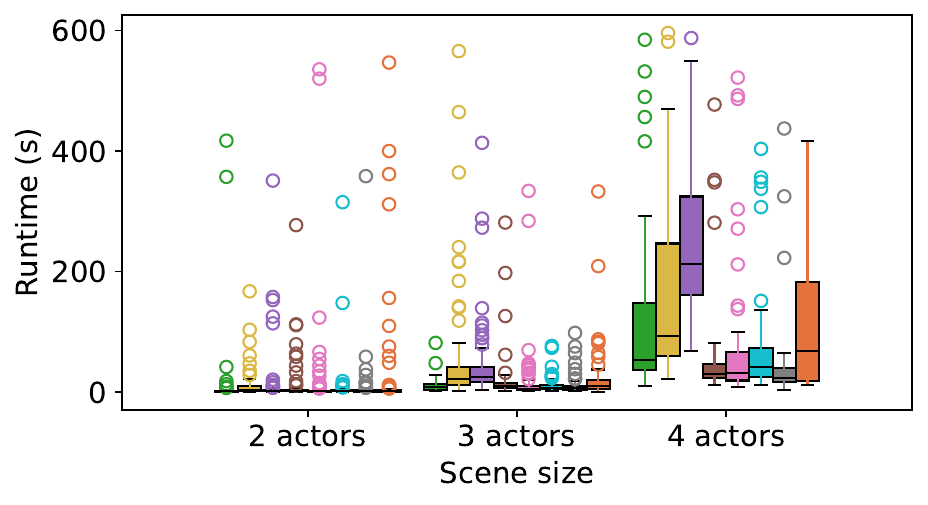} \\
        \includegraphics[valign=c, margin=0pt 1ex 0pt 1ex, width=\linewidth-2\tabcolsep-1.3333\arrayrulewidth]{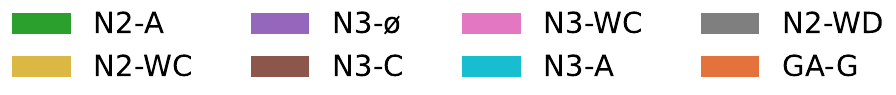} 

    \end{tabular}
    \captionsetup{justification=centering}
    \caption{Scene-level success rate and runtime measurements for various \moopt{} configurations on the \domtram{} map} 
    \label{fig:eval-rq0}
\end{figure}

\renewcommand\tabularxcolumn[1]{m{#1}}
\newcommand{\tablefigwidth}[0]{0.32\textwidth}
\begin{figure*}[htp]
  \begin{tabularx}{\linewidth}{l@{}c@{}c@{}c}
  & \domcarla & \domzala &  \domtram\\

  \rotatebox{90}{\rquestion{2.1}} &\raisebox{-0.5\totalheight}{\includegraphics[width=\tablefigwidth]{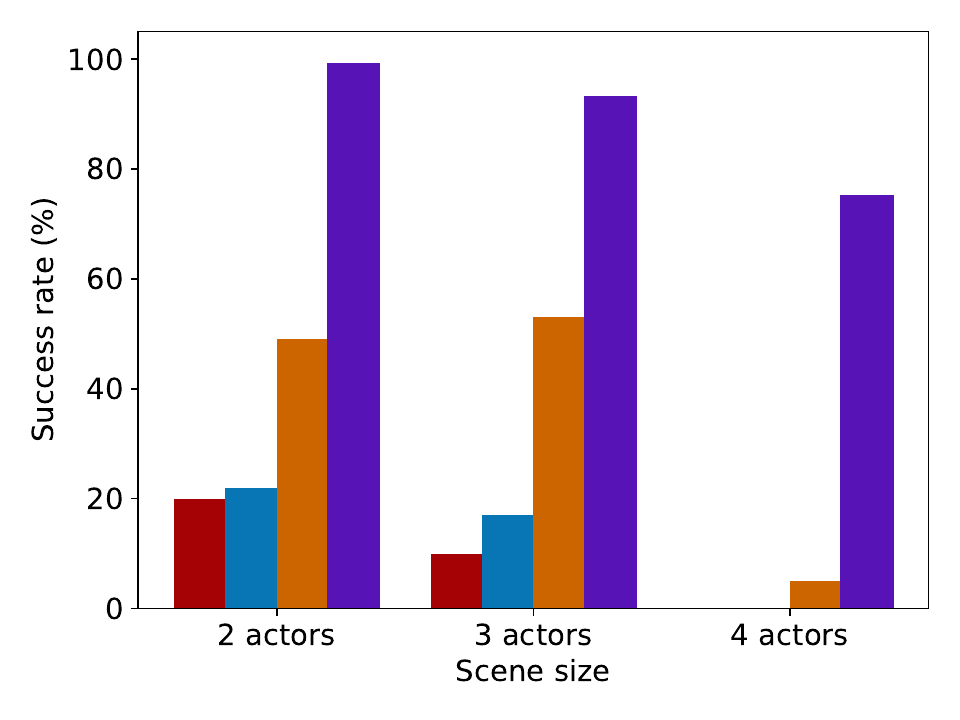}}
  &\raisebox{-0.5\totalheight}{\includegraphics[width=\tablefigwidth]{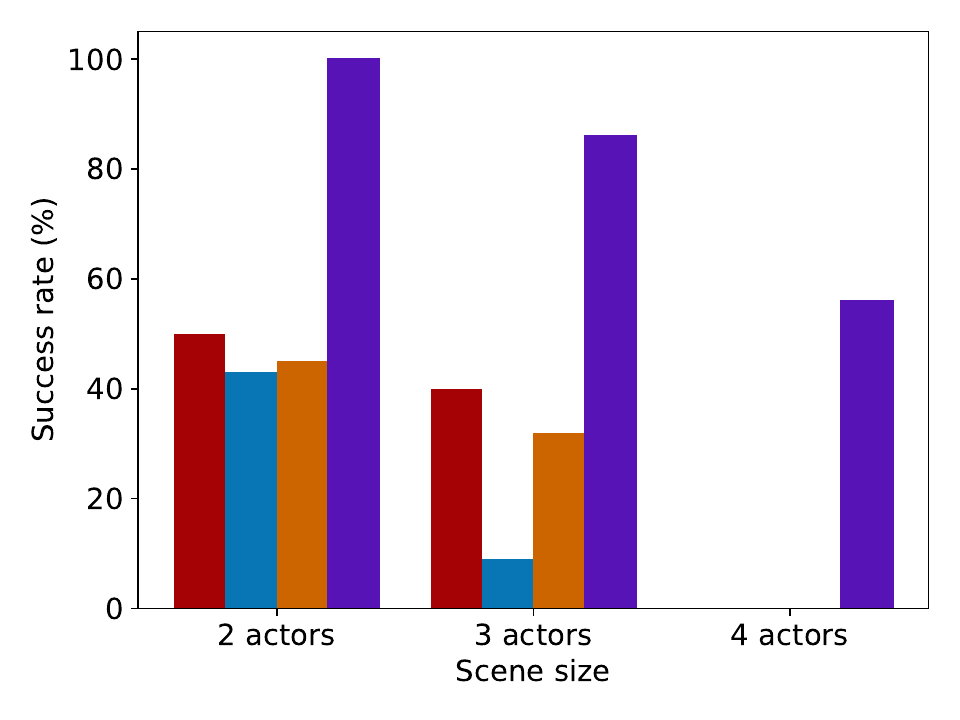}}
  &\raisebox{-0.5\totalheight}{\includegraphics[width=\tablefigwidth]{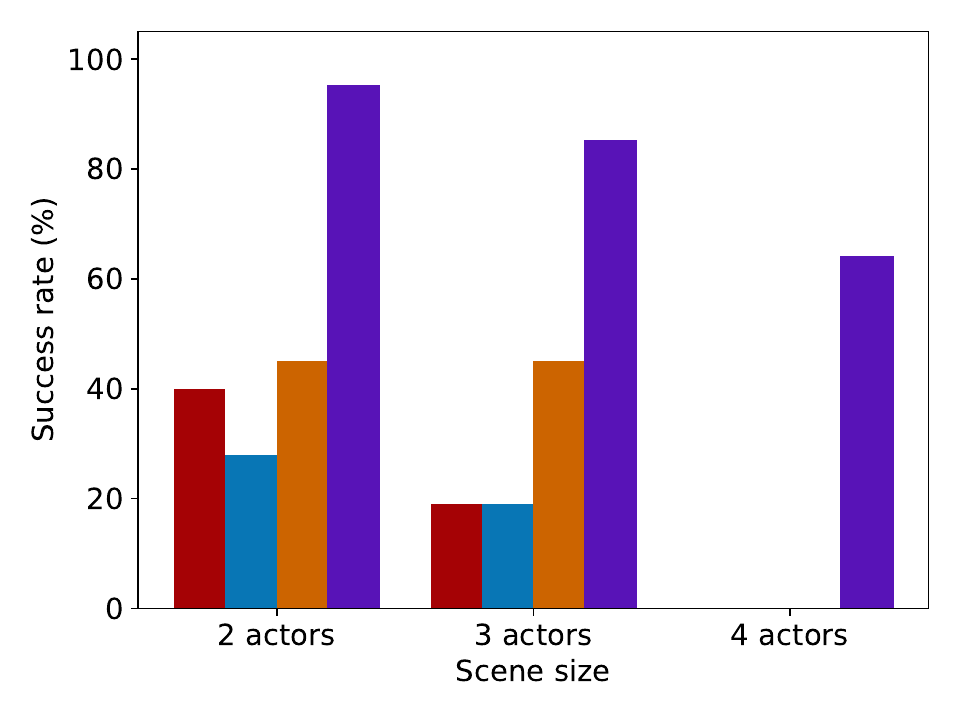}}\\
  
 \rotatebox{90}{\rquestion{2.2}} &\raisebox{-0.5\totalheight}{\includegraphics[width=\tablefigwidth]{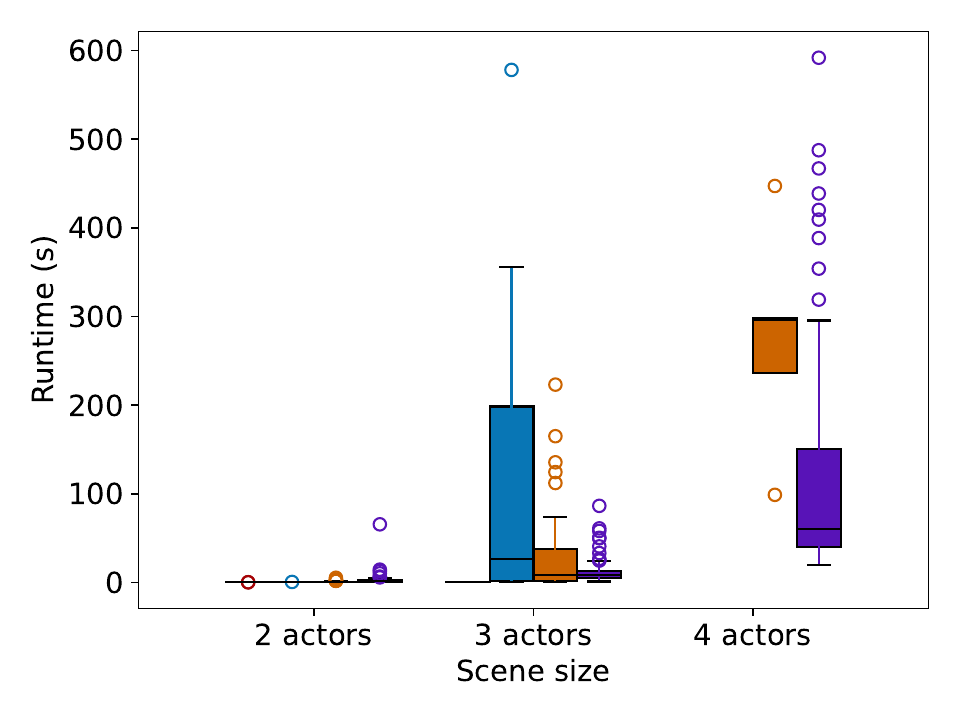}}
  &\raisebox{-0.5\totalheight}{\includegraphics[width=\tablefigwidth]{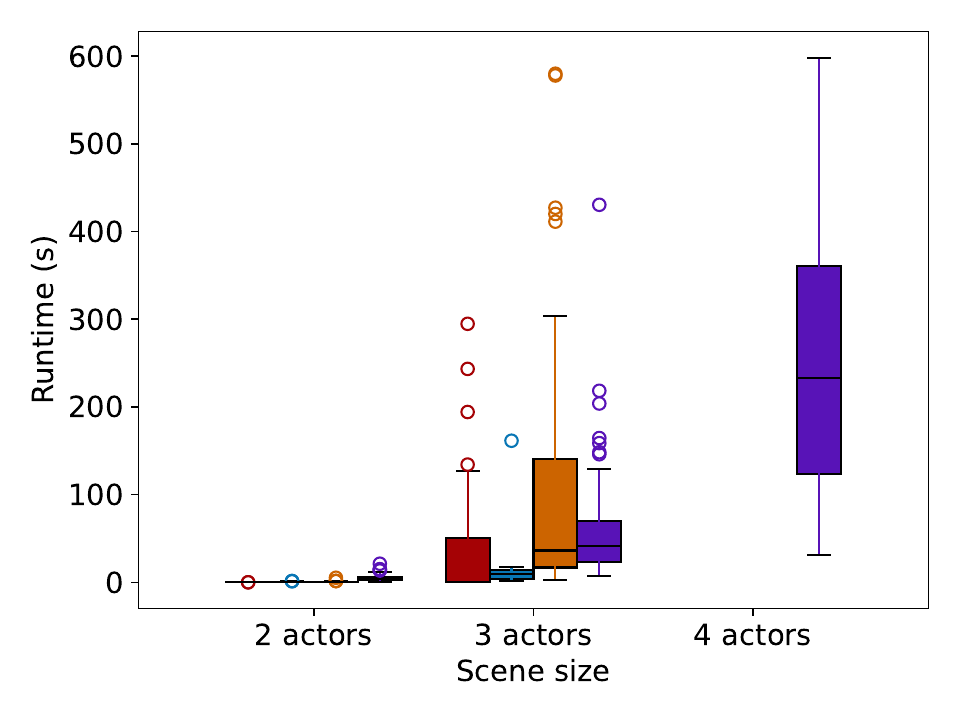}}
  &\raisebox{-0.5\totalheight}{\includegraphics[width=\tablefigwidth]{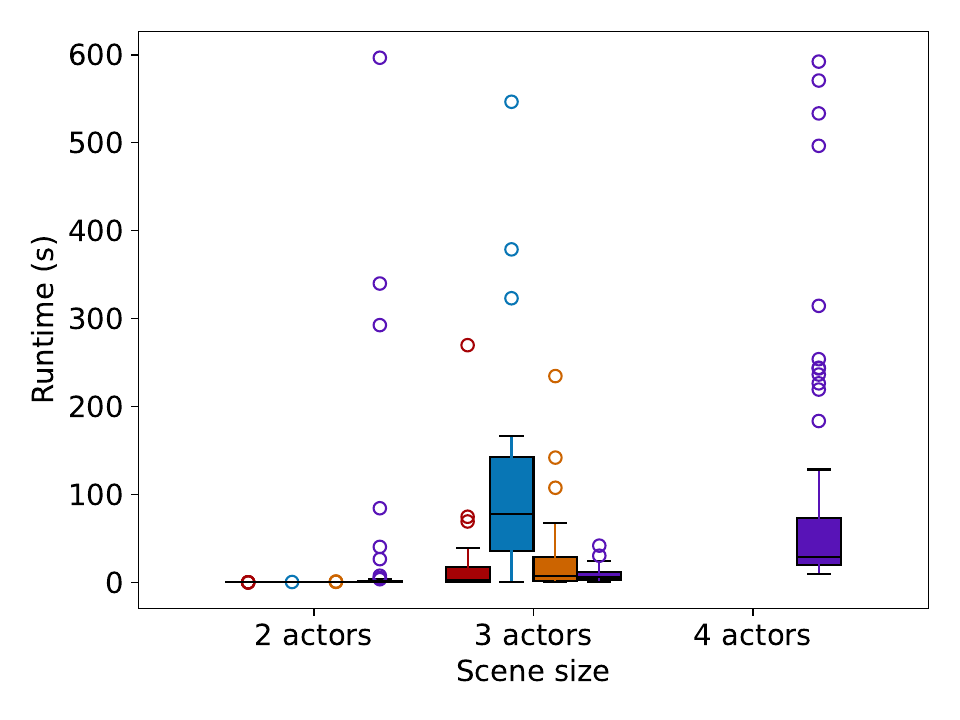}}\\
  
  \midrule
  & 2 actors & 3 actors & 4 actors\\

  \rotatebox{90}{\rquestion{2.3}}
  &\raisebox{-0.5\totalheight}{\includegraphics[width=\tablefigwidth]{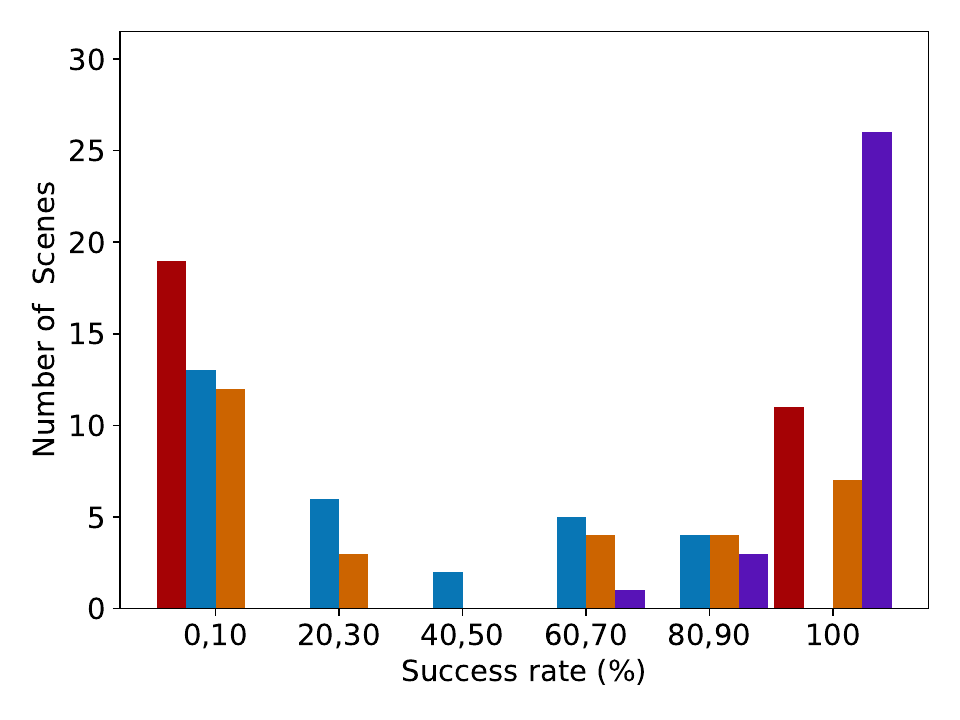}}
  &\raisebox{-0.5\totalheight}{\includegraphics[width=\tablefigwidth]{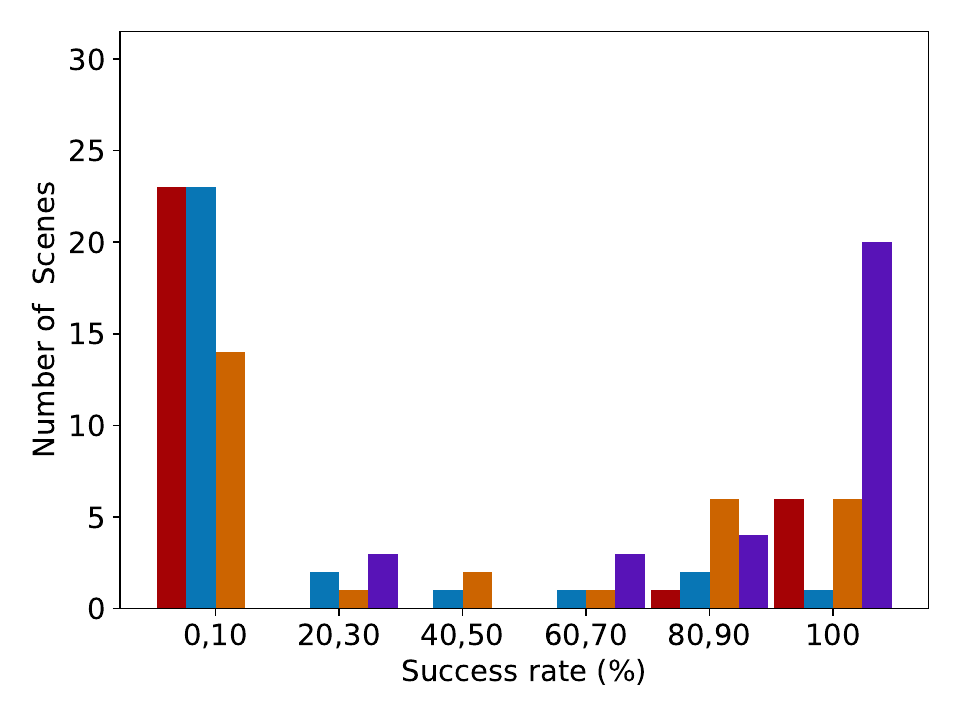}}
  &\raisebox{-0.5\totalheight}{\includegraphics[width=\tablefigwidth]{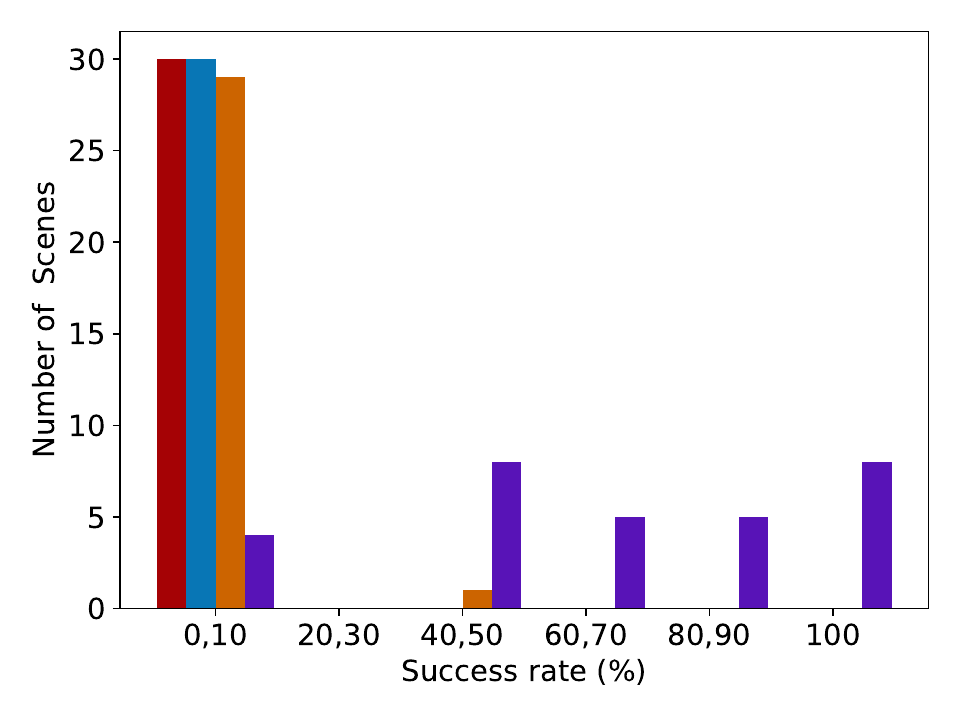}}\\
  \multicolumn{4}{c}{\includegraphics[width=.6\linewidth]{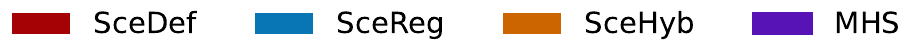}}
 
  \end{tabularx}
  \captionsetup{justification=centering}
  \caption{Measurement data for success rate (\rquestion{2.1}) and  runtime (\rquestion{2.2}) wrt. different maps,\\ and  success rate wrt. increasing number of actors (\rquestion{2.3})}
\label{fig:eval-rq1}
\end{figure*}


    

\subsection{RQ1: Comparison of \moopt{} configurations}
\label{sec: new-RQ1}

\textbf{Measurement setup:}
\textit{This experiment aims to determine which optimization algorithm and objective function aggregation strategy combination provides the best results when implementing our proposed approach.}
We perform measurements over the \domtram{} map (which is shown in \rquestion{2} to be of intermediate difficulty for the evaluated approaches) using scenes with 2, 3 and 4 actors (size) to compare 8 \moopt{} configurations.
We set target scene sizes aligned with the level of scalability offered by many recent publications \cite{Riccio2020,Abdessalem2018TestingVisionBased,Abdessalem2016TestingADAS,Haq2023Reinforcement,Zhong2021Fuzzing,Abdessalem2018TestingFeatureInteractionsBriandNsgaForConcreteScenes,calo2020GeneratingAvoidableCollision,Wu2021,Klischat2019} that handle local traffic constraints over individual actors (such as the ones proposed in our paper). 

Each evaluated configuration has an objective function aggregation strategy and a \moopt{} algorithm selected accordingly.
Single-objective optimization \evolGA{} is used with \aggGlob{}, which yields one objective function.
Multi-objective optimization \evolNTwo{} is used for at most 3 objective functions: \aggAct{}, \aggWeiImp{}.
Many-objective optimization \evolNThree{} is used for more than 3 objective functions: \aggCat{}, \aggWeiCat{} and \aggNOne{}.
Moreover, we evaluate the \evolNTwo{}-\aggWeiCat{} configuration (as initial measurements provided promising results) and the \evolNThree{}-\aggAct{} configuration (as the number of objective functions increases with the number of actors).

For each scene size, we randomly generate 10 \fss{}s as inputs (see \autoref{sec:res-genmeasset}), and run each approach 10 times, for a total of 100 runs.
A 10-minute time-out  is set for each run.

\textbf{Analysis of results:} Success rate and runtime measurements comparing the \moopt{} configurations are provided in \autoref{fig:eval-rq0}.
Each configuration is depicted with a uniquely colored box.
For each configuration, we evaluate scene-level aggregate success rate: we derive a cumulative success rate for each of the 10 \fss{}s (each \fss{} is subject to 10 runs) and depict their distribution.

We determine which configuration is best suited for our proposed approach by evaluating the statistical significance of our results.
For \emph{success rate measurements}, we perform the \textit{Fisher exact test} \cite{Fisher1992FishersExactTest} to determine p-value and measure the \textit{odds ratio} \cite{Haddock1998OddsRatio} for effect size, as suggested in existing guidelines \cite{Arcuri2011} for comparing algorithms with dichotomous outcomes (i.e. success or failure).
Analysis shows that (despite underperforming for 3-actor scenes), the \evolNTwo{}-\aggAct{} configuration provides better success rate with statistical significance ($p < 0.05$) compared to all other configurations except for \evolNTwo{}-\aggWeiCat{} and \evolNThree{}-\aggNOne{}.
However, effect size is not large for our measurements (between 1.845 and 4.378).

For \emph{runtime measurements}, we perform the \textit{Mann-Whitney U-test} \cite{Mann1947MannWhitneyU} to determine \textit{p-value} and measure the Vargha and Delaney's \textit{$\hat{A}_{12}$} \cite{Vargha2000VarghaDelaneyA12} for effect size.
Additionally, following existing guidelines \cite{Arcuri2011}, we only consider the times of successful runs.
Analysis shows that the \evolNTwo{}-\aggAct{} configuration provides better runtime with statistical significance ($p < 0.05$) than the \evolNTwo{}-\aggWeiCat{} and \evolNThree{}-\aggNOne{} configurations for all three scene sizes.
Effect size is medium to large for our measurements (between 0.662 and 0.853).

Considering these results, we identify \evolNTwo{}-\aggAct{} as the best configuration for our experimental setup and use it for comparison with \ascenic{} approaches in \rquestion{2}.
Furthermore, we select \evolNTwo{}-\aggAct{}, \evolNTwo{}-\aggWeiCat{} and \evolNThree{}-\aggCat{} as the top three promising configurations to be evaluated in scalability measurements (\rquestion{3} and \rquestion{4}).
Despite its slightly worse success rate, we select \evolNThree{}-\aggCat{} over \evolNThree{}-\aggNOne{} due to its significantly faster runtimes.

\ranswer{1}{
Out of the 8 evaluated \moopt{} configurations, \evolNTwo{}-\aggAct{} either provides significantly better success rate or better runtime compared to all other configurations.
}

\subsection{RQ2: Comparing \moopt{} with Scenic approaches}
\label{sec: newRQ2}

\textbf{Measurement setup:} 
\textit{This experiment aims to determine which approach completes scene concretization in reasonable time (\rquestion{2.1}, \rquestion{2.2}). Given a particular scene specification as input, we also determine (\rquestion{2.3})  which approach is most likely to succeed in concretizing it.}
We perform measurements over the 3 road maps using scenes with 2, 3 and 4 actors to compare the 4 concretization approaches.
As in \rquestion{1}, we exclude larger scene sizes to enable cross-approach comparisons with higher success rates.
Additionally, as discussed in \rquestion{1}, \ansga{} refers to the \evolNTwo{}-\aggAct{} configuration of our approach.

For each size and map, we randomly generate 10 \fss{}s as inputs (see \autoref{sec:res-genmeasset}), and run each approach 10 times (each with a time-out of 10 minutes), for a total of 100 runs.

\textbf{Analysis of results (\rquestion{2.1}):} 
We compare the overall success rate  wrt. different maps in the top row of \autoref{fig:eval-rq1}.
Each figure contains cumulative success rate results for all four approaches, depicted in the corresponding color (scene-level success rate is addressed in \rquestion{2.3}).
For \ascenic~approaches, a run is considered to be successful if the approach provides a solution to the input partial problem that also satisfies the removed relations (i.e. the provided solution solves the complete problem).

Among the \ascenic~approaches, \asceB~consistently provides relatively high success rates.
Nevertheless, for all maps, and for all scene sizes, the success rate is dominated by \ansga.
Particularly for 4-actor scenes, \ascenic~approaches are generally unable to provide any solutions, while the success rate of \ansga~varies between 56-75\%.


\begin{table}[htp]
    \footnotesize
    \centering
    \captionsetup{justification=centering}
    \caption{Statistical test results comparing success rates: \ansga~vs. the best \ascenic~approach wrt. map and scene size (\textit{p}: \textit{p-value},  \textit{e}: \textit{effect size}). Effect size is \textit{large} for all data points.}
    \setlength{\tabcolsep}{4pt}
    \begin{tabularx}{\linewidth}{|r||c|Z||c|Z||c|Z|}
      \hline
      \multirow{3}{*}{Map} & \multicolumn{6}{c|}{Scene size}\\
      \cline{2-7}
      & \multicolumn{2}{c||}{2 actors} &
       \multicolumn{2}{c||}{3 actors} &
       \multicolumn{2}{c|}{4 actors}\\
      \cline{2-7}
       & \textit{p} & \textit{e} & \textit{p} & \textit{e} & \textit{p} & \textit{e}\\
      \hline
      \domcarla & 5.22e-18 & 103.0 & 1.01e-10 & 11.8 & 2.26e-26 & 57.0\\
      \domzala & 4.45e-19 & $\infty$ & 1.35e-11 & 9.2 & 4.94e-22 & $\infty$\\
      \domtram & 1.37e-15 & 23.2 & 3.55e-09 & 7.0 & 2.33e-26 & $\infty$\\
      \hline
    \end{tabularx}

    \label{fig:eval-p-values}
\end{table}

We evaluate statistical significance of our results according to existing guidelines \cite{Arcuri2011} as detailed in \autoref{sec: new-RQ1}.
For each configuration (map and scene size), we evaluate the statistical difference between \ansga~and the \ascenic~approach with the highest success rate.
The statistical test results are shown in \autoref{fig:eval-p-values}.
\textit{p-values} are lower than 0.05 for all configurations, and the lowest \textit{effect size} is 7.0.
Hence, the success rates are significantly higher (with large effect size) for \ansga~compared to \ascenic~approaches. 
\vspace{-6pt}
\ranswer{2.1}{
For success rates, \ansga~dominates all \ascenic~approaches with statistical significance.
\ascenic~approaches reach their scalability limit at 4 actors with close to 0 success rate, while the success rate of \ansga~ is still 56-75\%.
}

\textbf{Analysis of results (\rquestion{2.2}):}
We compare the runtimes wrt. different maps in the middle row of \autoref{fig:eval-rq1}.
Each figure contains measurement results for scenes with up to 4 actors.
Results for all four of the approaches are depicted as color-coded box plots.


\textit{Is \ascenic~faster than \ansga?}
For scenes with 2 or 3 actors, the \ansga~approach is generally slower than the \ascenic~approaches.
We evaluate the statistical significance of our results according to existing guidelines \cite{Arcuri2011}, see \autoref{sec: new-RQ1}.

For each configuration (map and scene size), we evaluate the statistical difference between the \ansga~approach and each \ascenic~approach (pairwise).
For 2-actor scenes, all pairwise comparisons show a  statistically significant difference ( $p <0.05$) with large effect size ($\hat{A}_{12} > 0.85$) in favor of the \ascenic~approaches.
For 3-actor scenes, there is only a statistically significant difference in favor of \asceA{} for the \domcarla{} and \domzala{} maps and of \asceC{} for the \domzala{} map (with large effect size, $\hat{A}_{12} > 0.7$).

\textit{How much faster is \ascenic?}
To evaluate how much faster \ascenic~approaches are, we multiply the \ascenic~runtimes by a constant factor $c$, then we evaluate the statistical difference between the new runtimes and the \ansga~runtimes (pairwise).
If no statistically significant difference is detected, we conclude that the given \ascenic~approach is at most $c$ times faster than the \ansga~approach.

For 2-actor scenes, our results show that \asceA~is 141 times faster, \asceC~is 19 times faster, and \asceB~is 13 times faster than \ansga.
However, while \ascenic~approaches are often very fast (1-10 milliseconds),
\ansga~also provide results in reasonable time.

For 3-actor scenes, \ascenic~approaches are at most 2.7 times faster than \ansga~approaches.
An exception is for the \asceA~approach applied to the \domcarla~map, where \asceA~is 122 times faster.
However, in this case, there is a significant difference in success rate between the two approaches, favoring \ansga. 

For 4-actor scenes, very few data points exist for \ascenic~approaches as they failed to provide a solution in most runs.
\ansga~data shows that the median runtimes of successful runs are 29.1s for \domtram, 60.0s for \domcarla~and 232.8s for \domzala.
As such, we notice that the \domzala~ map provides the most challenging scene concretization problem for \ansga{} in terms of runtime, while also providing comparable or lower success rates as other maps (see \rquestion{2.1}).
This is attributed to the large map size and complex structure (containing many unusual road segments), which affects the search space for \ansga.

\ranswer{2.2}{
For scenes with 2 or 3 actors, \ascenic~approaches are 1-2 orders of magnitude faster than \ansga, which can still provide results in reasonable time (with better success rates).
For 4-actor scenes, only \ansga~is successful, with a median runtime of at most 232.8s (reported for the \domzala{} map).
}

\textbf{Analysis of results (\rquestion{2.3}):} 
Success rates wrt. increasing number of actors are shown in the last row of \autoref{fig:eval-rq1}.
Each figure \emph{aggregates data for all measurements} (i.e. for all maps and scenes) performed for \emph{a given number of actors}.

For this research question, we evaluate scene-level success rate, as in \rquestion{1}.
Specifically, each bar in these figures represents the number of scenes where the associated success rate corresponds to the x-axis label.
For instance, the bottom-left subfigure of \autoref{fig:eval-rq1} shows that there are 19 (2-actor) scenes where \asceA~provides a success rate of 0\% or 10\% (the leftmost bar).
Similarly, there are 26 scenes where \ansga~provides a success rate of 100\% (the rightmost bar).



For 2-actor and 3-actor scenes, (1) \asceA~provides distributions skewed towards lower and higher success rates, (2) \asceC~and \asceB~provide more uniform distributions, and (3) \ansga~provides a distribution skewed towards higher success rates.
This shows that certain scenes cannot be concretized by \ascenic~approaches, while \ansga~provides at least 20\% success rate for every 2-actor or 3-actor scene.

For 4-actor scenes, \ansga~provides a uniform distribution of success rates, while \ascenic~approaches most often cannot solve the concretization problems.
In fact, \ansga~was able to provide at least 1 solution (i.e. at least 10\% success rate) for \textit{every scenes} with 4 actors.





\ranswer{2.3}{
The \ansga~approach provides at least one solution (i.e. a 10\%+ success rate) for \textit{every input scenes} (i.e. concretization problems) with increasing number of actors.
For an arbitrary practical scene concretization problem, \ansga~is more likely to find a solution than  \ascenic~approaches.
}



\renewcommand\tabularxcolumn[1]{m{#1}}
\newcommand{\tablefigwidthTwo}[0]{0.45\textwidth}
\begin{figure*}[htp]
  \begin{tabularx}{\linewidth}{l@{}c@{}c}
  & Success Rate & Runtime \\
  \midrule

  \rotatebox{90}{\rquestion{3}} 
  &\raisebox{-0.5\totalheight}{\includegraphics[width=\tablefigwidthTwo]{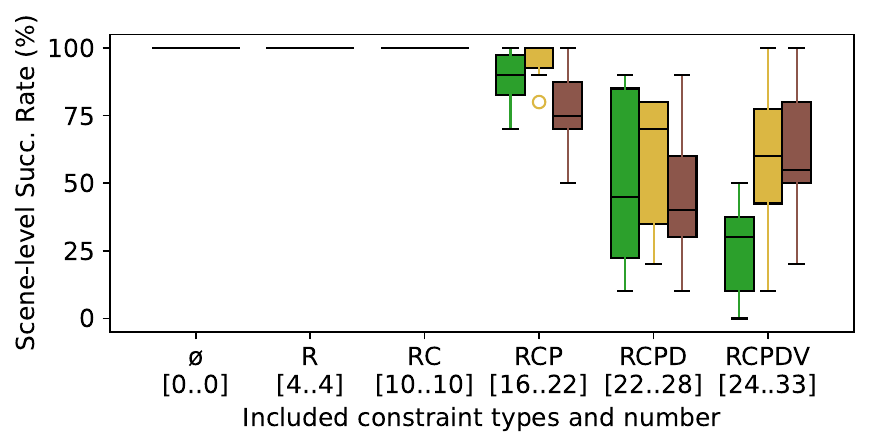}}
  &\raisebox{-0.5\totalheight}{\includegraphics[width=\tablefigwidthTwo]{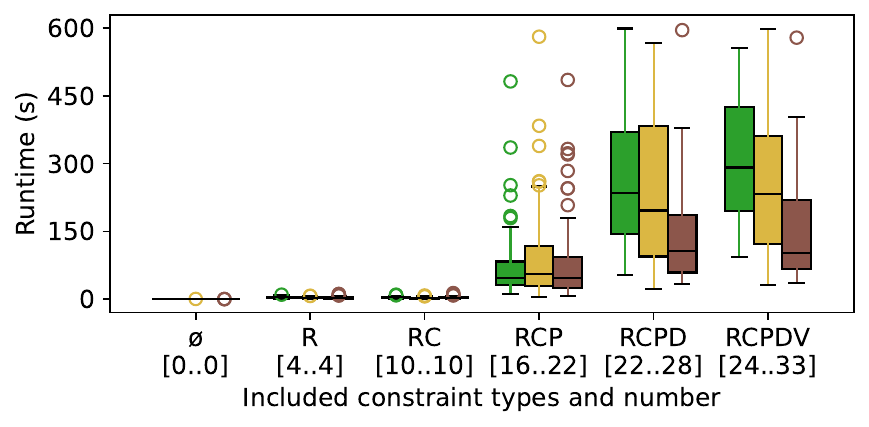}}\\

  \midrule
 \rotatebox{90}{\rquestion{4}} 
  &\raisebox{-0.5\totalheight}{\includegraphics[width=\tablefigwidthTwo]{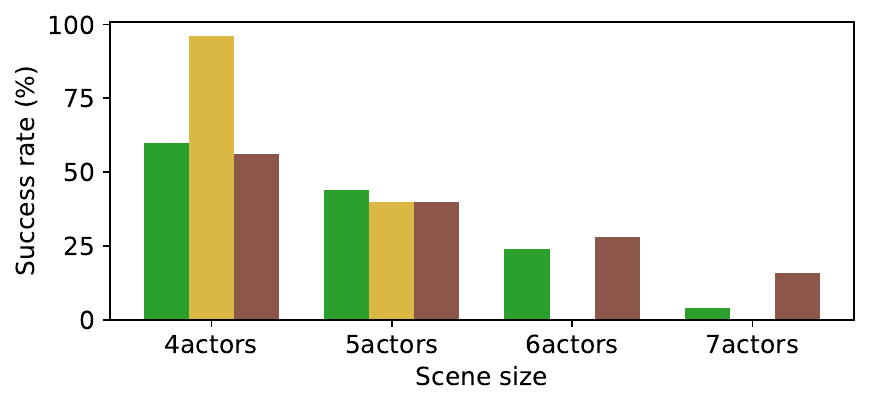}}
  &\raisebox{-0.5\totalheight}{\includegraphics[width=\tablefigwidthTwo]{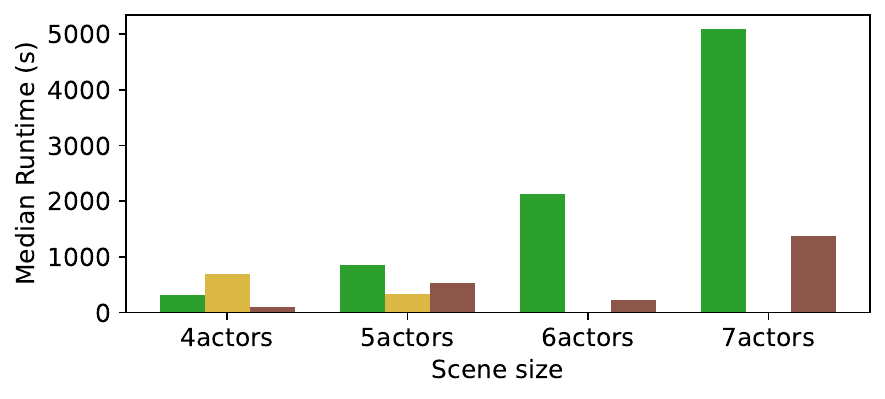}}\\
  
  \midrule
  \multicolumn{3}{c}{\includegraphics[width=.4\linewidth]{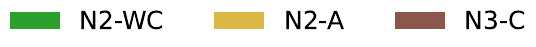}}
 
  \end{tabularx}
  \captionsetup{justification=centering}
  \caption{Scalability results (for \ansga{} runs on the \domzala{} map) wrt. constraints (for 4 actors) (\rquestion{3}) and wrt. number of actors (\rquestion{4}). We report measurement data for success rate and runtime.}
\label{fig:eval-new-rq3-rq4}
\end{figure*}

\subsection{RQ3: Scalability analysis wrt. constraints}
\label{sec: new-RQ3}

\textbf{Measurement setup:} 
\textit{This experiment aims to determine how the inclusion of additional constraints influences runtime and success rate for three promising \ansga{} configuration.
}
For this research question, our measurements are restricted to the most challenging 4-actor \domzala~configuration (see \rquestion{2}).
As discussed in \rquestion{1}, we evaluate the \evolNTwo{}-\aggWeiCat{}, \evolNTwo{}-\aggAct{} and \evolNThree{}-\aggCat{} configurations of our  approach.
Furthermore, since \ascenic~approaches failed to handle scenes with 4 actors, we exclude them from our scalability measurements.

We use the same 10 scenes used for \rquestion{2}, however, we gradually build up the scenes by including all constraints of a certain type one by one and then performing measurements after adding each constraint type.
Specifically, we start with scenes containing no constraints ({\o}), then we gradually add road placement (R), collision avoidance (C), positional (P), distance (D) and visibility (V) constraints until we reach the complete scene specification.

The order of adding constraint types is based on their dependencies: (1) collision avoidance (C) is irrelevant for realistic initial scene generation if vehicles are not placed on roads (R), (2) distance (D) and position (P) cannot be measured if vehicles are overlapping (C), and (3) visibility (V) is based on position (P).
While, for each constraint type, the exact number of added constraints may vary between scenes, adding constraints by type ensures that, for a given scene, the number of constraints gradually increases.

As in \rquestion{1} and \rquestion{2}, we perform 10 iterations per scene with a 10-minute time-out.

\textbf{Analysis of results:}
\definecolor{bl}{HTML}{0020A2}
\definecolor{gr}{HTML}{2F9E00}
Measurement results for \rquestion{3} are shown in the top row of \autoref{fig:eval-new-rq3-rq4}.
As in \rquestion{1}, a distribution of scene-level success rate is depicted.
For \evolNTwo{}-\aggWeiCat{}, the gradual inclusion of constraint types results in a decrease in success rate and an increase in median runtime.
Similar trends are also observed for \evolNThree{}-\aggCat{} and for \evolNTwo{}-\aggAct{}, however, only up to the inclusion of distance (D) constraints.
In all cases, these trends become particularly noticeable  after including positional (P) constraints, which constitute \emph{complex constraints}  involving \emph{multiple vehicles} (unlike e.g. road placement constraints).

For \evolNThree{}{}-\aggCat{}, the further inclusion of visibility (V) constraints results in an \textit{increase} in success rate and a \textit{stagnation} in runtime.
Although these results might seem counter-intuitive, they are in line with the behavior of popular SAT solvers where 
the increasing number of constraints does not necessarily correlate with the increasing complexity of the underlying constraint satisfaction problem \cite{Katebi2011AnatomyOfSAT}.

For \evolNTwo{}-\aggAct{}, the further inclusion of visibility (V) constraints results in a \textit{stagnation} in both success rate and median runtime, which is attributed to the use of \aggAct{} as the objective function aggregation strategy.
\aggCat{} and \aggWeiCat{} both produce a new objective function for each newly added constraint type, which explains the observed variation in success rate and runtime throughout the experiment.
However, \aggAct{} produces a constant number of aggregation functions regardless of the included constraints.
Furthermore, visibility constraints have a similar formalization as \textcolor{blue2}{\qaFront{}()} positional (P) constraints.
As such, according to our random \fss{} generation approach described in \autoref{sec:res-genmeasset}, scenes that contain visibility constraints likely also contain \textcolor{blue2}{\qaFront{}()} constraints.
Considering that \textcolor{blue2}{\qaFront{}()} constraints are already handled with the inclusion of positional (P) constraints, the addition of  visibility constraints should not significantly influence success rate and runtime, as shown in our results.


\ranswer{3}{
Gradual inclusion of constraint types generally results in a performance decrease for the \ansga{} approaches, particularly with the addition of positional constraints.
Furthermore, the choice of objective function aggregation strategy influences scalability results wrt. constraints.
}


\subsection{RQ4: Scalability analysis wrt. actors}
\label{sec: RQ3}

\textbf{Measurement setup:} 
\textit{This experiment aims to determine the maximum size of a scene specification that our approach can  successfully concretize in reasonable time.
}
For this research question, we measure the scalability of the \ansga~approach by introducing more than 4 actors.
As in \rquestion{3}, we evaluate the \evolNTwo{}-\aggWeiCat{}, \evolNTwo{}-\aggAct{} and \evolNThree{}-\aggCat{} configurations on the most challenging \domzala~map.

We perform measurements over 5 randomly generated input scene specifications with increasing size up to the point where the \ansga~approach is predominantly failing (i.e. under 10\% success rate).
We run concretization 5 times for each scene specification, for a total of 25 runs per scene size per \ansga{} configuration.
However, we increase the time-out to 2 hours for each run and measure the success rate and runtime of the concretization runs.


\textbf{Analysis of results:}
\definecolor{bl}{HTML}{0020A2}
\definecolor{gr}{HTML}{2F9E00}
Measurement results for \rquestion{4} are shown in the bottom row of \autoref{fig:eval-new-rq3-rq4}.
Despite a high success rate for 4-actor scenes, the \evolNTwo{}-\aggAct{} configuration is only capable of concretizing scenes with up to 5 actors.
For \evolNTwo{}-\aggWeiCat{}, scalability is limited to 6-actor scenes.
\evolNThree{}-\aggCat{} is the most scalable configuration and can concretize scenes with up to 7 actors (our results show a 16\% success rate for 7-actor scenes).
Additionally, despite the 2-hour time-out, the median runtime for the \evolNThree{}-\aggCat{} configuration for 7-actor scenes is 1375s (23 minutes).


Note that increasing the number of actors by one represents an exponential increase in the overall complexity of the scene concretization problem.
Consider a scene containing $m$ actors defined over a set of $n$ directed binary relation symbols.
The size of the entire search space is estimated (i.e. over-approximated) as ${(2^n)}^{m(m-1)}$, where (i) $2^n$ is the number of possible relation combinations over an ordered pair of actors, and (ii) $m(m-1)$ is the number of ordered actor pairs in the scene.
As such, the search space complexity is $\mathcal{O}(2^{nm^2})$.
In particular, the largest search space handled by \ansga~(for 7 actors) is $2^{360}$ times (over 100 orders of magnitude) 
larger than the 3-actor space handled by \ascenic~approaches.

\ranswer{4}{
The scalability of \ansga~approaches is limited to scenes with 7 actors over the \domzala~map which solves a search space with $2^{420}$ states.
As such, \ansga~can handle an exponentially (over 100 orders of magnitude) larger search space compared to \ascenic~approaches.
}



\subsection{Towards testing vision-based ML components for semantic segmentation}

Although our initial scene concretization approach is a first step towards complete scenario-based testing of AVs, concrete scenes are commonly used for testing various components of AVs, such as cameras and LiDAR sensors.
As a proof of concept, we provide initial results for the testing of the semantic segmentation functionality of a computer-vision component
by integrating our proposed approach with the realistic CARLA simulator \cite{Dosovitskiy2017CarlaSimulator}, which includes a simple AV stack that may be tested.
Additionally, in an ongoing work, we use our scene concretization technique to evaluate three computer-vision components, namely \textit{SegFormer} \cite{Xie2021SegFormer}, \textit{ANN} \cite{Zhu2019ANN} and \textit{BiSeNet V2} \cite{Yu2020BiSeNetV2}.
However, the complete evaluation of these components is outside of the scope of this paper, which primarily focuses on the core scene concretization technique.

\begin{figure}[htp]
     \centering
\captionsetup{justification=centering}
        
        \begin{subfigure}[b]{0.49\columnwidth}
            \centering
            \includegraphics[width=\columnwidth]{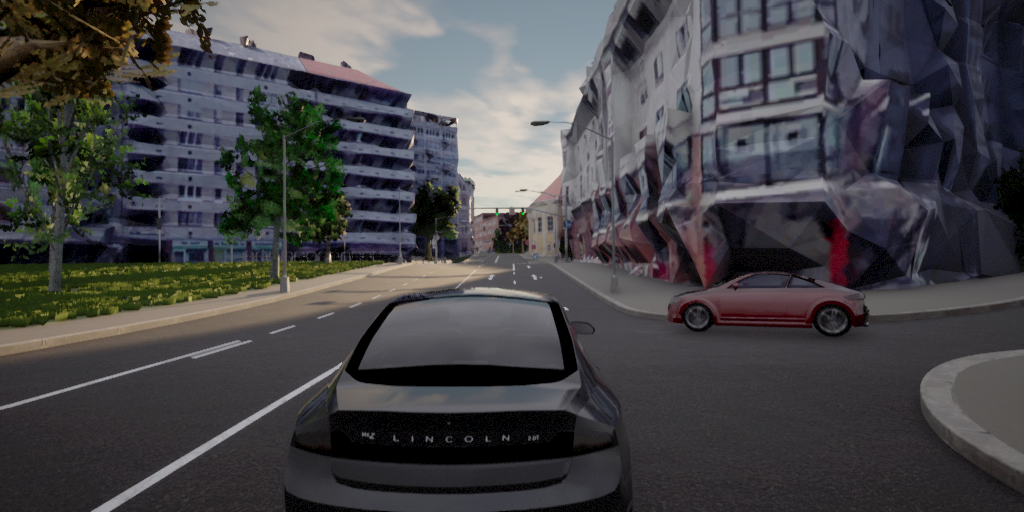}
            \caption[]{RGB image as simulated in CARLA}%
            \label{fig:SimRGB}
        \end{subfigure}
        \begin{subfigure}[b]{0.49\columnwidth}
            \centering
            \includegraphics[width=\columnwidth]{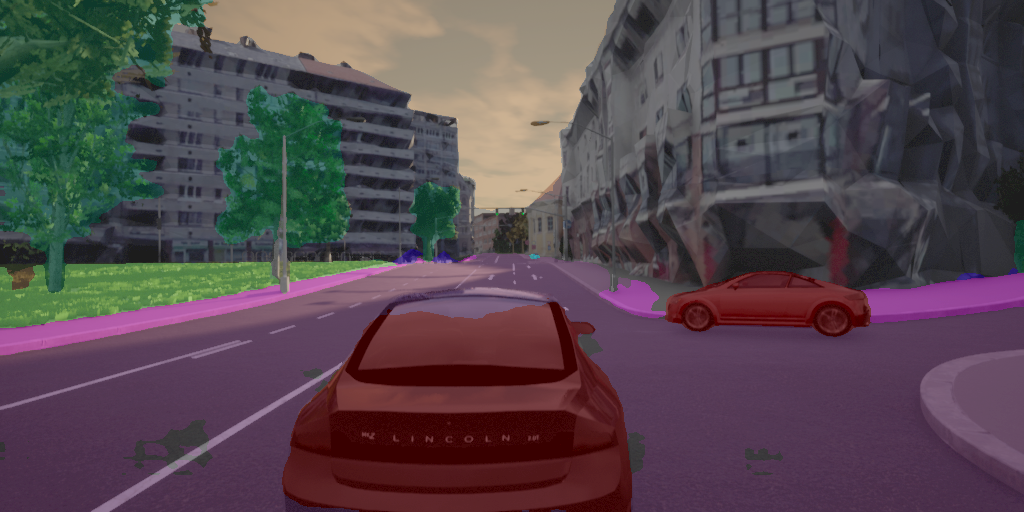}
            \caption[]{Predicted SEG overlayed on \autoref{fig:SimRGB}}      
            \label{fig:SimOv}
        \end{subfigure}
        \begin{subfigure}[b]{0.49\columnwidth}
            \centering
            \includegraphics[width=\columnwidth]{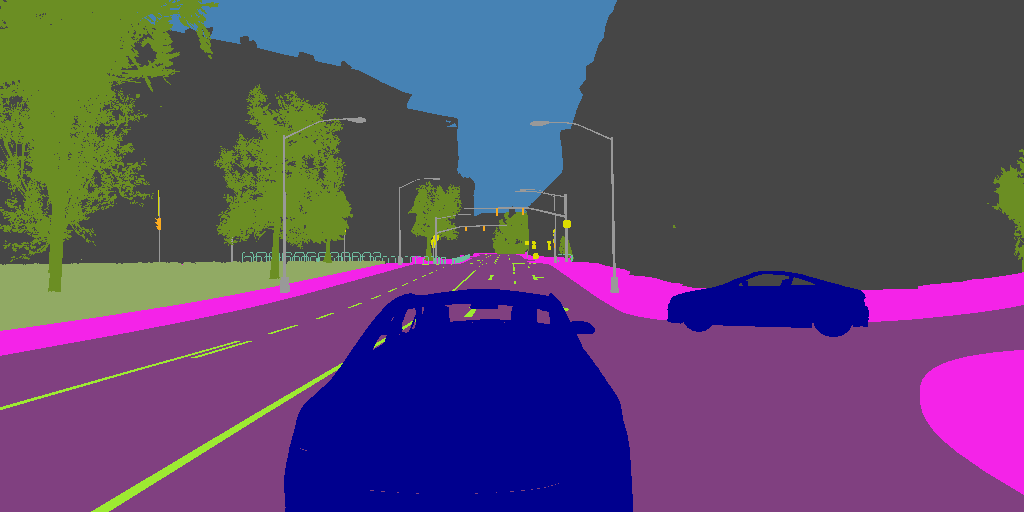}
            \caption[]{Ground truth semantic segmentation}      
            \label{fig:SimGT}
        \end{subfigure}
        \begin{subfigure}[b]{0.49\columnwidth}
            \centering
            \includegraphics[width=\columnwidth]{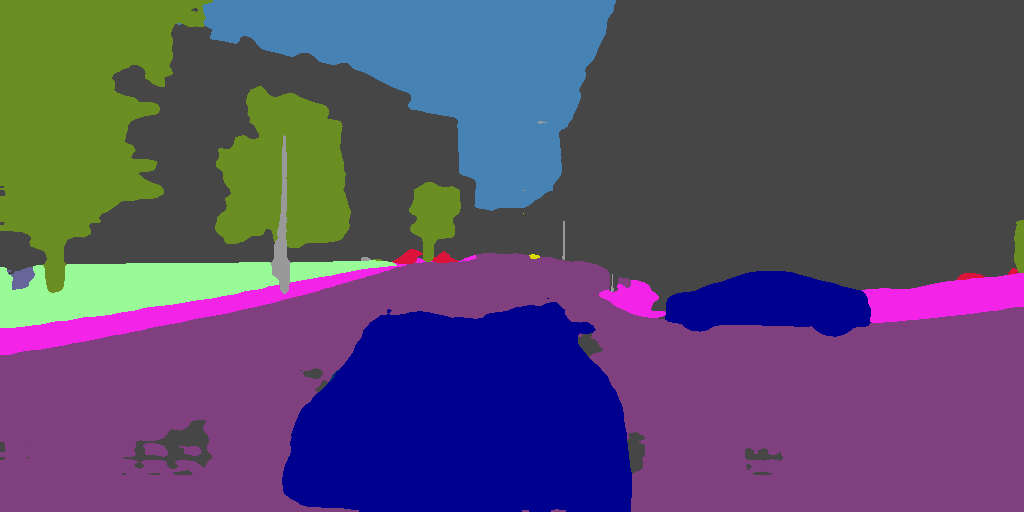}
            \caption[]{Predicted semantic segmentation using \textit{ANN}}      
            \label{fig:SimNN}
        \end{subfigure}
        
        \begin{subfigure}[b]{\columnwidth}
            \footnotesize
            \centering
            \vskip 3pt
            \begin{tabular}{|r||c|c|c|}
              \hline
              Class & Precision & Recall & Intersection-over-union \\
              \hline
              Vehicles & 0.991 & 0.941 & 0.934 \\
              Sidewalks & 0.946 & 0.488 & 0.474 \\
              Traffic Signs & 0.6 & 0.094 & 0.089 \\
              \hline
            \end{tabular}
            \caption{Metrics measurement data}
            \label{fig:SimMet}
        \end{subfigure}
        
        \caption{Artifacts derived from using a generated scene to test the semantic segmentation capabilities of \textit{ANN}
        }
        
        \label{fig:Sim}
\end{figure}

\autoref{fig:Sim} shows an example where a 3-actor scene generated by our approach is used as a test case to evaluate a computer-vision component.
For such scenes, we provide:
\begin{enumerate}
    \item an RGB image of the scene (\autoref{fig:SimRGB}) from the ego vehicle's viewpoint (dashcam footage), 
    \item ground truth semantic segmentation (SEG) of the RGB image as provided by CARLA (\autoref{fig:SimGT}),
    \item predicted SEGs of the RGB image using the three listed computer vision components (a sample SEG using \textit{ANN} is shown in (\autoref{fig:SimNN}),
    \item an overlay of the predicted SEGs over the RGB image (\autoref{fig:SimOv}), and
    \item initial measurement data for \textit{precision}, \textit{recall} and \textit{intersection-over-union} (\textit{IoU}) \cite{Everingham2009IoU} for different classes of objects in the scene (\autoref{fig:SimMet}).
        This measurement data is used to evaluate the performance of the computer-vision component under test.
\end{enumerate}

The preliminary measurements shown in \autoref{fig:Sim} indicate that \textit{ANN} most successfully detects vehicles in the test scene. \textit{ANN} also successfully identifies two of the three sidewalk segments in the image, while it performs poorly in the identification of traffic signs. As such, our preliminary results show that the depicted scene may act as a good test case for traffic sign and sidewalk detection. As future work, we plan to build upon these preliminary results to guide scene generation towards challenging object detection test cases.

In addition to the previously listed artifacts, we also include (online on the publication page):
\begin{enumerate}
  \setcounter{enumi}{5}
    \item a \ascenic~description containing exact positions of each vehicle (for reproducibility), and
    \item a video of the scene running with the default AV stack included in CARLA.
\end{enumerate}

Further initial results (for 30 scenes located on the \domzala~map containing 2, 3 or 4 actors) are also available on the publication page.
These scenes are derived such that all non-ego vehicles can be seen by the ego vehicle, which increases the relevance of our auto-generated scenes for the testing of vision-based ML components (as each object has an impact on the test).


\subsection{Threats to validity}

\textbf{Construct validity.}
Our approach represents a traffic scene as a set of pre-defined abstract relations.
In this paper, we excluded certain relations associated to, e.g., vehicle size and orientation, but our proposed approach may be generalized to include those relations.
Additionally, we use various approximations (i.e. actors are modeled as rectangles) in our definition of numeric constraints to simplify the scene concretization task.
Nevertheless, we provide a complete conceptual basis for scene concretization with an \textit{extensible specification language} and \textit{customizable numeric constraints}.

We compare various \moopt{} algorithms and objective function aggregation strategies to solve the derived \mom~problem.
We selected population size and number of offsprings based on preliminary measurements.
Default values are used for all other parameters.

\textbf{Internal validity.}
To strengthen internal validity, we explicitly call the garbage collector between scene concretization runs.
Additionally, we derive the input functional scenes in such a way to ensure its feasibility on the tested map (to avoid contradicting relations).
Our input scene derivation approach is only limited by an enforced maximum relative distance between actors, which does not restrict the diversity of the derived scenes.

\textbf{External validity.}
We mitigate threats to external validity by performing measurements over 3 maps with different characteristics, derived from different sources (including a real test track and a real world location).
We perform comparative evaluation of up to 8 \moopt{} configurations, composed of 3 \moopt{} algorithms and of 6 objective function aggregation strategies.
We also compare our proposed approach to three variations of the state-of-the-art baseline approach, until their scalability limits are reached.

For \rquestion{1} and \rquestion{2}, we perform a thorough evaluation (100 concretization runs per approach/configuration, per map, per scene size) to adequately compare the approaches and to cover a diverse set of scenes for each setting.
We accompany our measurement with statistical significance analysis in accordance with the best practices.
We also perform thorough evaluation for \rquestion{3} (100 concretization runs per \ansga{} configuration, per set of constraint types).
Our scalability analysis for \rquestion{4} is restricted to \moopt{} approaches (as the baseline \ascenic{} approaches failed to provide solutions in \rquestion{2}). Here, 
we limit our evaluation to 25 concretization runs, over 5 scenes per size, which is still 
sufficient to illustrate the scalability of our \moopt{} approach.

\section{Discussion}
\label{sec:discussion}
In this section, we discuss the practical implications of the proposed approach and we situate it within a longer-term research strategy towards rigorous certification of AVs.

\subsection{Practical implications}

\textbf{Complexity of modeling:}
Despite handling large search spaces with up to $2^{420}$ states, the complexity of modeling within our approach is only internal. In other words, the burden on users of our tool is minimal: users only need to provide a compact model to define functional scene specifications at a high-level of abstraction (\autoref{sec:pre-lang}). 

The large, complex 
internal models are observed only while concretizing scenes and are handled by (third party) \moopt{} algorithms (in \autoref{sec:elab-concretization}). 
Furthermore, the large internal complexity of modeling is a necessary consequence of our highly expressive input language. As a core contribution,
our proposed approach handles a wide breadth of possible input scenarios, which justifies the complexity of the internal models that our approach can handle. 

\textbf{Scalability:}
Our proposed approach reliably generates scenes with up to 4 actors (and even up to 7 actors, with a longer timeout). In the context of state-of-the-art AV testing, this level of scalability seems to be the standard. Our paper clearly shows that our proposed approach dominates Scenic, which has been used in various practical applications. Additionally, many recent publications that handle local traffic constraints over individual actors (such as the ones proposed in our paper) are limited to scenes with, 1 actor \cite{Riccio2020}, 2 actors \cite{Abdessalem2018TestingVisionBased,Abdessalem2016TestingADAS}, 3 actors \cite{Haq2023Reinforcement,Zhong2021Fuzzing,Abdessalem2018TestingFeatureInteractionsBriandNsgaForConcreteScenes,calo2020GeneratingAvoidableCollision,Wu2021} or 4 actors for an intersection testing scenario \cite{Klischat2019}.

Naturally, there may be other practically applicable AV testing approaches that require significantly more actors. For instance, test scenarios on busy highways (also presented in \cite{Klischat2019}), or in bumper-to-bumper traffic. However, these test cases do not require the handling of local constraints over specific vehicles such as the ones proposed in this paper (they may handle global constraints, e.g. traffic density).

\textbf{Customizability and extensibility:}
As defined in \autoref{sec:elab-mapping}, our proposed mapping is \emph{customizable}: mathematical formulae used to represent a functional relation may be adjusted (i.e. its constants may be adjusted) according to use case requirements. For instance, the  visibility angle may vary according to the type of scenarios (e.g. field of view is reduced for high-speed scenarios).

As discussed in \autoref{sec:elab-benefits}, our mapping is \emph{extensible}: experts may define any new (arbitrary) functional relations (by qualitative abstractions), along with the corresponding numeric constraints and fitness functions.
Our framework can automatically apply \moopt{} over the defined fitness functions while evaluating all other fitness functions from the input functional scene specification. 

Our framework also provides support for manual validation of newly added functional relations. For that purpose, a developer may (1) implement the new functional relation (e.g. a \qaVeryFarPred{\actorFun{A}}{\actorFun{B}} distance constraint), (2) create a functional scene specification containing only the newly implemented constraint, (3) run the scene generation, (4) after observing the generated scene, visually validate whether the involved actors are indeed satisfying the newly implemented constraint (i.e. if two actors are indeed very far from each other, which would be easy to observe), and (5) make adjustments if necessary. For further validation, developers may also follow a similar approach to address the negation of the newly implemented constraint. Such validation approaches are common for constraint handling tools such as Alloy \cite{Jackson2002}.

\subsection{Research outlook}
Our scene concretization and two-step scenario generation
approach 
can contribute to a
long-term strategy for rigorous AV certification to ensure AV safety criteria related to all possible (practically relevant) scenarios.



\textbf{Existing research:} 
Given the high-level AV certification challenge presented in \autoref{sec:intro}, 
one way to address this objective is to automatically derive a suite of relevant test scenarios. As an initial step towards
AV certification,
existing approaches
generate tests to investigate a \textit{specific} scenario configuration,
which entails (1) a given target maneuver for the AV-under-test (2) at a given location (3) with a given placement of actors (defined as a set of abstract relations) and a corresponding maneuver assignment. 

For instance, Calo \textit{et al.} \cite{calo2020GeneratingAvoidableCollision} generate scenarios at a 4-way intersection where “the [ego vehicle] is proceeding on its lane and two cars are crossing the main road from left to right”. 
Other approaches \cite{Zhong2021Fuzzing,Abdessalem2018TestingFeatureInteractionsBriandNsgaForConcreteScenes} generate scenarios on a straight road leading up to an intersection which feature an ego vehicle on the straight road, a leading non-ego vehicle and a pedestrian that is crossing the intersection.

The practical benefit of existing approaches is to derive scenarios where the AV is subject to dangerous situations to expose bugs in AV behavior. However, it is unclear how such test scenario generation approaches customized for a given configuration (provided a priori) would generalize (out-of-the-box) to satisfy broader certification criteria that involve other, unrelated scenarios. For instance, it is unclear whether an approach for test generation at intersections \cite{calo2020GeneratingAvoidableCollision}
would be applicable to e.g. overtaking scenarios on a highway without further customization. As such, existing approaches address \textit{a part of the certification objective} mentioned above by \textit{fixing a specific scenario configuration}. 

\textbf{Our contribution:} Rigorous certification of AVs (potentially over arbitrary scenarios) requires an approach that \textit{inherently handles} arbitrary scenarios. It is not scalable to list all possible scenarios and to design customized scenario generation approaches for each of them. Our approach addresses \textit{a different fragment of the certification challenge} by inherently handling \textit{arbitrary scenes} and \textit{arbitrary locations}. 

We believe that the contributions of the paper, particularly those related to handling arbitrary scenes over arbitrary map locations, form a \textit{novel and necessary} conceptual base towards AV certification with \textit{guarantees over the entire space of valid scenarios that can be represented with a given vocabulary of relations} (as opposed to guarantees limited to a given scenario, as offered by existing research). This fits into our \textit{long-term research strategy} that follows the \textit{divide-and-conquer principle to gradually approach rigorous certification}:
first, we address the challenge of handling \textit{arbitrary scenes at arbitrary locations}, and we later focus on the concepts required for adequate handling of arbitrary AV maneuvers.

\section{Related work}
\label{sec:relWork}

\newcommand{\li}[1]{\textit{(#1)}}
First, we overview existing \textit{abstract scenario specification languages} that  describe scenes and scenarios at a functional level (\autoref{tab:comparison-lang}).
We evaluate them with respect to \li{a} \textit{expressiveness} (to represent arbitrary scenarios and constraints), \li{b} \textit{extensibility} with custom traffic concepts, \li{c} handling of \textit{temporal}, e.g. behaviors, \li{d} support for static \textit{error detection}, and \li{e} prior use in scenario \textit{concretization}.

We then evaluate existing (concrete) \textit{scenario generation approaches} (\autoref{tab:comparison-gen}) (as a generalisation of scenario concretization approaches) to check if they can handle  \li{a} \textit{arbitrary abstract scenario specifications} as input (i.e. assumptions about interactions between actors are not hard-coded into the approach), \li{b} \textit{adjustable numeric constraints}, and \li{c} any underlying \textit{road map given as input}.

\subsection{Scenario specification languages}

Existing scenario specification languages often build upon a conceptual basis for traffic scenarios proposed by Ulbrich, \textit{et al.} \cite{Ulbrich2015Defining}, by Menzel, \textit{et al.} \cite{Menzel2018ScenariosForDevelopment} and by Steimle, \textit{et al.} \cite{steimle2021}.
These approaches describe the various components and abstraction levels required for scenario specification.
At a conceptual level, scenarios are often defined through multi-layer representations \cite{Schuldt2018,Bagschik2018,Scholtes2021,Urbieta2021} which provide a hierarchy for traffic scenario components.

\newcommand{\colwidth}{0.8em}
\begin{figure}[tb]
	\centering
	\footnotesize
	\setlength{\tabcolsep}{4pt}
    \begin{subfigure}[b]{\columnwidth}
    \centering
    	\begin{tabular}{ l@{}l c c c c c }
    		&& \rot[40][\colwidth]{Expressive}
    		& \rot[40][\colwidth]{Extensible}
    		& \rot[40][\colwidth]{Temporal}
    		& \rot[40][\colwidth]{Err. det.}
    		& \rot[40][\colwidth]{Concret.} \\
    		\toprule
    		Ont.s \& models &\cite{Geyer2014,Klueck2018,Bagschik2018}
    		& \yes & \yes & \yes & \no   & \no    \\
    		Temporal scen.s &\cite{Queiroz2019GeoScenarioAn,Schutt2020,Hempen2017}
    		& \maybe  & \maybe    & \yes   & \no & \maybe   \\
    		Gener. appr.s &\cite{Abdessalem2018TestingFeatureInteractionsBriandNsgaForConcreteScenes,Majumdar2021ParacsmJournal,Althoff2018}
            & \no   & \no   & \maybe & \no   & \yes   \\
    		\scenic &\cite{Fremont2019ScenicLanguage}
    		& \maybe   & \maybe   & \yes & \no   & \yes   \\
    		\midrule
    		Our approach      &  & \yes   & \yes   & \no & \yes   & \yes   \\
    		\bottomrule
    	\end{tabular}
	\caption{Comparison of \textit{abstract scenario specification languages}}
	\label{tab:comparison-lang}
    \end{subfigure}
    

    \begin{subfigure}[b]{\columnwidth}
    \centering
    	\begin{tabular}{ l@{}l c c c }
    		&& \rot[45][\colwidth]{Arbitrary scen.}
    		& \rot[45][\colwidth]{Adj. num. cons.}
    		& \rot[45][\colwidth]{Map as input} \\      
    		\toprule
    		Search-based &
    		\cite{Abdessalem2018TestingFeatureInteractionsBriandNsgaForConcreteScenes,Babikian2021dReal,Wu2021} &
    		\maybe & \no & \no \\
    		Sampling-based &
    		\cite{Majumdar2021ParacsmJournal,Rocklage2018AutomatedScenarioGeneration,OKelly2018ScalableEndToEnd} &
    		 \maybe  & \maybe    & \no \\
    		Path Planning &
    		\cite{Althoff2018,Klischat2019} &
    		 \no   & \maybe   & \maybe \\
    		\scenic &      
    		\cite{Fremont2019ScenicLanguage} &
    		 \yes   & \maybe   & \yes \\
    		\midrule
    		Our approach &         & \yes   & \yes   & \yes  \\
    		\bottomrule
    	\end{tabular}\hspace{0.5cm}
    \caption{Comparison of \textit{scenario generation approaches}}
	\label{tab:comparison-gen}
    \end{subfigure}
    \captionsetup{justification=centering}
	\caption{Comparison of our approach with the existing state of the art. Notation -- \yes: yes, \maybe: to a certain extent, \no: no. }
	\label{tab:comparison-both}
	
\end{figure}


\textbf{Ontologies and models}:
\textit{Ontologies} \cite{Geyer2014,Klueck2018,Bagschik2018,Urbieta2021} can provide a formal basis for functional scenario specification.
A similar level of formality is provided by \textit{model-based scenario} specification approaches \cite{Bach2016ModelBasedScenario,Gonzalez2018}.
Conceptually, these approaches represent an expressible and extensible specification language, but they are not often used in existing research as inputs for scenario concretization yet.

\textbf{Temporal scenario concepts}:
Other specification approaches use \textit{temporal concepts} as building blocks for scenarios.
Such temporal concepts include (1) reasoning over vehicle paths \cite{Queiroz2019GeoScenarioAn}, (2) sequence of vehicle behaviors \cite{Schutt2020} or (3) conditional state transitions between scenes \cite{Hempen2017}. The expressivity and extensibility of these approaches are limited, 
but initial concretization results are often provided.

\textbf{Input languages for generation approaches}:
Existing scenario generation approaches (detailed in \autoref{sec:rw-concretization}), often define functional scenarios with a custom specification language that may include temporal concepts.
However, such input languages are often tailored to represent a specific type of scenario (e.g. with a single maneuver decided a priori).
As such, they lack in expressiveness and extensibility.
An exception is \scenic~\cite{Fremont2019ScenicLanguage}, which is thoroughly discussed in this paper.
Despite its expressiveness limitations, \scenic~handles arbitrary scenario specifications as input.

\subsection{Scenario generation approaches}
\label{sec:rw-concretization}

\textbf{Search-based approaches} 
are most commonly used for scenario generation.
Many-objective search can be used to test feature interactions in AVs \cite{Abdessalem2018TestingFeatureInteractionsBriandNsgaForConcreteScenes}, to perform efficient \textit{online} testing \cite{Haq2022} and to address the branch coverage of test suite generation approaches \cite{Panichella2015}.
Additionally, Ben Abdessalem, \textit{et al.} rely on multi-objective search \cite{Abdessalem2016TestingADAS}, and learnable evolutionary algorithm \cite{Abdessalem2018TestingVisionBased} to guide scenario generation towards \textit{critical scenarios}.
Critical scenarios have also been derived using a weighted search-based approach \cite{calo2020GeneratingAvoidableCollision} and genetic algorithm \cite{Wu2021}.
Similarly, \textsc{DeepJanus} \cite{Riccio2020} combines evolutionary and novelty search to derive test inputs at the behavioral frontier of AVs, while Babikian, \textit{et al.} \cite{Babikian2021dReal} use a hybrid, graph and numeric solver-based approach to concretize a limited-visibility pedestrian crossing scenario.
Note that different search-based approaches may represent domain-specific constraints differently in the underlying algorithm \cite{Fan2017} (e.g. objectives vs. hard constraints). 
  
Existing search-based approaches are often used to guide scenario generation towards test cases with particular characteristics.
Our approach can complement such existing work as discussed in \autoref{sec: complement}.
Nevertheless, existing search-based approaches as standalone components are limited in various aspects.
Such approaches are often designed for the concretization of a specific scenario type, which is selected upfront and then hard-coded into the search problem.
In particular, numeric constraints are hard-coded according to (1) the specific interactions between actors and (2) the fixed underlying map structure.
While our measurements clearly highlighted that the actual map location has a substantial influence on the de facto complexity of a scenario concretization problem, \emph{all existing search-based approaches are limited to a pre-defined map location}, i.e. the underlying map is not given as input.
Search-based approaches exist for map generation \cite{Gambi2019AsFaultFullPaper}, but no actors are involved in such cases.

\textbf{Sampling-based approaches} have also been used for scenario generation. 
Certain approaches \cite{Majumdar2021ParacsmJournal,Rocklage2018AutomatedScenarioGeneration} sample over a parametric (discrete) representation of arbitrary functional input scenarios, but they often avoid numeric (continuous) constraints (i.e. the logical scenario).
O'Kelly, \textit{et al.} \cite{OKelly2018ScalableEndToEnd} use Monte Carlo sampling (over a continuous domain) to simulate scenarios with rare events, thus reasoning directly at the logical scenario level.
However, all these sampling-based approaches are limited to a fixed map location. 

The \scenic~framework \cite{Fremont2019ScenicLanguage} also provides sampling-based concretization, but it improves on other sampling-based approaches by handling any road map as an input parameter.
Limitations of the input language and of the underlying functional-to-numeric constraint mapping are discussed in \autoref{sec:prel-staticAnal} and in \autoref{sec:elab-benefits}, respectively.

\textbf{Path-planning approaches} \cite{Althoff2018,Klischat2019} address scenario generation directly at the level of numeric constraints.
As such, they cannot handle abstract constraints as input.
These approaches often use formalizations of safety requirements as guiding metrics for scenario generation.
Furthermore, despite the lack of experimental results, such path-planning approaches are in principle adaptable to any road maps.

\section{Conclusions}
\label{sec:conclusion}

In this paper, we proposed a traffic scene concretization approach that leverages metaheuristic search to place vehicles on an arbitrary road map (given as input) such that they satisfy a set of input constraints.
Our approach handles traffic scenes on three different levels of abstractions in compliance with safety assurance best practices for autonomous vehicles.
The input is a functional  scene specification (represented in a novel scene specification language with 4-valued partial model semantics) that captures the scene concretization problem by abstract relations, and enables early detection of inconsistent specifications.
Then, the functional scene specification is mapped to a complex numeric constraint satisfaction problem on the logical level.
Finally, we use metaheuristic search with customizable objective functions and constraint aggregation strategies to solve the numeric problem in order to derive a concrete scene that can be investigated in the popular CARLA simulator \cite{Dosovitskiy2017CarlaSimulator}. 

We carried out a detailed experimental evaluation comparing eight configurations of our proposed approach over three realistic road maps to assess success rate, runtime and scalability.
Our results show that despite higher runtimes, our approach provides significantly better success rate and scalability than the state-of-the-art \scenic~tool,  while traversing a search space with $2^{420}$ states.

(Initial) scene concretization is a subproblem of the complex challenge of scenario-based testing of AVs.
As such, our future work aims to integrate the handling of behaviors and dynamic constraints, potentially representable through temporal logic languages.
Moreover, we plan to address the systematic synthesis of abstract functional scene specifications with certain coverage guarantees over the auto-generated specification suites.



\section*{Acknowledgements}
We would like to thank (1) Krist{\'o}f Marussy for his help in formalizing the logic structures, 
(2) Catriona McIntosh for her help in running preliminary measurements, as well as
(3) Attila Ficsor and Boqi Chen for their work in vision-based ML component testing.

This paper has been partially supported by
the NSERC RGPIN-2022-04357 project,
the NSERC PGSD3-546810-2020 scholarship,
the NRDI Fund based on the charter of bolster issued by the NRDI Office under the auspices of the Ministry for Innovation and Technology,
and by the \'{U}NKP-21-4 New National Excellence Program of the Ministry for Innovation and Technology from the source of the National Research, Development and Innovation Fund, the Wallenberg AI, Autonomous Systems and Software Program (WASP), Sweden, and an Amazon Research Award. 
During the development of the achievements, we took into consideration the goals set by the Balatonfüred System Science Innovation Cluster and the plans of the ``BME Balatonf\"ured Knowledge Center'', supported by EFOP 4.2.1-16-2017-00021.

\bibliographystyle{IEEEtran}
\bibliography{bib/bib-mendeley}



\newpage

\input{insbox}
\newcommand{\bio}[3]{
\InsertBoxL{0}{\includegraphics[height=38mm,keepaspectratio]{#1}}[0]
\noindent\textbf{#2} #3
\vspace{5mm}
}


\bio{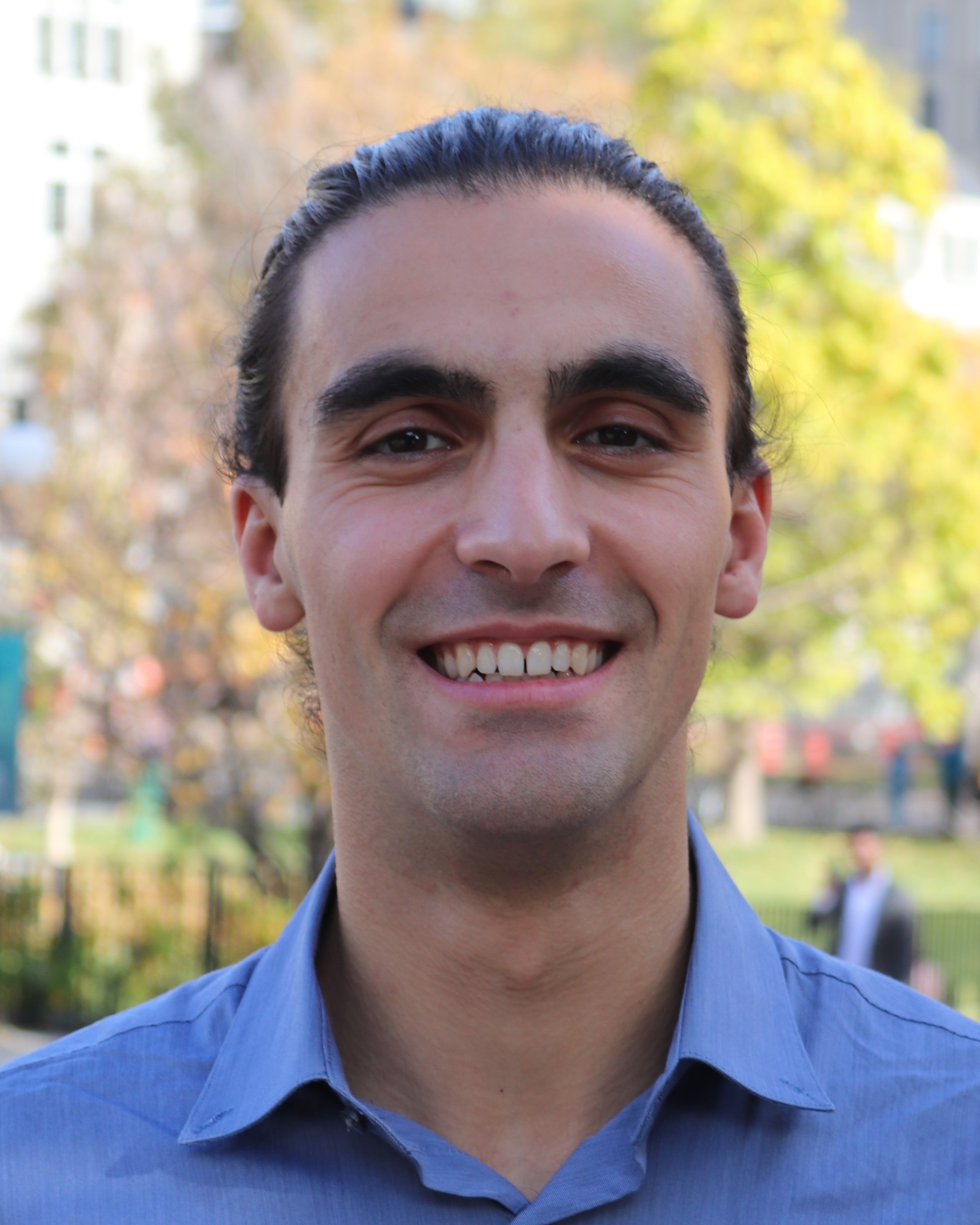}{Aren A. Babikian}{ is a PhD student at the Department of Electrical and Computer Engineering at McGill University. His research focuses on using model generation techniques for the safety assurance of autonomous vehicles. He has published a related research paper at the international FASE 2020 conference and SOSYM 2021 journal. He has also done related internships at NVIDIA and at AWS.}

\bio{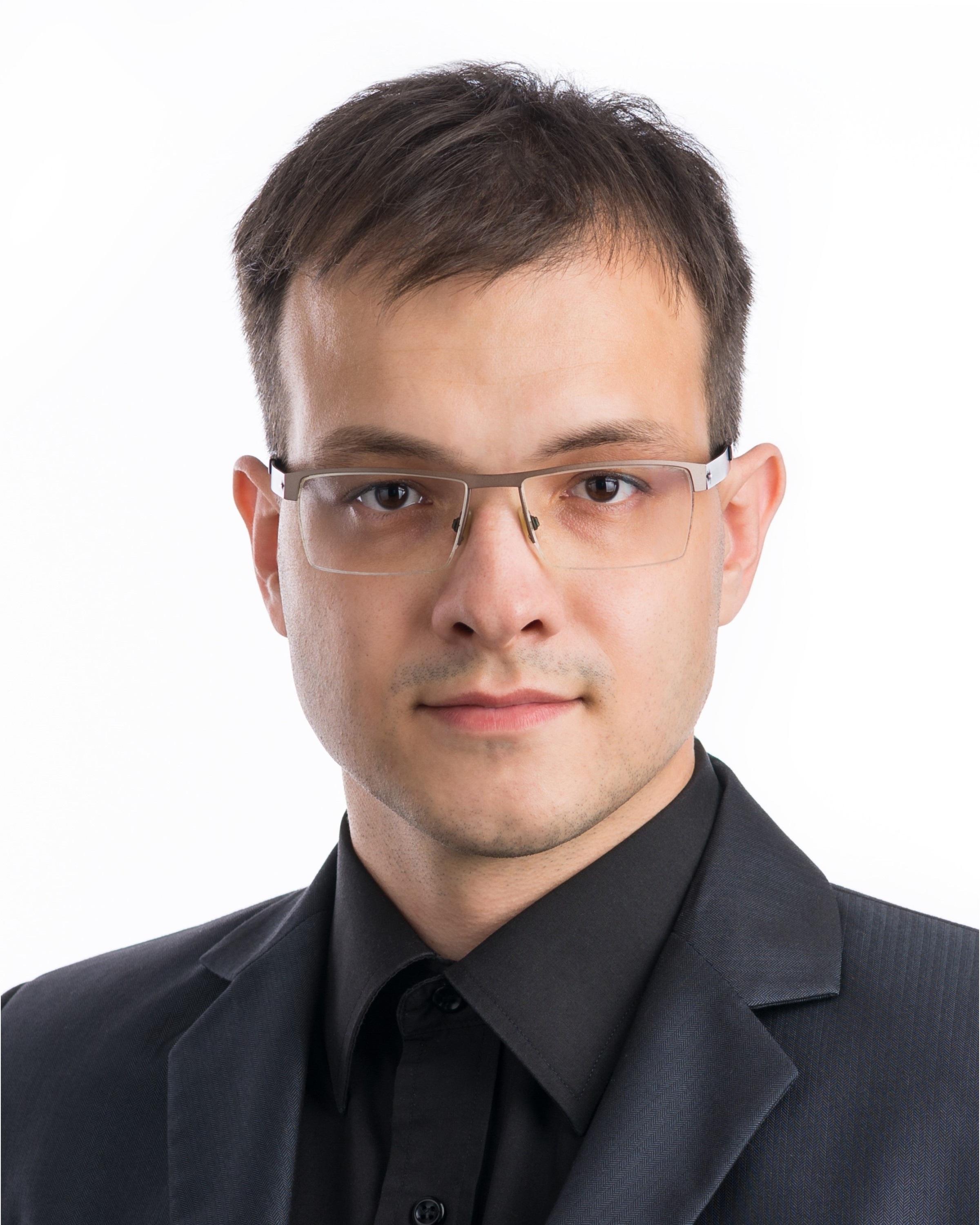}{Oszk\'ar Semer\'ath}{ is an assistant professor at Budapest University of Technology and Economics. His research focuses on modeling technologies and logic solvers.
He won IEEE/ACM best paper award at \mbox{MODELS} conference, Young Researcher Award from the National Academy of Science, John George Kemeny Award from John von Neumann Computer Socienty, Amazon Research Award, and the New National Excellence Program scholarship three times.}

\bio{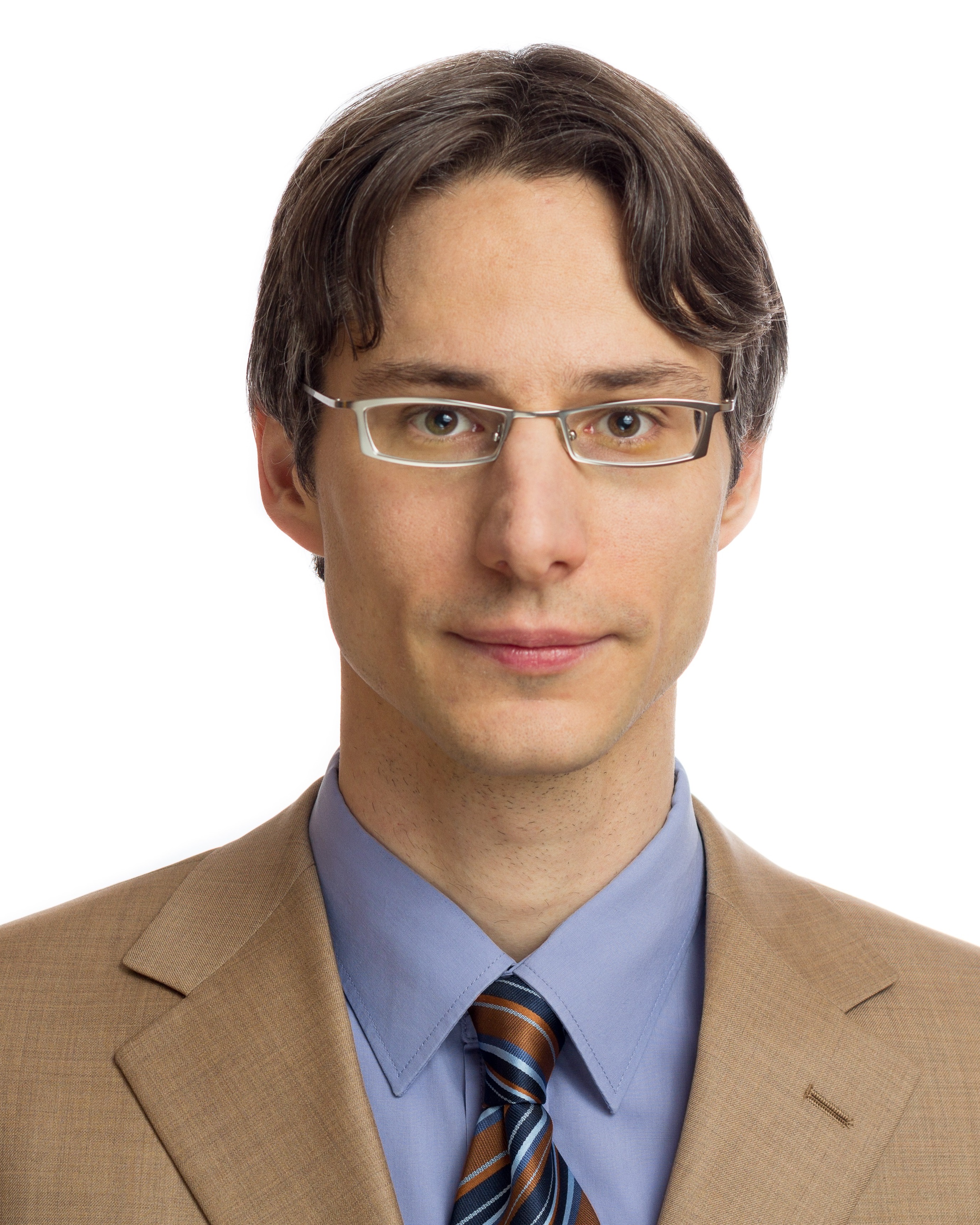}{D\'aniel Varr\'o}{Daniel Varro is a full professor at Linköping University and an adjunct professor at McGill University and at BME.  He is a co-author of more than 180 scientific papers with seven Distinguished Paper Awards, and three Most Influential Paper Awards. He serves on the editorial board of Software and Systems Modeling, and served as a program co-chair of MODELS 2021, SLE 2016, ICMT 2014, FASE 2013 conferences. 
He is a co-founder of the VIATRA open-source model transformation framework, and IncQuery Labs, a technology-intensive company.}


\clearpage
\appendix

\label{appendix}

\noindent\textbf{Soundness of a numeric solution.}

We demonstrate the soundness of a numeric solution by showing that:

\theoremA{}
where $N = \langle \actorSetI{N}, \constrSetI{N}, \roadmapI{N}, \dimensionsI{N} \rangle$, $P = \logicModelTwo{\dslObjectSetI{P}}{\interpretationFun{P}}$ and $s_N: \actorSetI{N} \rightarrow \mathbb{R}^5$ is formalized as in \autoref{sec:elab-num-formal}.

A key component of our soundness proof is the partial model refinement operation, denoted by $P \sqsubseteq P_{s}$, as defined in \cite{Marussy2020JOT}.
Informally, this means that (1) all \unk~relations in $P$ are mapped to a \true~or \false~relation in $P_{s}$, and (2) no other relations are modified.

\autoref{fig:square} provides a visual representation of our proof:
\begin{figure}[htp]
    \begin{center}
    \captionsetup{justification=centering}
      \includegraphics[width=0.3\linewidth]{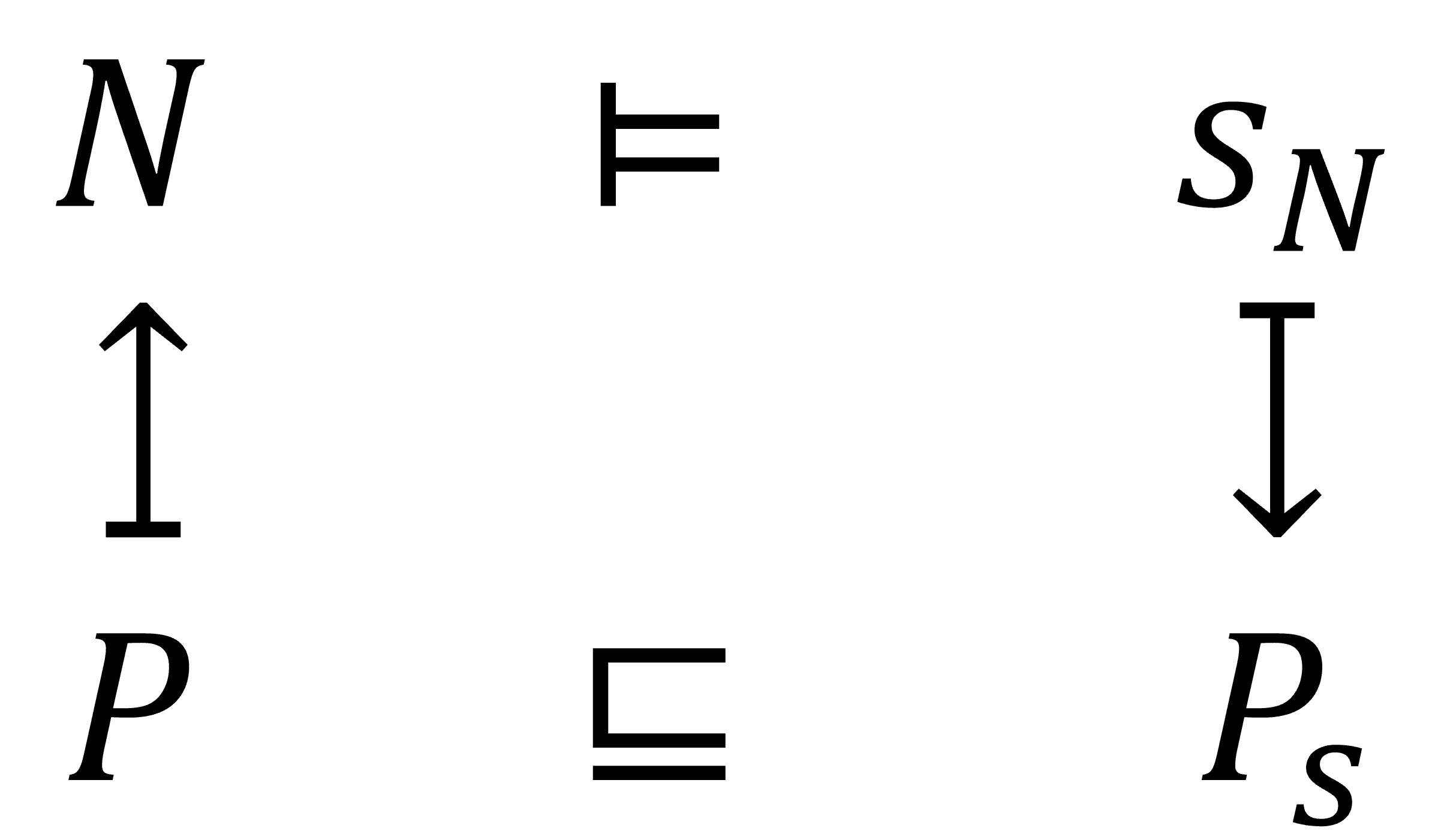}
      \caption{A solution $s_N$ of a numeric problem $N = \funtolog(P)$ corresponds to a refinement $P_{s}$ of $P$.}
      \label{fig:square}
    \end{center}
\end{figure}


\begin{proof}

\begin{enumerate}

    \item By construction, $P$ maps to $N$ such that each positive or negative relation in $P$ maps to a corresponding numeric constraint in $\constrSetI{N}$.
    
    \item By definition, $s_N$ is a solution to $N$ iff it satisfies all numeric constraints in $N$.
    This means that it also satisfied the corresponding relations in $P$ (i.e. all positive and negative relation in $P$)
    
    \item Given the numeric solution $s_N$, a corresponding partial model $P_{s}$ is derived by
    (1) enumerating abstract relations $r_i(\actorFunNoMath{a}, \actorFunNoMath{b})$ from every possible abstract relation symbol $r_i \in \Sigma$ over every pair of actors $\actorFunNoMath{a}, \actorFunNoMath{b} \in \dslObjectSetI{P}$,
    (2) deriving a numeric constraint $c_i$ for each abstract relation $r_i(\actorFunNoMath{a}, \actorFunNoMath{b})$, and 
    (3) evaluating $c_i$ over $s_N$.
    If $s_N$ satisfies $c_i$, then $r_i(\actorFunNoMath{a}, \actorFunNoMath{b})$ holds in $P_{s}$ (formally, $\interpretationFun{P_{s}}(r_i)(\actorFunNoMath{a}, \actorFunNoMath{b}) = \true$).
    Otherwise, $\interpretationFun{P_{s}}(r_i)(\actorFunNoMath{a}, \actorFunNoMath{b}) = \false$.
    
    \item
    (1) Considering that $s_N$ satisfies all positive and negative relation in $P$, then their truth value is unchanged in $P_s$.
    Furthermore, (2) all \unk~relations in $P$ may or may not hold in $P_{s}$ (i.e. the corresponding numeric constraint may or may not be satisfied in $s_N$).
    In any case, these relations are refined either to a \true~or to a \false~value in $P_{s}$.
    This shows that $P \sqsubseteq P_{s}$ holds for any numeric problem $N$ and solution $s_N$.
    
\end{enumerate}

\end{proof}

\noindent\textbf{Soundness of a \moopt~solution.}

We demonstrate the soundness of a \moopt~solution by showing that:

\theoremB{}
where \(M_{min} = \langle \moovarsI{M}, \mooobjsI{M} \rangle\), $N = \langle \actorSetI{N}, \constrSetI{N}, \roadmapI{N}, \dimensionsI{N} \rangle$, and $s_{min}$ is a value assignment (within the specified range) for each $v_i \in \moovarsI{M}$.

\begin{proof}


A solution $s_{min}$ to $M_{min}$ is defined such that all objective functions $OF_i \in \mooobjsI{M}$ applied over $s_{min}$ evaluate to a value below a threshold $\epsilon > 0$.
Note that $OF_i$ is constructed as a weight function applied over the sum of a set $D$ of non-negative distance functions.
Considering that a weight function returns non-negative values and does not modify the minima, we can deduce that $OF_i < \epsilon$ holds iff all the distance functions in $D$ return a value below a threshold $\epsilon' > 0$.

By definition, distance functions explicitly return $0$ iff the corresponding numeric constraint $c_i \in \constrSetI{N}$ holds for the candidate solution it is measured on.
Since all distance functions applied to $s_{min}$ return 0 (with precision $\epsilon'$), then all corresponding constraints (i.e. all constraints in $\constrSetI{N}$) are satisfied by the numeric solution $s_N$ represented by $s_{min}$.

\end{proof}

\end{document}